\documentclass[useAMS,usenatbib]{mn2e}

\usepackage{graphics}
\usepackage{graphicx}
\usepackage{amsxtra}
\usepackage{journals}
\usepackage{url}
\usepackage{latexsym}
\usepackage{amssymb}

\usepackage{times}
\usepackage{amsmath}
\usepackage{subfigure}
\usepackage{natbib}
\usepackage[colorlinks=true, linkcolor=blue, citecolor=blue, urlcolor=blue]{hyperref}
\usepackage{txfonts}

\usepackage{ulem}

\usepackage{color}

\newcommand{\ser}{S\'ersic}

\begin{document}
\title
[Galaxies with warps]
{Galaxies with conspicuous optical warps
\thanks{E-mail: v.reshetnikov@spbu.ru}
\thanks{Partly based on observations obtained with the 6-m telescope of the
Special Astrophysical Observatory of the Russian Academy of
Sciences.}
}
\author[V. Reshetnikov et al.]
{
Vladimir P. Reshetnikov,$^1$ Aleksandr V. Mosenkov,$^{2,1,3}$ Alexei V. Moiseev,$^{4,5}$
\newauthor
Sergey S. Kotov,$^1$
and Sergey S. Savchenko$^1$
\\ \\
$^1$St.Petersburg State University, 7/9 Universitetskaya nab., St.Petersburg, 199034 Russia\\
$^2$Sterrenkundig Observatorium, Universiteit Gent, Krijgslaan 281, B-9000 Gent, Belgium\\
$^3$Central Astronomical Observatory, Russian Academy of Sciences, 65/1 Pulkovskoye chaussee, St. Petersburg, 
196140 Russia\\
$^4$Special Astrophysical Observatory, Russian Academy of Sciences, Nizhnii Arkhyz, 
Karachaevo-Cherkesskaya Republic, 369167 Russia \\
$^5$ Sternberg Astronomical Institute, Moscow M.V. Lomonosov State University, Universitetskij pr., 13,  
Moscow, 119992, Russia
}

\date{\today}

\pagerange{\pageref{firstpage}--\pageref{lastpage}} \pubyear{2016}

\maketitle

\label{firstpage}

\begin{abstract}

In this paper, we present results of a photometric and kinematic study for a sample of
13 edge-on spiral galaxies with pronounced integral-shape warps of their stellar discs. The global structure of the galaxies is analyzed on the basis of the Sloan Digital Sky Survey (SDSS) imaging, in the $g$, $r$ and $i$ passbands. Spectroscopic observations are obtained with the 6-m Special Astrophysical Observatory telescope.
In general, galaxies of the sample are typical bright spiral galaxies satisfying
the Tully-Fisher relation. Most of the galaxies reside in dense spatial environments
and, therefore, tidal encounters are the most probable mechanism for generating 
their stellar warps. We carried out a detailed analysis of the galaxies and their
warps and obtained the following main results: 
(i) maximum angles of stellar warps in our sample are about 20$^{\rm o}$; 
(ii) warps start, on average, between 2 and 3 exponential scale lengths of a disc;
(iii) stronger warps start closer to the center, weak warps start farther;
(iv) warps are asymmetric, with the typical degree of asymmetry of about several 
degrees (warp angle); 
(v) massive dark halo is likely to preclude the formation of strong and asymmetric warps.

\end{abstract}


\section{Introduction}

According to the prevailing point of view, disc galaxies are usually highly 
thin and flat. This is true, but only to a certain extent. 
Peripheral parts of galactic discs often exhibit deviations from the
united plane and demonstrate global warps. This phenomenon has been 
revealed in the neutral gas component, through HI observations 
(e.g. \citealp{san1976}, \citealp{bosma1981}, \citealp{briggs1990}, 
\citealt{gsk2002}), and in a lesser extent through optical and infrared 
observations (e.g. \citealp{ss1990}, \citealp{grijs1997}, 
\citealp{rc1998, rc1999}, \citealp{schd2001},
\citealp{ss2003}, \citealp{ap2006}, \citealp{saha2009}, \citealp{gui2010}).

Typically, warps appear in the outskirts of optical discs, and careful investigation 
of edge-on galaxies have revealed that a significant fraction 
(e.g. $\approx$50\% -- \citealp{ss1990}, $\approx$40\% -- \citealp{rc1998},
$\approx$50\% -- \citealp{ap2006}) 
of stellar discs shows integral-shape with typical amplitudes of a few degrees. This high
percentage of observed warps would suggest that the majority of galaxies are warped, 
since the projection effects should hide large fraction of warps, whose line of 
nodes is perpendicular to the line of sight. Optical warps were also common in the past, 
with even greater amplitudes at $z \sim 1$ (\citealp{rbcj2002}).

Several theoretical mechanisms have been proposed to explain the formation and 
maintenance of warped discs (e.g. \citealp{bin1992}, \citealp{kgr2001}, \citealp{sell2013}
and references therein). Among the proposed scenarios,  
discrete modes of bending in a self-gravitating disc (\citealp{toomre1983}, 
\citealp{spcas1988}), misaligned dark halos (\citealp{dubk1995}),
galaxy interactions and accretion of satellites (e.g. \citealp{hucar1997},
\citealp{schd2001}, \citealp{kim2014}), 
direct accretion of intergalactic matter in the outskirts of galaxies 
(e.g. \citealp{repf2001}, \citealp{vdk2007}, \citealp{ros2010}), 
extragalactic magnetic fields (\citealp{bat1990}), and others.
This large variety of proposed mechanisms and their modifications probably
indicates that there is no single mechanism responsible for all observable warps in 
galaxies. The current situation looks like the largest warps are mostly caused by tidal 
distortions (\citealp{schd2001}, \citealp{ap2006}), whereas relatively small
warps are triggered and supported by a variety of mechanisms.

Most previous studies of optical warps have been devoted to receiving a quantitative 
description of this phenomenon by collecting statistics for their properties and frequency depending on galaxy morphology, while thorough investigations of warped discs for 
individual galaxies were quite rare. (This is partly explained by 
the weakness of the phenomenon: warps are usually seen in the periphery of stellar
discs, and their amplitudes are often small.) Some representative examples of galaxies with the
detailed studies of their optical warps are M~33 \citep{san1980}, NGC~5907 
(warp angle $\psi \approx 4^{\rm o}$ -- \citealp{sasaki1987}), 
Mrk~176 ($\psi \approx 19^{\rm o}$ -- \citealp{resh1989}), NGC~4013, NGC~4565, 
NGC~6504 (\citealp{flo1991}), MCG~06-30-005 
($\psi \approx 15^{\rm o}$ -- \citealp{kemp1993}), 
UGC~3697 ($\psi \approx 22^{\rm o}$ -- \citealp{ann2007}).
Stellar discs of the Milky Way (\citealp{reed1996}), LMC (\citealp{os2002}), and 
the Andromeda galaxy (\citealp{inn1982}) were found to be also warped.

The main goal of this paper is to perform a detailed photometric and kinematic 
study of thirteen edge-on spiral galaxies with integral-shaped stellar discs in 
order to derive observational characteristics of these galaxies and their warps. 
This information is very important for understanding the mechanisms responsible for the generation 
and maintenance of these prominent galaxy features. Until now, a detailed analysis of warped galaxies has been carried out for a few objects. In this paper, we are about to make a new contribution in this area.

This paper is organized as follows. In the next section we present our sample. In Section~\ref{Preparation}, we introduce our decomposition technique, as well as the preparation and fitting of optical galaxy images. In Section~\ref{spectro}, we describe our own spectroscopic observations. The results of our investigation are presented in Section~\ref{Results}. We summarize our main findings and conclusions in Section~\ref{Conclusions}. 

Throughout this article, we adopt a standard flat $\Lambda$CDM
cosmology with $\Omega_m$=0.3, $\Omega_{\Lambda}$=0.7, $H_0$=70 km\,s$^{-1}$\,Mpc$^{-1}$.

\section[]{The sample}
\label{s_samples}

We selected our sample of galaxies based on the view of their non-flat stellar 
discs in the Sloan Digital Sky Survey (SDSS, \citealp{2015ApJS..219...12A}). 
Most of them were visually selected during the creation of the  Sloan-based 
Polar Ring Catalogue \citep[SPRC][]{2011MNRAS.418..244M} and the catalogue 
of Edge-on disc Galaxies In SDSS \citep[EGIS][]{biz2014}.
Our sample contains 13 objects and it is biased to galaxies with strong optical warps.

Selected galaxies belong to different environments --
from group members to relatively isolated objects (this information is taken 
from the NASA/IPAC Extragalactic Database, NED\footnote{http://ned.ipac.caltech.edu/} 
hereafter). An overview of the basic 
properties of the sample galaxies is given in Table~\ref{Table1}. 
Fig.~\ref{SDSS_images} 
shows SDSS thumbnail images of the galaxies. Some brief notes on each object are 
given below.

{\bf IC~194} (UGC~1542, RFGC~439, PGC~7812) is seen almost exactly edge-on. 
Optical disc warping is not strong: edge-on disc tips have a barely detectable 
$S$-shape. This galaxy has no nearby bright companions, but a loose
group of galaxies PPS2~136 with the close redshift is located at a projected 
distance of 170~kpc from IC~194 (\citealp{tb1998}). 

{\bf 2MFGC~6306} is classified within the framework of the GalaxyZoo
project \citep{2008MNRAS.389.1179L, 2011MNRAS.410..166L}, with the weighted 
fraction of votes out of all responses that this galaxy is edge-on $P=0.89$. 
Its stellar disc looks asymmetric and warped. 
2MFGC-6306 probably forms a group with three other galaxies with close redshifts 
-- SDSS~J075644.97+440536.7, CGCG 207-007, and 2MASSX~J07563889+4407408.

{\bf SPRC-192} is classified as a related to polar-ring galaxies (PRGs) object 
\citep{2011MNRAS.418..244M}. It is an early-type spiral with the visible 
dust lane along the disc major axis surrounded by an inclined ring-like structure. 
Morphologically, it is similar to the well-known polar-ring galaxy NGC~660.
Obviously, if this galaxy had a higher inclination, it would look like a strongly warped spiral.
The galaxy has no nearby companions of comparable sizes, but its
morphology may point to a profound external perturbation or an accretion event 
in the past.

{\bf UGC~4591} (RFGC~1430, PGC~24674) is an edge-on spiral galaxy, a member of 
the group of three galaxies WBL~193 \citep{1999AJ....118.2014W} (the brightest 
of them is the spiral galaxy IC~2394). The apparent disc bending starts at the 
half-radius of the galaxy.

{\bf MCG~+06-22-041} (RFGC~1674, PGC~28776) at $z=0.027$ is a thin late-type 
spiral galaxy without any presence of a bulge. It was classified in the GalaxyZoo as 
an edge-on galaxy ($P=1.0$). Similar redshifts of nearby objects
(elliptical galaxy CGCG~182-048 and 2MASSX~J09574267+3603307, both 
at $z=0.027$) suggest that MCG~+06-22-041 may be in a group.

{\bf NGC~3160} (UGC~5513, PGC~29830) is one of the most interesting objects 
in our sample. It was classified in the 
GalaxyZoo as an edge-on galaxy with $P=0.838$. NGC~3160 shows several 
prominent features such as strong disc warping and a contrast X-shaped structure 
in the central part of the galaxy. NGC~3160 is a member of the cluster of 
galaxies NRGb~78, with the brightest elliptical galaxy NGC~3158 at the centre.

{\bf UGC~5791} (SPRC-197, PGC~31697) is a blue peculiar galaxy for which 
S$^4$G (\citealp{s4g}) analysis has been 
done. It is the nearest galaxy in the sample. The inclination angle for 
this galaxy is barely measurable because of its peculiar shape. It is a 
pair member, together with the galaxy 
UGC~5798 \citep{1979ApJS...40..527P}. UGC~5791 is classified as a 
PRG-related object \citep{2011MNRAS.418..244M}.

{\bf NGC~3753} (UGC~6602, Arp~320, SPRC-203, PGC~36016) is an interacting galaxy, 
marked by \cite{2011MNRAS.418..244M} as an object related to PRGs. It belongs to 
the compact group Hickson~57 \citep{Hick1982} and was included in 
the catalogue of nearby poor clusters of galaxies \citep{1999AJ....118.2014W}. 
The warped dust disc has apparent edge-on orientation. The stellar disc is 
significantly warped and has an asymmetric view. The galaxy image gives 
us a hint that its structure may include a bar or an inner disc overlapping 
with a dust component. Two bright companions near NGC~3753 are the 
spiral galaxy NGC~3754 and the elliptical galaxy NGC~3750. 

{\bf UGC~6882} (SPRC-204, PGC~37372) is another galaxy related to 
PRGs \citep{2011MNRAS.418..244M}, 
which is probably close to edge-on orientation according to the classification from the 
GalaxyZoo ($P=0.575$). It also belongs to the 2MASS selected Flat Galaxy 
Catalog \citep[2MFGC,][]{2004BSAO...57....5M}. Some small nearby
galaxies are seen in the SDSS image, although they do not have 
optical spectra, and, thus, estimated redshifts. 

{\bf SDSS~J140639.64+272242.4} and {\bf SDSS~J153538.63+464229.5} are
the two most distant and small in angular size galaxies in our sample. 
Both galaxies have several nearby companions in projection, but, unfortunately, they do not
have measured redshifts.

{\bf UGC~10716} (RFGC~3242, PGC~59657) is a late-type spiral galaxy viewed almost 
edge-on ($P=0.933$). 
It was selected in the RFGC catalogue and in the catalogue of edge-on 
galaxies created by \cite{2006A&A...445..765K}. The galaxy forms a triplet
with UGC~10714 and SDSS~J170726.32+301316.6 (\citealp{ber2006}).

{\bf UGC~12253} (RFGC~4028, PGC~70040) is seen in almost exactly edge-on
orientation.  The sharp dust lane divides the main body 
of UGC~12253 almost ideally in two halves. The X-pattern at the centre suggests that this 
galaxy has a boxy/peanut-shaped bulge, a possible evidence of the presence of a bar \citep{2006MNRAS.370..753B}. There are two nearby galaxies with close redshifts (PGC~070044
and SDSS~J225556.61+124701.9) and, therefore, UGC~12253 can be a member
of a triplet or a group of galaxies.

\begin{table*}
 \centering
\begin{minipage}{150mm}
 \centering
\parbox[t]{150mm} {\caption{Basic properties of the sample galaxies.}
\label{Table1}}
\begin{tabular}{cccccccccccc}
\hline 
\hline
\# & Galaxy & RA & Dec & $D$ (Mpc) & $M$ (mag) & $a$ (\arcsec) & $q$ & $g-r$ & $r-i$ & $T$ & $i$ (deg) \\ 
(1) & (2) & (3) & (4) & (5) & (6) & (7) & (8) & (9) & (10) & (11) & (12) \\  \hline
1&IC 194&02:03:05&+02:36:51&83.8&-20.88&61.9&0.25&0.75&0.51&Sb & 87.3$\pm$1.2  \tabularnewline
2&2MFGC 6306&07:56:43&+44:05:49&192.0&-21.12&36.4&0.22&0.85&0.52&Sb & 88.0$\pm$1.6 \tabularnewline
3&SPRC 192&08:23:01&+32:00:54&266.5&-21.93&25.7&0.37&0.75&0.46&Sab & 80.0$\pm$5.1 \tabularnewline
4&UGC 4591&08:46:58&+28:14:17&91.5&-20.26&47.4&0.34&0.64&0.34&Scd & 89.7$\pm$0.4 \tabularnewline
5&MCG +06-22-041&09:57:43&+36:04:09&113.3&-18.85&29.9&0.22&0.05&-0.1&Sd & 89.8$\pm$0.6  \tabularnewline
6&NGC 3160&10:13:55&+38:50:34&98.2&-21.40&56.2&0.36&0.85&0.44&Sb & 89.2$\pm$1.7  \tabularnewline
7&UGC 5791&10:39:27&+47:56:50&14.6&-16.68&56.3&0.43&0.31&0.16&Sc & 85.0$\pm$5.0  \tabularnewline
8&NGC 3753&11:37:54&+21:58:53&124.0&-22.29&74.6&0.31&0.89&0.49&Sab& 84.3$\pm$2.7  \tabularnewline
9&UGC 6882&11:54:43&+33:32:12&135.1&-20.98&43.2&0.23&0.74&0.38&Sbc& 80.1$\pm$6.2  \tabularnewline
10&SDSS J140639.64+272242.4&14:06:40&+27:22:42&303.5&-20.89&24.3&0.18&0.64&0.49&---& 88.6$\pm$1.3\tabularnewline
11&SDSS J153538.63+464229.5&15:35:39&+46:42:30&271.4&-20.17&20.7&0.22&0.41&0.18&---& 84.4$\pm$1.9\tabularnewline
12&UGC 10716&17:07:44&+30:19:35&132.5&-20.61&39.9&0.23&0.66&0.41&Sb & 87.9$\pm$1.3 \tabularnewline
13&UGC 12253 &22:56:02&+12:45:59&101.4&-20.61&49.8&0.27&0.84&0.42&Sb& 89.8$\pm$0.5  \tabularnewline

\hline\\
\end{tabular}
\end{minipage}
   
\parbox[t]{150mm}{ Columns: \\
(1) Designation number in the sample, \\
(2) first name from NED, \\
(3), (4) J2000 coordinates from NED,\\
(5) 3K CMB distance from NED (except for 2MFGC 6306 for which the distance 
was calculated from the radial velocity taken from the HyperLeda database), \\
(6) absolute magnitude in the $r$ band inside of the isophote of 
25.5 mag/arcsec$^2$. Galactic extinction 
(according to \citealp{2011ApJ...737..103S}) and K-correction 
(using the NED K-Correction calculator based on \citealp{2012MNRAS.419.1727C}) was taken into account, \\
(7), (8) semi-major axis and galaxy flattening of the 25.5 mag/arcsec$^2$ isophote in the $r$-band, \\ 
(9), (10) colours corrected for Galactic extinction and K-correction, \\ 
(11) morphological type from NED. For SDSS~J140639.64+272242.4 and SDSS~J153538.63+464229.5, there is no morphological classification in NED or HyperLeda databases,\\
(12) galaxy disc inclination (see Appendix A in \citealp{2015MNRAS.451.2376M}). 
For UGC~4591 and MCG~+06-22-041, we applied the first approach from 
\cite{2015MNRAS.451.2376M} using the 3D disc decomposition available in 
the \textsc{IMFIT} code. Inclination estimate for UGC~5791 is found on the basis 
of the apparent axial ratio of the central part of the galaxy. For the remaining 
galaxies, we estimated the inclination by the orientation of the dust lane (the second 
method from \citealp{2015MNRAS.451.2376M}).
}
\end{table*} 

\begin{figure*}
\includegraphics[width=3.5cm, angle=0, clip=]{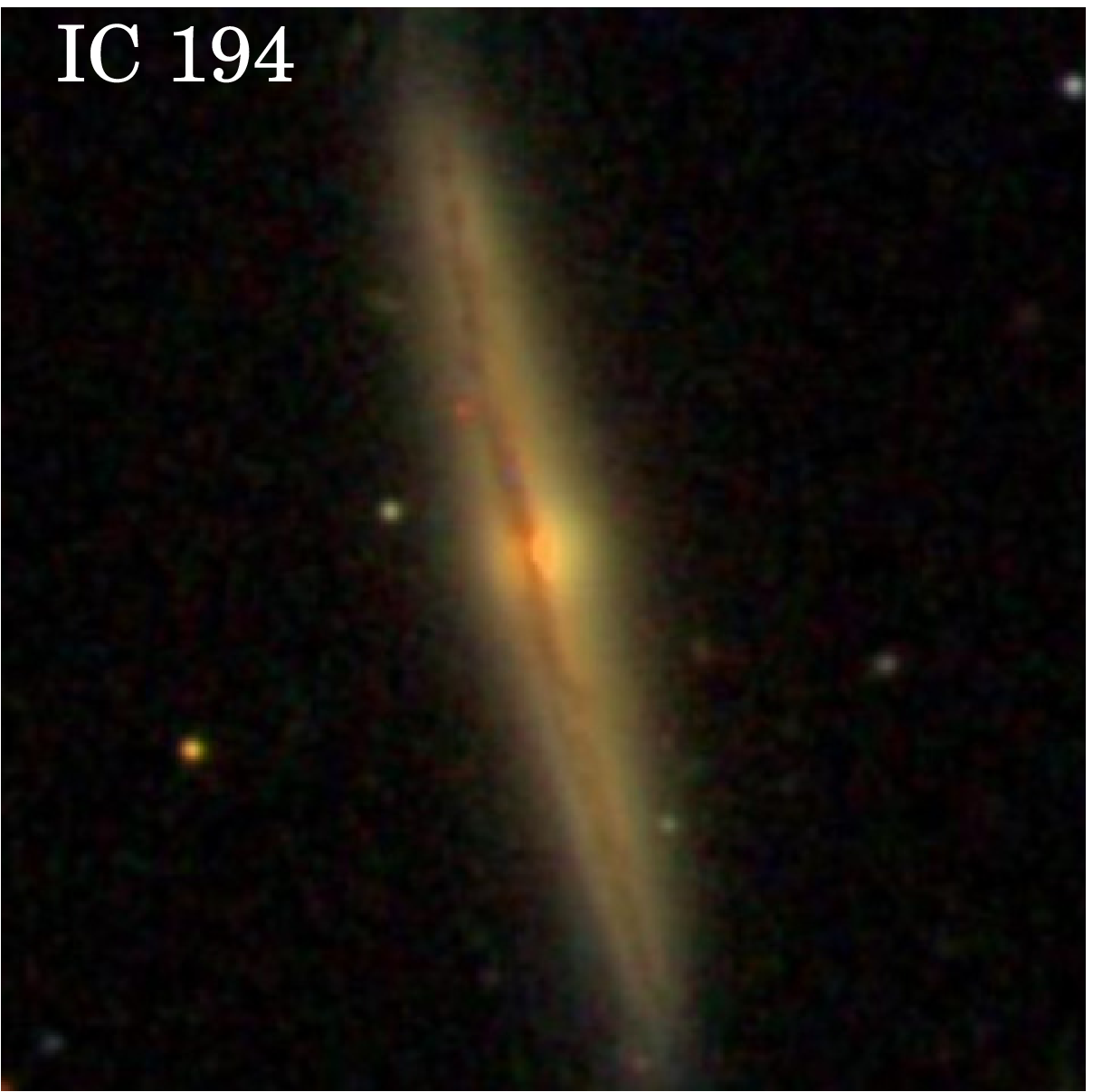}
\includegraphics[width=3.5cm, angle=0, clip=]{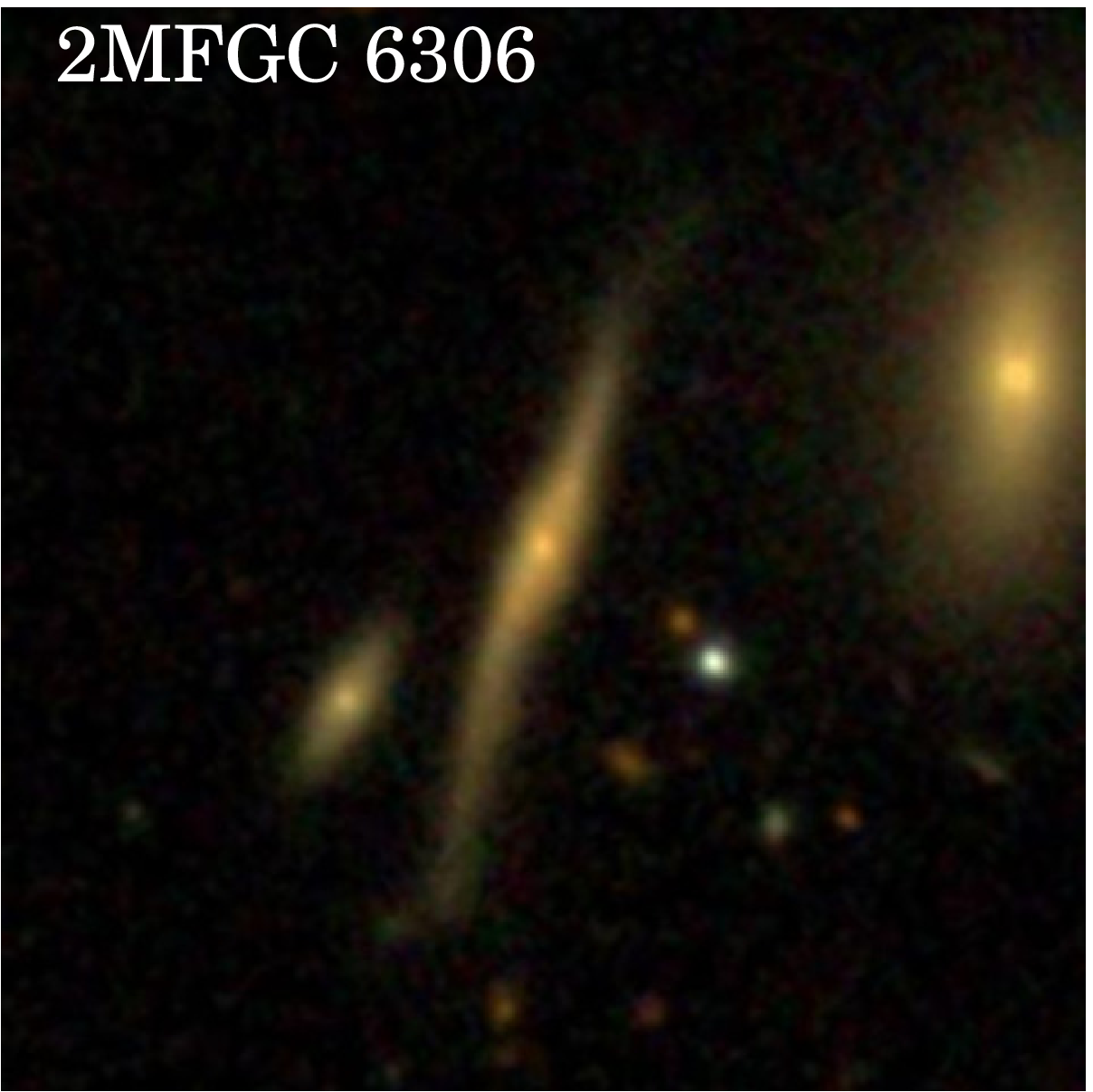}
\includegraphics[width=3.5cm, angle=0, clip=]{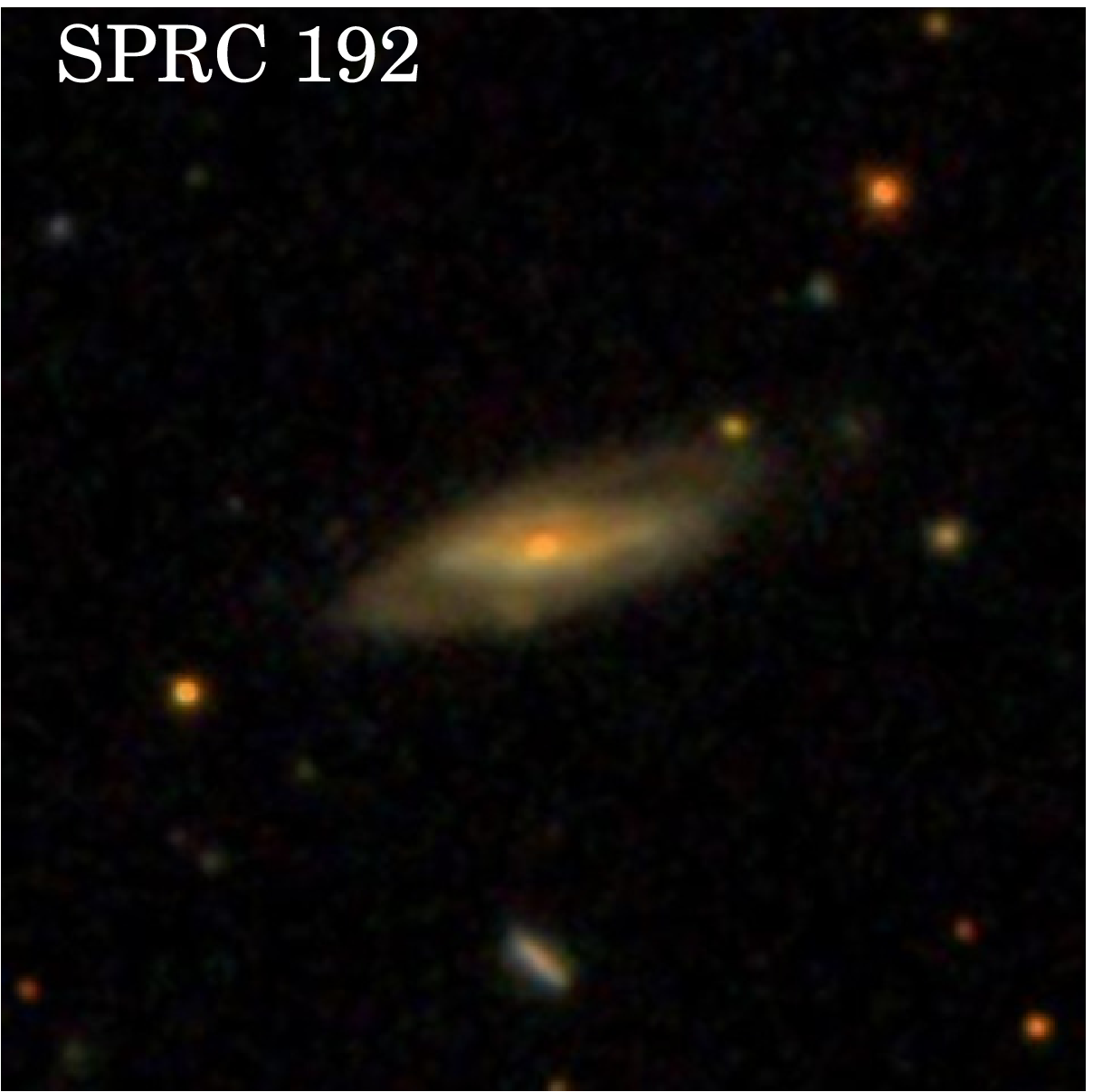}
\includegraphics[width=3.5cm, angle=0, clip=]{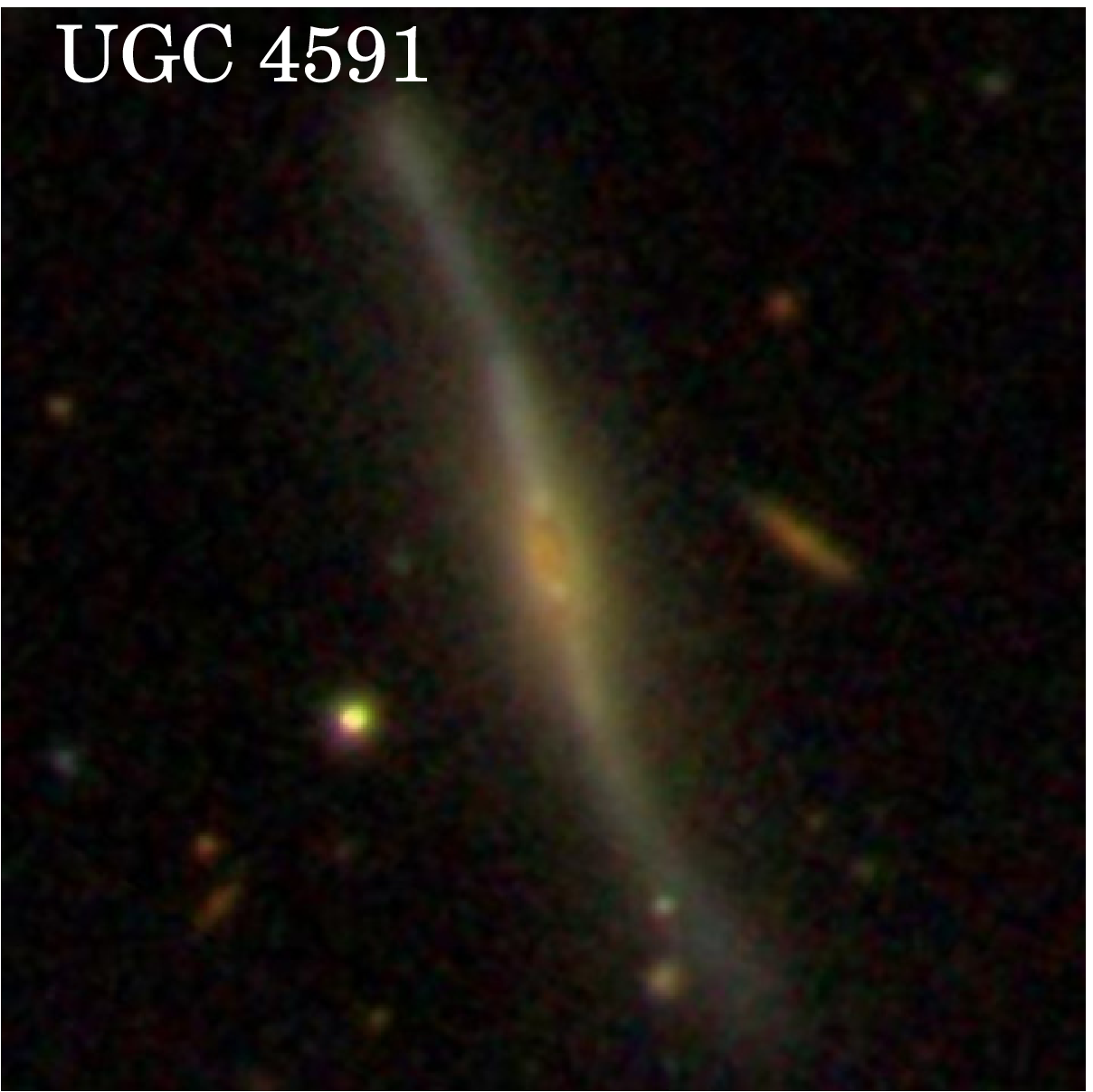}
\includegraphics[width=3.5cm, angle=0, clip=]{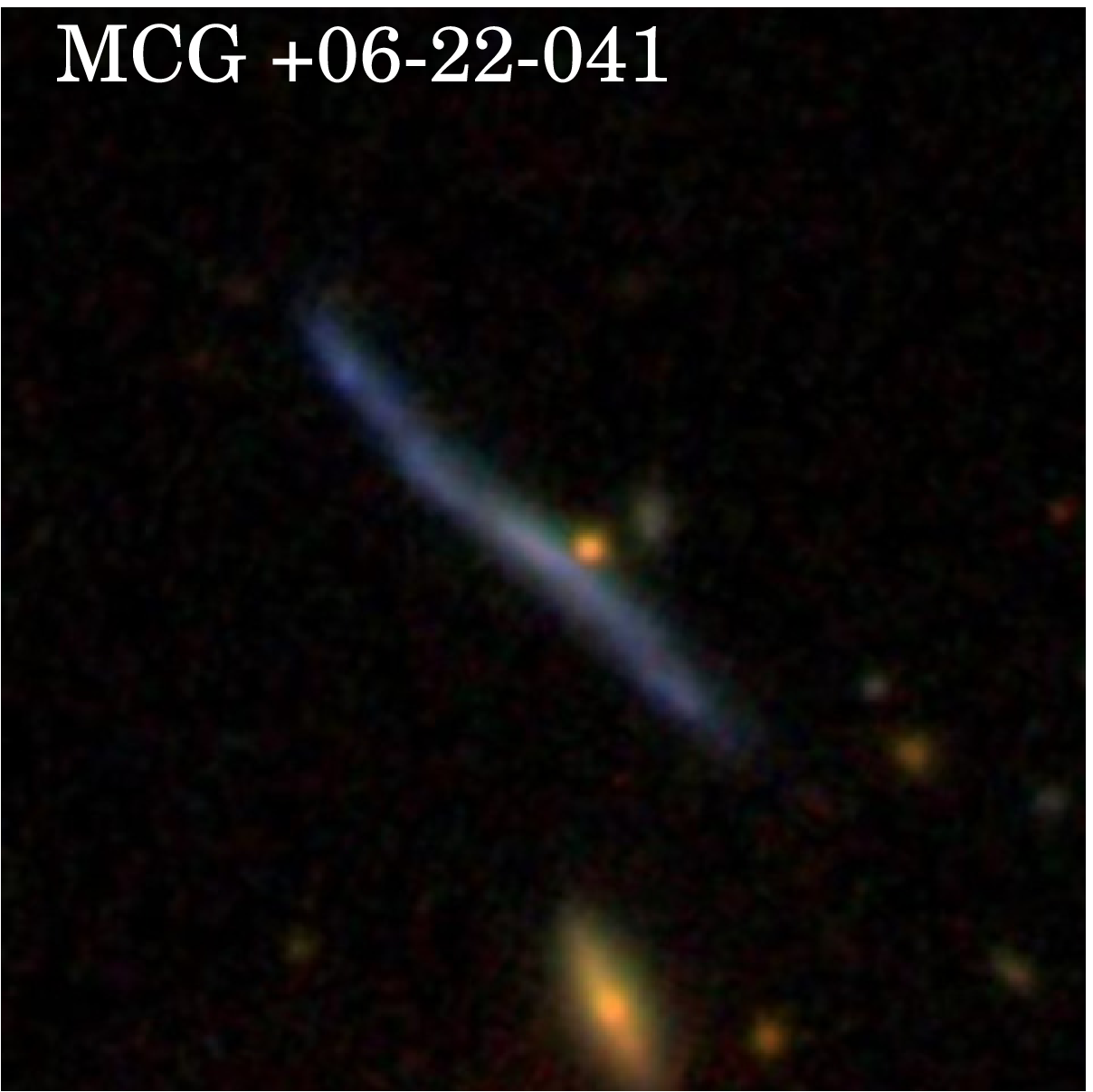}
\includegraphics[width=3.5cm, angle=0, clip=]{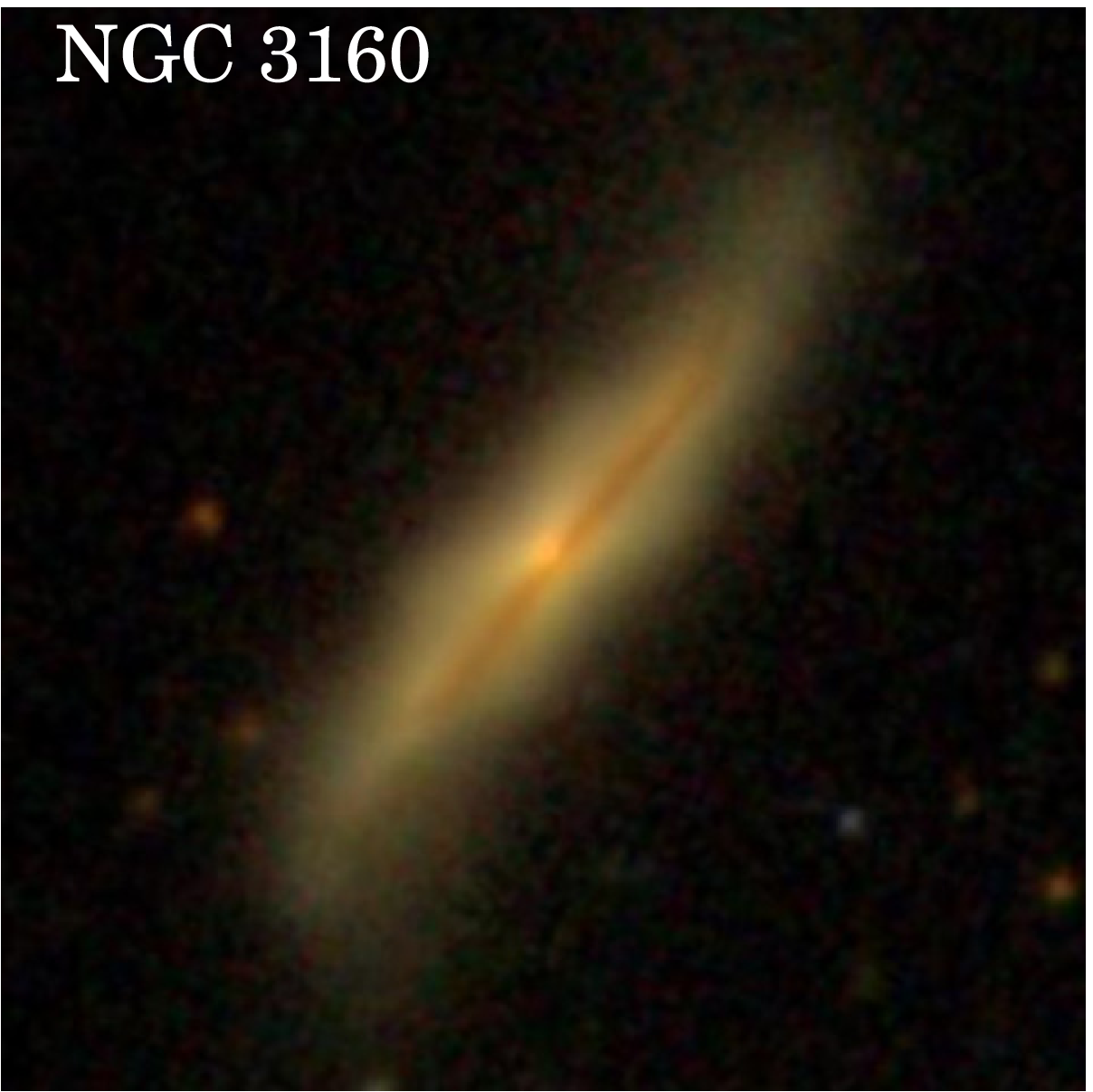}
\includegraphics[width=3.5cm, angle=0, clip=]{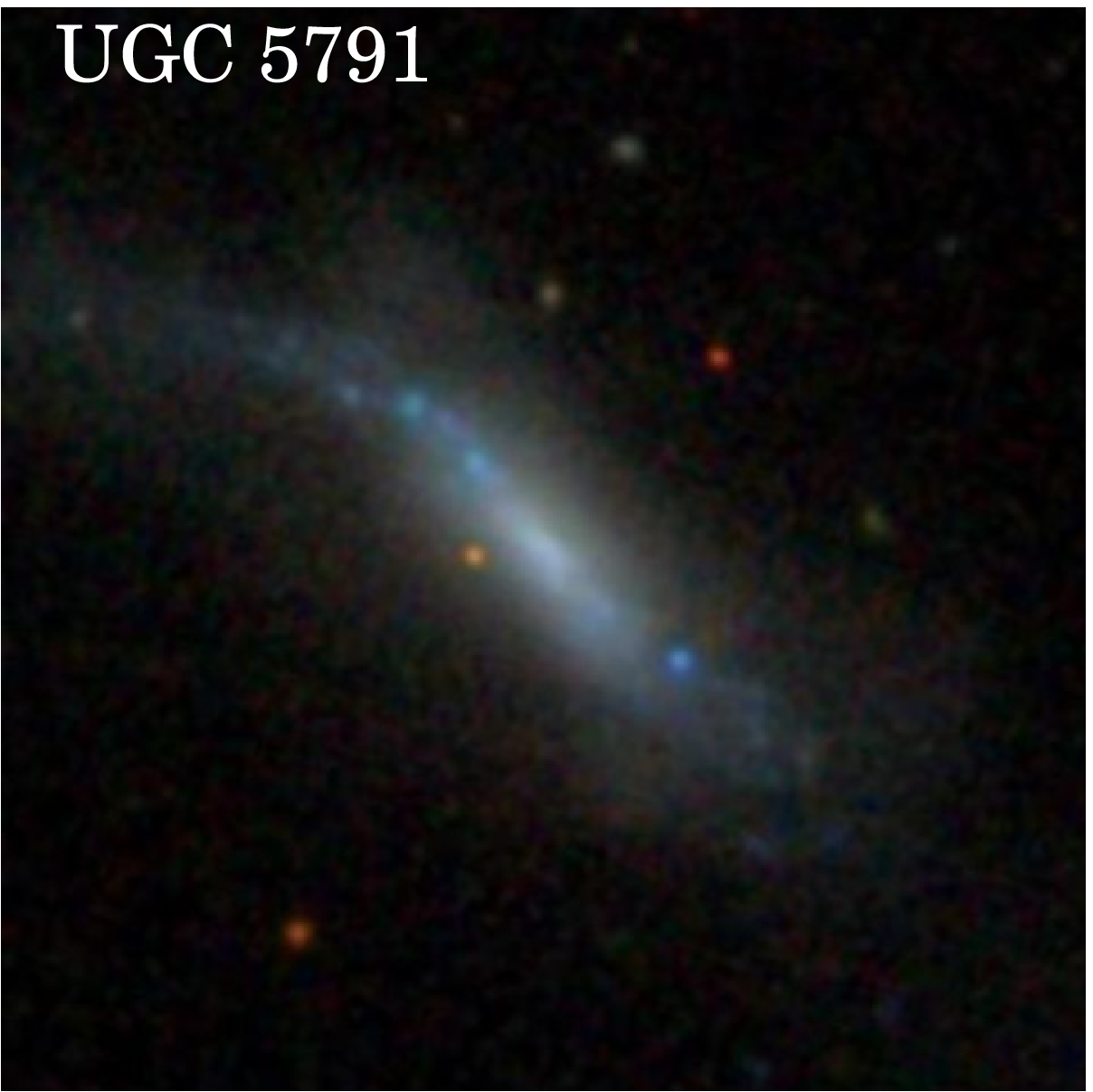}
\includegraphics[width=3.5cm, angle=0, clip=]{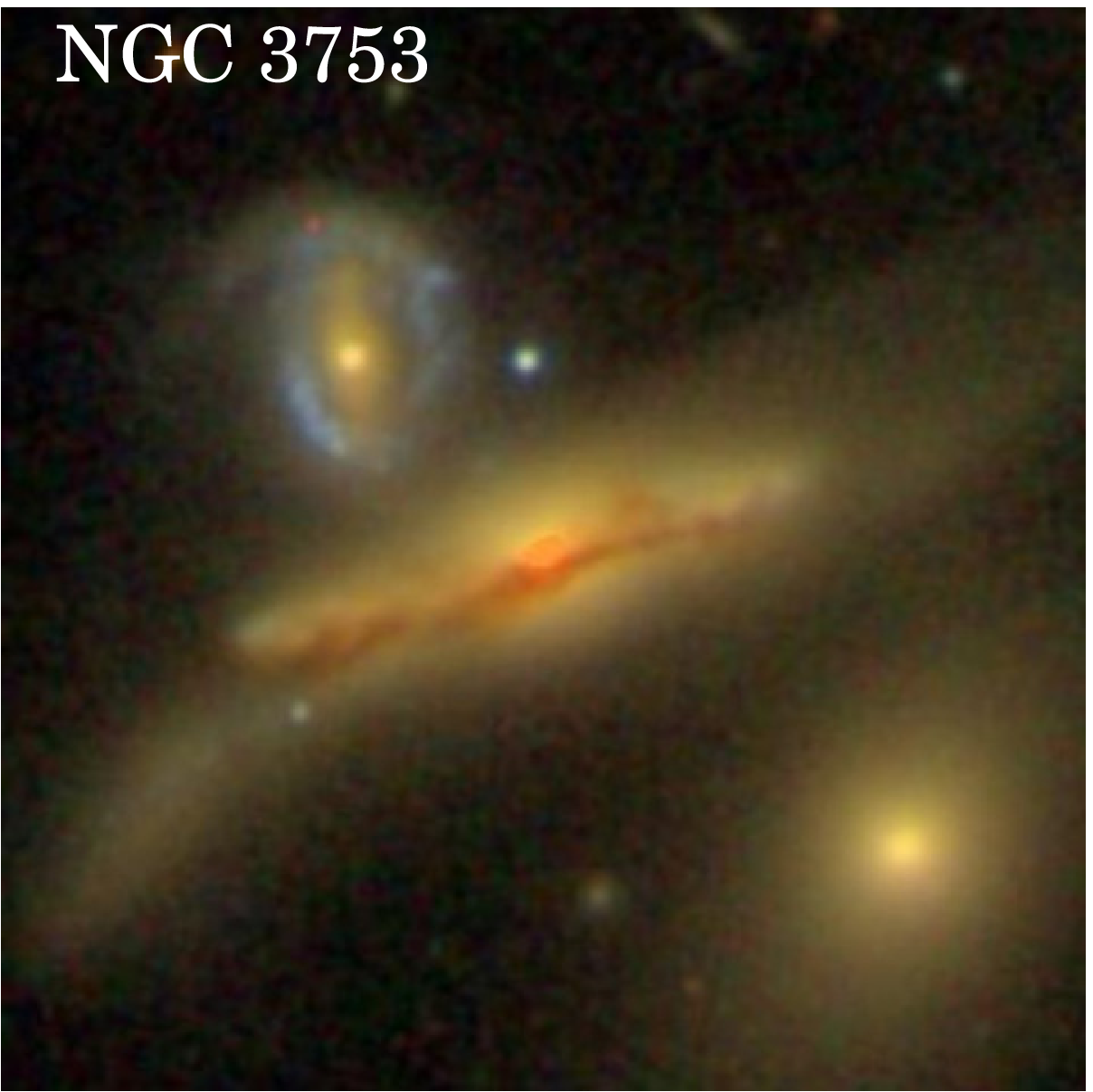}
\includegraphics[width=3.5cm, angle=0, clip=]{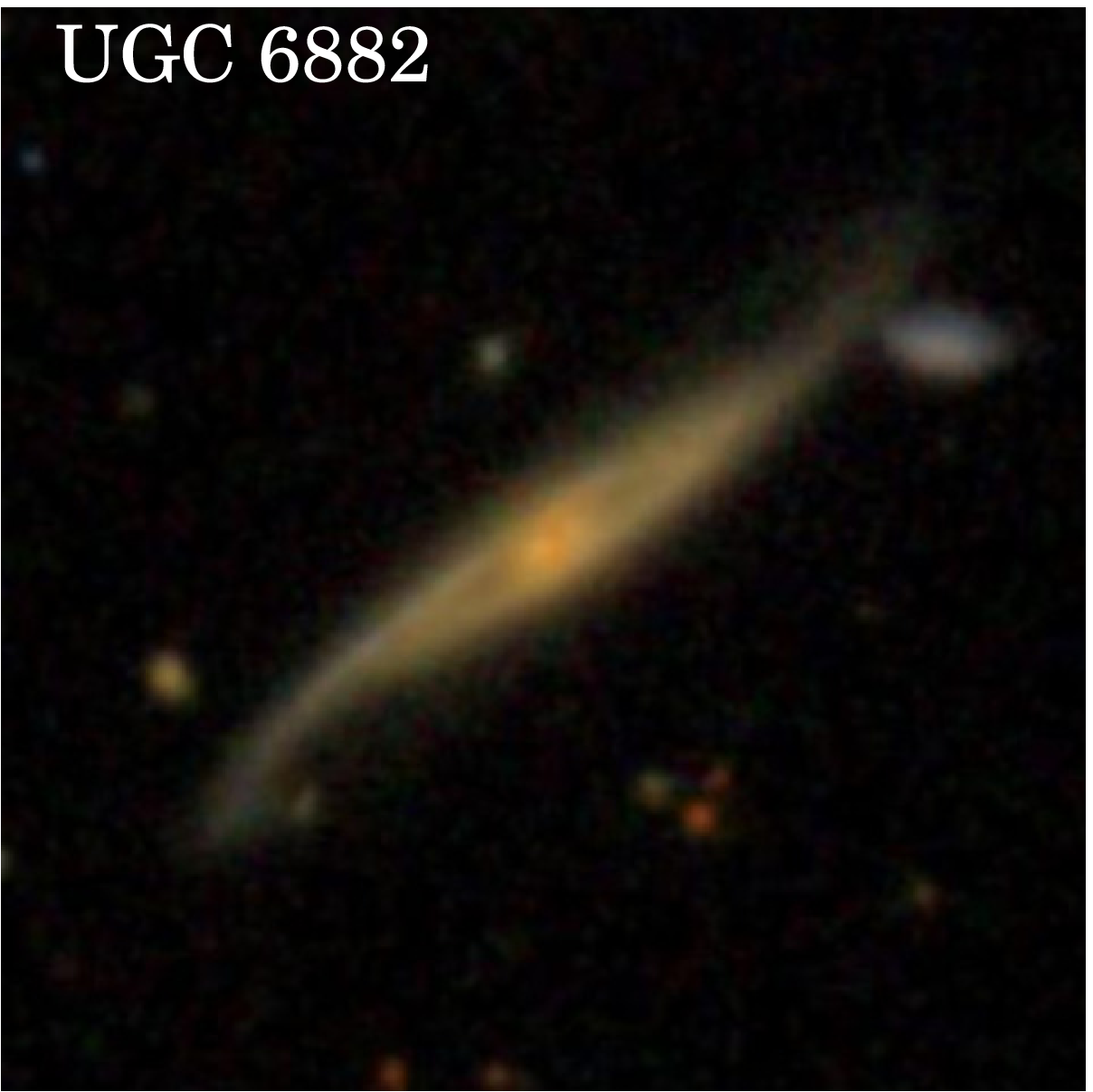}
\includegraphics[width=3.5cm, angle=0, clip=]{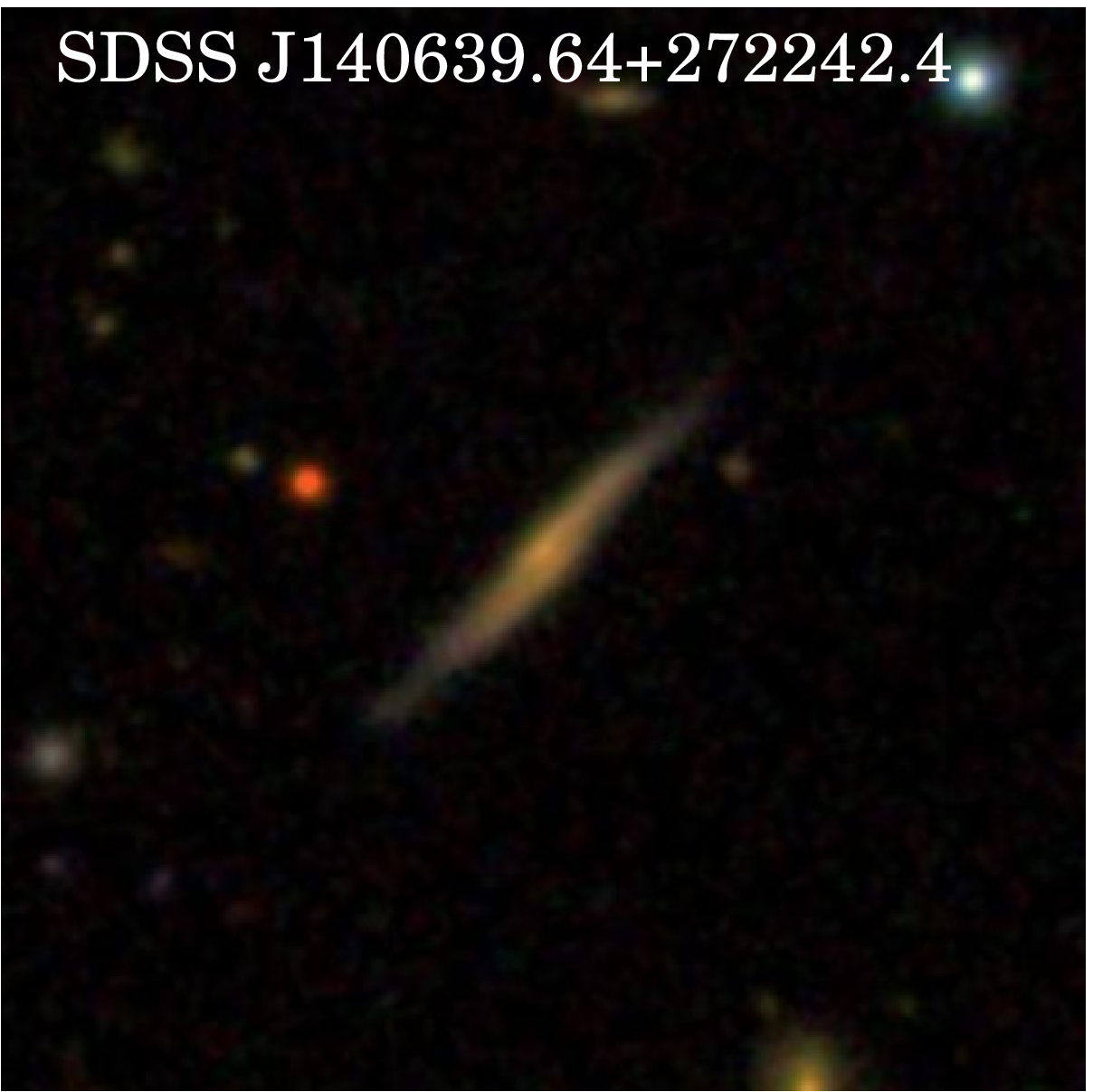}
\includegraphics[width=3.5cm, angle=0, clip=]{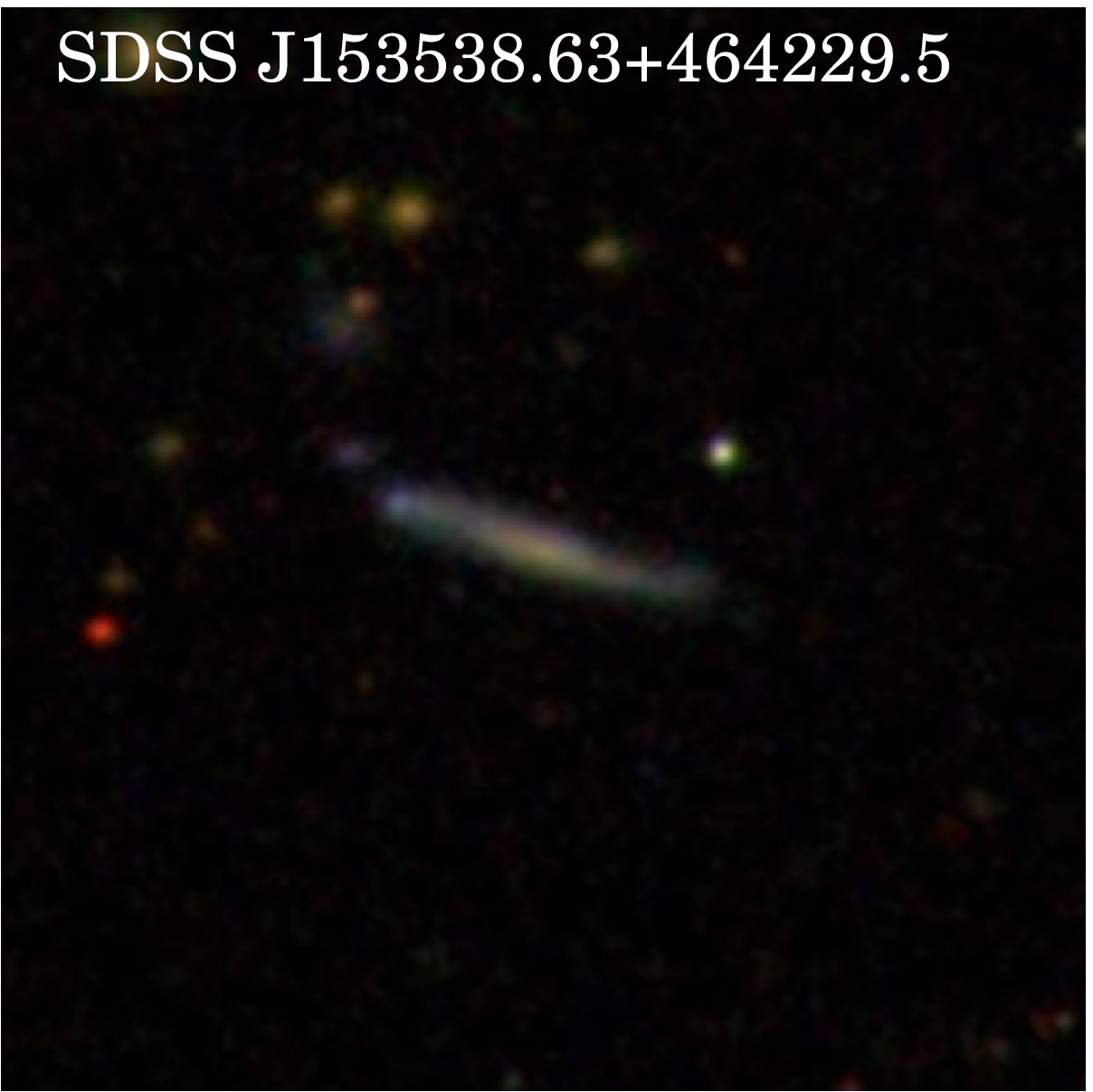}
\includegraphics[width=3.5cm, angle=0, clip=]{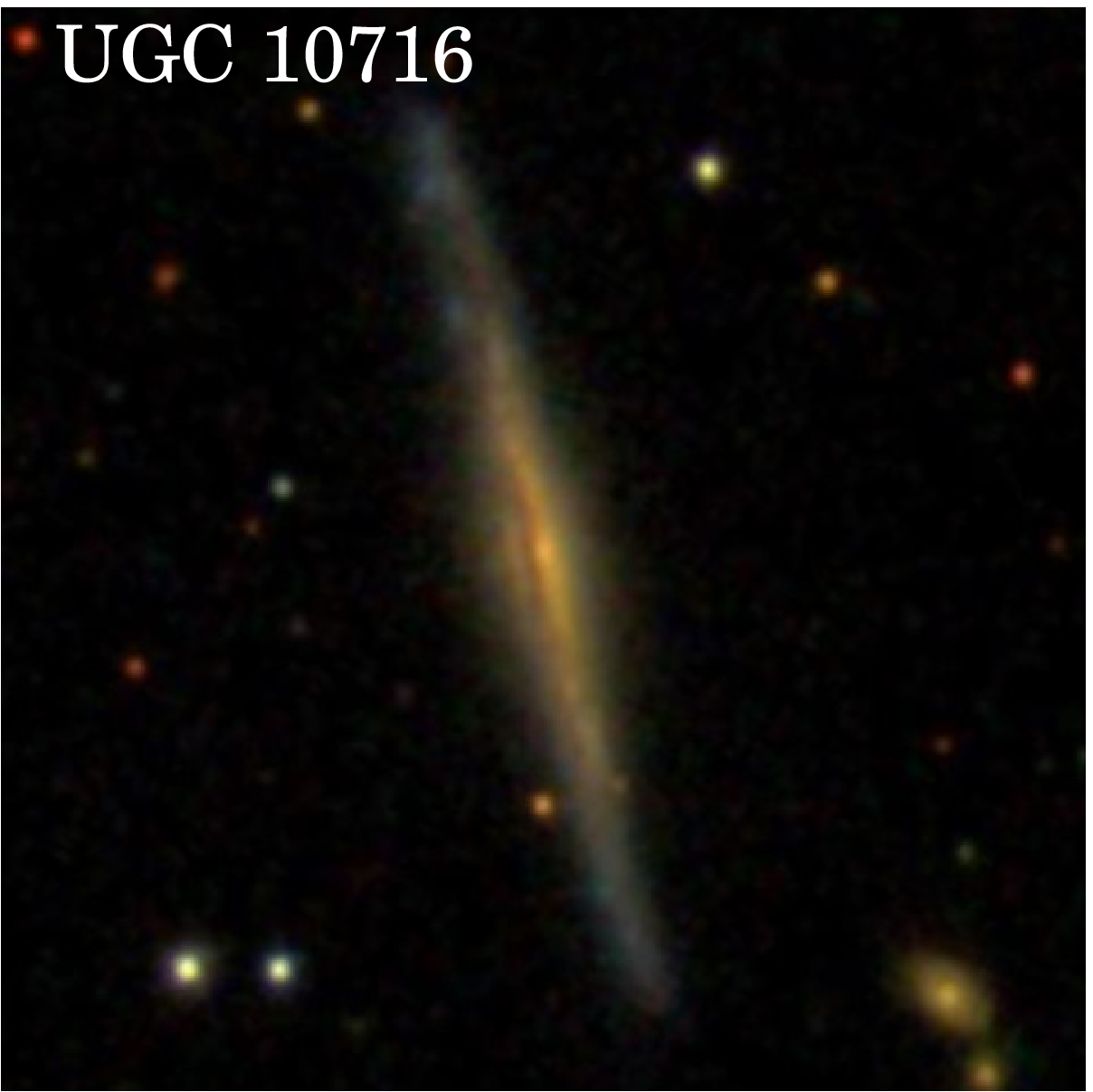}
\includegraphics[width=3.5cm, angle=0, clip=]{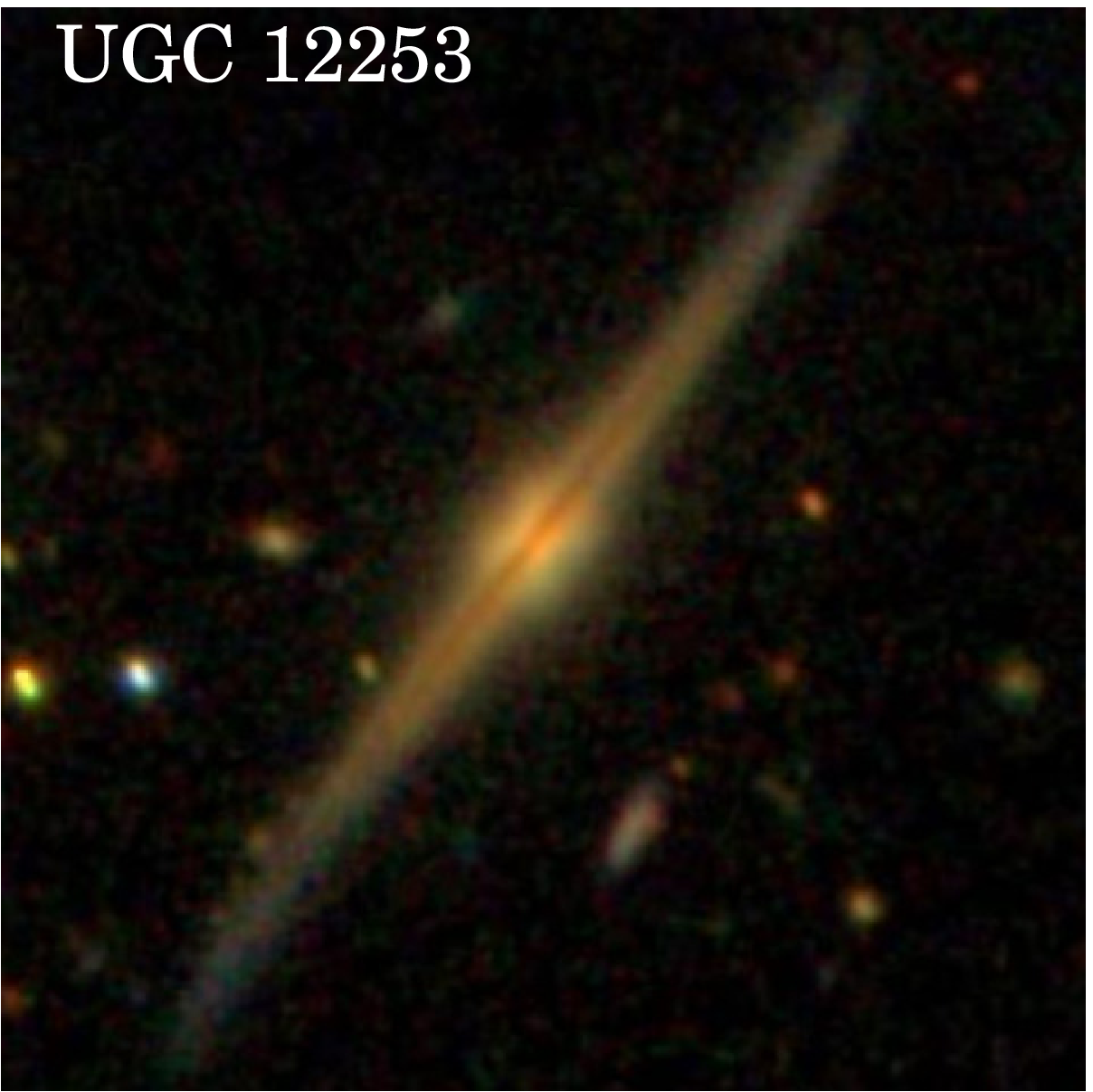}

\caption{SDSS thumbnail images for the 13 galaxies in our sample with the field of view of 90$\arcsec$ by 90$\arcsec$.}
\label{SDSS_images}
\end{figure*}

\section{Preparation and decomposition of galaxy images}
\label{Preparation}

\subsection{Photometric decomposition technique}
\label{Decomposition_Technique}

An ordinary way to the determine structural parameters of spiral galaxies implies
performing the decomposition of their 1D- or 2D-brightness distributions
into their two main components: a spherical bulge and a flat disc. The main goal of the
decomposition analysis is to separate contributions of different components
in the overall brightness distribution and to build a photometric model of the galaxy. 

The common empirical model for the bulge representation (in mag/$\Box''$) is
the \ser~model (\citealp{ser1968}):
\begin{equation}
\mu(r) = \mu_\mathrm{0,b} + \frac{2.5\,\nu_\mathrm{n}}{\ln 10} \left( \frac{r}{r_\mathrm{e,b}} \right)^{\frac{1}{n}},
\label{eqSersicMu}
\end{equation}
where $\mu_\mathrm{0,b}$ is the central surface brightness, $r_\mathrm{e,b}$ is the effective
radius, and $n$ is the so-called \ser~index with $\nu_\mathrm{n}$ to be a function
depending on $n$ \citep{caon+1993}.

The disc is usually described by the exponential law,
which in the logarithmic scale (in mag/$\Box''$) turns into a simple linear 
function of the radius (\citealt{Pat1940, Freeman1970}):
\begin{equation}
\mu(r) = \mu_\mathrm{0,d} + 1.086\, \frac{r}{h},
\label{eqExpDiskMu}
\end{equation}
where $\mu_\mathrm{0,d}$ is the disc central surface brightness and $h$ is its exponential
scale length.

The distribution of the surface brightness in the radial $r$ and vertical $z$ 
directions for the transparent ``exponential'' disc observed at the edge-on 
orientation is described by the following expression (in intensities):
\begin{equation}
I(r,z) = 
  I(0,0)
  \displaystyle \frac{r}{h} \, K_1\left(\frac{r}{h}\right) 
  \mathrm{sech}^2(z/z_0) , \,
\label{form_DSB}
\end{equation}
where $I(0,0)$ is the disc central intensity, $h$ is the radial 
scale length, $z_0$ is the ``isothermal'' scale height of the disc 
\citep{Spitzer1942, vanderKruit1981a, vanderKruit1981b, vanderKruit1982a, vanderKruit1982b},
and $K_1$ is the modified Bessel function of the first order. 
To find the edge-on disc central surface brightness, we should 
calculate $I_{0}^{edge-on}=\int_{-\infty}^{\infty} I(r,0)\,dr$. 
The central surface brightness reduced to the face-on is found 
as $I_{0}^{face-on}=I_{0}^{edge-on}\,z_0/h$. 
Further, we will use its logarithmic value $\mu_\mathrm{0,d}$ in mag/$\Box''$. 

The above and other models are implemented into many open-access software tools designed 
to perform photometric decomposition of galaxy images on several components: 
{\sevensize budda} (\citealt{desouza+2004}),
\textsc{gim2d} (\citealt{Simard+2002}), 
\textsc{galfit} (\citealt{peng2010}),
\textsc{imfit} (\citealt{2015ApJ...799..226E}),
\textsc{deca} (\citealt{2014AstBu..69...99M}), etc. 

In our study, we will use a non-standard approach of applying genetic 
algorithms (GA) to find the best fit model for a galaxy \citep{Goldberg1989}. 
This approach allows us to perform 
decomposition in a more robust way and avoid local minima of $\chi^2$, 
whereas a standard way to find an optimal model is to apply the gradient 
descent (Levenberg-Marquardt) or downhill simplex (Nelder-Mead) methods. 
Also, since genetic algorithms are random and a parameter space is vast, 
the fitting procedure can be repeated multiple times without resulting in 
the exact the same solution. This gives us an estimate of the uncertainty 
for the fit parameters. Other advantages and robustness of GA for 
decomposition of galaxy images are presented 
in \cite{2013A&A...550A..74D, 2014MNRAS.441..869D}. Below we provide 
a brief description of GA properties used in the current work.  

GA is an effective computational algorithm for searching solutions of an 
optimisation problem. It simulates
the natural selection process and represents possible solutions
of the problem as ``organisms'', with ``genes'' as free parameters
of a model. During the optimisation procedure these ``organisms'' evolve
through ``breeding'' and ``mutation'' processes.

We use \textsc{galfit} (and \textsc{imfit} for the galaxy SPRC-192, see details 
in Sect.~\ref{Decomposition}) to generate model images of the galaxies 
convolved with the point spread function (PSF) images (``organisms''), 
and find $\chi^2$ value (the fitness which characterizes the quality of the 
model). $\chi^2$ is calculated internally in \textsc{galfit} 
and \textsc{imfit} which take into account the mask (to ignore all objects 
which will affect our simple bulge/disc decomposition) and the weight 
images of the galaxy. The values of each gene of each
organism of zero-generation are being set randomly. The ranges of possible 
values for each gene are chosen such that they undoubtedly overlap the true 
value of the corresponding model parameter. Although using the larger range decreases 
the rate of the method convergence, we
set the parameter ranges as large as possible, to make sure that the true values 
of the parameters are inside these ranges. 
The zero generation and any further generation comprises 200 organisms. 
The total number of generations is chosen to be 100. To retrieve estimates of uncertainties (their lower bounds)
for each free parameter, we repeat GA ten times. The best model with 
minimal $\chi^2$ is then taken as an initial guess for the subsequent 
\textsc{galfit} (or \textsc{imfit}) decomposition. This is done in order 
make the solution found by GA more precise: \textsc{galfit} and \textsc{imfit} 
implement the Levenberg-Marquardt algorithm which is suitable for searching the 
global minimum of $\chi^2$ when a solution, found with GA, is close to the true one. 

\subsection{Data handling}
\label{Data_handling}
For our photometric analysis of galaxies, we use the data release DR12 of SDSS.
The corrected frames and PSF images (prefix ``psField'') for three $g$, $r$, $i$ 
bands were retrieved directly from the SDSS Science Archive Server (SAS). 
Although downloaded frames have been bias subtracted and flat-fielded, we 
need to do some additional preparation of the galaxy images for the further 
decomposition and analysis of galactic warping. All the images were sky-subtracted 
(sky background was fitted with the $2^\mathrm{nd}$ polynomial), rotated 
(such as the plane of the galaxy would be horizontal) and cut out from the 
original image (the final galaxy image should have a horizontal extent 
of $1.5a$ and a vertical extent of $1.5b$, where $a$ and $b$ are the major 
and minor axes respectively). Foreground stars and other contaminants 
were detected with the Source Extractor \citep{1996A&AS..117..393B}, 
revisited by eye, and then properly masked\footnote{These steps were done 
with the special \textsc{python} package https://github.com/latrop/pipeline.}. 

\subsection{Photometric decomposition}
\label{Decomposition}
Once the data preparation is done, we can perform photometric decomposition of the sample galaxies in the
$g$, $r$, and $i$ bands. For this, we used the GA approach described in 
Sect.~\ref{Decomposition_Technique}. In addition to the masks built in the 
previous step (Sect.~\ref{Data_handling}), we also masked by hand some prominent features of the galaxies 
which could affect the results of the fitting: dust lanes, which were detected 
in the $g$ band images, and disc warps at radii where the disc bending begins to
appear (see Sect.~\ref{Warp_Analysis}). 

As the $r$ band is the deepest band in SDSS, we will mainly use the decomposition results 
in this band for our further analysis of the galaxy structure. However, we carried out decomposition in two other bands as well. This information is useful to estimate the model colours $g-r$ and $r-i$ 
of the bulge and the disc, and compare them for different galaxies. Also, the disc scale length as well as the disc scale height can be, in principal, different in different passbands, and it is interesting to compare radial colour gradients for the galaxies of the sample. Some important details on the image decomposition are given below.

All the galaxies (except UGC~6882) were fitted to the edge-on disc plus bulge 
model (\ref{form_DSB}). The galaxy UGC~6882 looks to be not exactly edge-on, therefore 
for this galaxy we used an exponential non-edge-on disc model (\ref{eqExpDiskMu}) 
with the apparent disc flattening of $0.15$. However, we fitted the minor axis 
surface brightness distribution of this galaxy with the edge-on disc function 
and estimated its scale height to be $2.8\arcsec$, or 1.87~kpc. Since this 
galaxy is not purely edge-on, we can assume that its vertical scale is 
somewhat overestimated, and the found value should be used as its upper limit.

Five galaxies (UGC~4591, MCG~+06-22-041, UGC~5791, SDSS~J140639.64+272242.4, 
SDSS~J153538.63+464229.5) do not exhibit the 
presence of a well-defined bulge, and thus they were fitted to the single disc model. 
SPRC-192 is a galaxy with inclined ring-like structure, 
for which the ring was fitted 
with the ``GaussianRing'' function using the \textsc{imfit} code (\textsc{galfit} 
does not offer this function). We found that the ring has the diameter of 
12.6~kpc and its major axis is oriented at the angle of about $15\deg$ 
relative to the edge-on disc plane. 

We should note that for the half of the galaxies the dust lane is quite 
prominent, and, therefore, the dust extinction can severely affect the 
results of our fitting. In those cases where the dust lane was masked, 
the model of the bulge can become unreliable since the number of unmasked 
pixels of the compact bulge can be significantly reduced. For this reason, 
we do not rely on the retrieved bulge parameters, and only investigate 
the disc model. The best fit parameters and their uncertainties for the $r$ band are 
presented in Table~\ref{Table2}. The bulge-to-total luminosity ratio ($B/T$)
can be somewhat underestimated and should be used with caution.

\begin{table*}
\centering
\begin{minipage}{150mm}
\parbox[t]{150mm} {\caption{Structural parameters of the disc for the $r$-band 
and colours of the disc and the bulge (the D- and B-suffix respectively). 
The disc central surface brightness and the colours are corrected for 
Galactic extinction and $K$-correction.}
\label{Table2}}
\begin{tabular}{cccccccccc}
\hline
\hline
  \# & Name & \multicolumn{1}{c}{$\mu_\mathrm{0,d}$} & \multicolumn{1}{c}{$z_0$} & \multicolumn{1}{c}{$h$}   & $B/T$ & $(g-r)_\mathrm{D}$& $(r-i)_\mathrm{D}$& $(g-r)_\mathrm{B}$& $(r-i)_\mathrm{B}$ \\
         &      & \multicolumn{1}{c}{($\mathrm{mag}/\square''$)}& \multicolumn{1}{c}{(kpc)}   & \multicolumn{1}{c}{(kpc)} &       & (mag)  & (mag) & (mag) & (mag)    \\
      \hline
1&IC 194&21.98$\pm$ 0.12&1.50 $\pm$ 0.06&6.55 $\pm$ 0.65&0.27&0.64&0.44&0.91&0.59 \\
2&2MFGC 6306&20.71 $\pm$ 0.10&1.25 $\pm$ 0.04&3.95 $\pm$ 0.21&0.09&1.12&0.45&0.38&0.52 \\
3&SPRC 192&20.69 $\pm$ 0.10&3.36 $\pm$ 0.10&4.84 $\pm$ 0.51&0.45&0.68&0.28&1.09&0.76 \\
4&UGC 4591&21.22 $\pm$ 0.01&1.25 $\pm$ 0.01&3.25 $\pm$ 0.04&0.0&0.55&0.29&---&--- \\
5&MCG +06-22-041&22.18 $\pm$ 0.02&0.65 $\pm$ 0.01&2.87 $\pm$ 0.07&0.0&0.04&-0.11&---&--- \\
6&NGC 3160&20.62 $\pm$ 0.18&2.00 $\pm$ 0.13&4.77 $\pm$ 0.59&0.07&0.80&0.39&0.65&0.88 \\
7&UGC 5791&21.48 $\pm$ 0.28&0.34 $\pm$ 0.04&0.66 $\pm$ 0.13&0.0&0.26&0.15&---&--- \\
8&NGC 3753&20.70 $\pm$ 0.34&3.25 $\pm$ 0.50&7.00 $\pm$ 1.89&0.20&0.80&0.44&1.06&0.57 \\
9&UGC 6882&22.34 $\pm$ 0.05&1.87 $\pm$ 0.49&7.15 $\pm$ 0.03&0.17&0.28&0.72&1.15&0.57 \\
10&SDSS J140639.64+272242.4&20.17 $\pm$ 0.13 &1.61 $\pm$ 0.14 &5.96 $\pm$ 0.51 & 0.0 &0.27&0.20&---&--- \\
11&SDSS J153538.63+464229.5&20.97 $\pm$ 0.12 &1.73 $\pm$ 0.11 &6.15 $\pm$ 0.49 & 0.0 &0.20&-0.13&---&---\\
12&UGC 10716&22.89 $\pm$ 0.13&1.18 $\pm$ 0.11&7.47 $\pm$ 0.91&0.49&0.40&0.28&1.13&0.49 \\
13&UGC 12253&22.45 $\pm$ 0.10&1.41 $\pm$ 0.09&6.33 $\pm$ 0.63&0.37&0.71&0.38&0.89&0.41 \\
      \hline
    \end{tabular}
  \end{minipage}
\end{table*}

\subsection{Analysis of warps}
\label{Warp_Analysis}

In order to describe the structure of optical warps, we apply the following general technique to each galaxy image.

First, in each band we build an isophote map from 20.5 up to 25.5 
mag/$\Box''$ level. (Due to complex morphology of 2MFGC~6306, we use isophotal
level of 24.5 mag/$\Box''$ for it.) To plot the isophotes, we use the free astronomical 
application \textsc{SAOImage DS9}\footnote{http://ds9.si.edu/}, with the smoothness 
factor of 5. We verified that this smoothing does not affect the geometry of the warps.
Masked pixels are excluded from the analysis. 

After that, we find the center-line 
for each isophote using the ``skeletonization'' process to obtain a skeletal remnant. 
This topological skeleton of the isophote is equidistant to its boundaries, and therefore 
determines the center-line of the isophote. For this purpose we use the 
\textsc{scikit-image}\footnote{http://scikit-image.org/} collection of algorithms for image processing. All 
center-lines are then averaged that results in the final center-line of the galaxy, 
which goes from the central region to its optical outskirts. 

The found center-line then can be fitted with a piecewise linear function describing three different 
parts of the line: left (representing the left warp of the disc), middle part 
(the plane of the galaxy) and the right part (the right warp of the disc). 
As each segment has its own slope and intercept, we can obtain two points, 
where the warps begin ($R_w$), and their orientation relative to the central segment. 
Here, we define the warp angle $\psi$ as an angle measured from the galaxy centre, 
between the plane (middle segment of our centre-line) and line from centre to 
tips of the outer 25.5 isophote (see \citealp{ss1990}, \citealp{rc1998}). 
As warps are not always symmetric, we calculate the warp parameters
on either side of each galaxy independently. 
Notice here that the measurement of the warp angle is independent of how well 
we horizontally aligned our galaxy. 

\begin{figure*}
\includegraphics[width=4.4cm, angle=0, clip=]{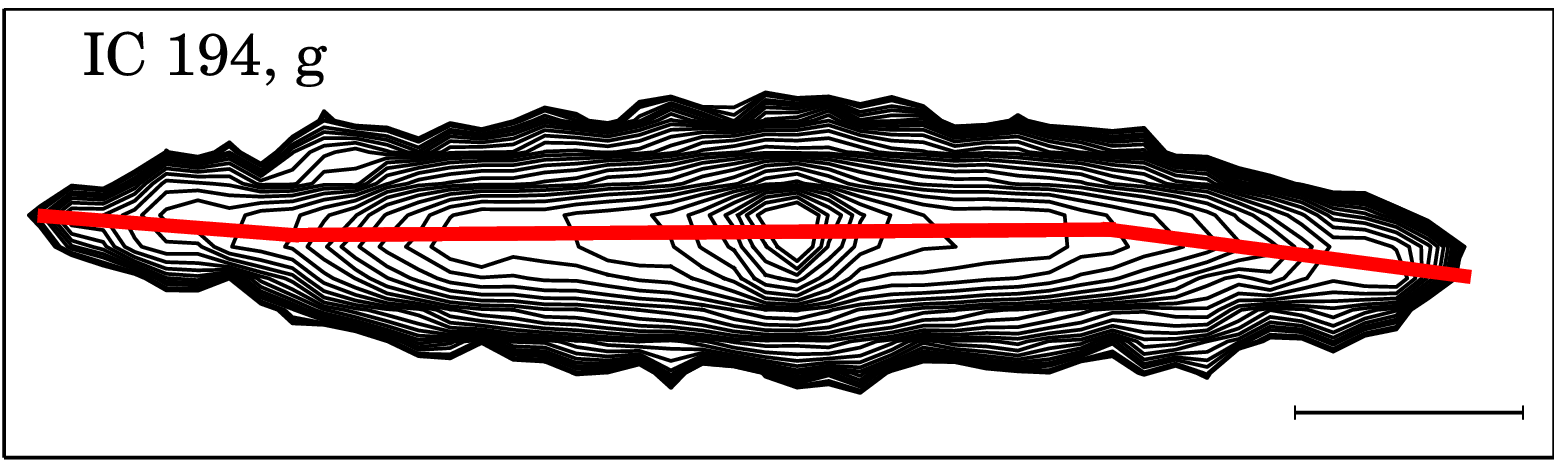}
\includegraphics[width=4.4cm, angle=0, clip=]{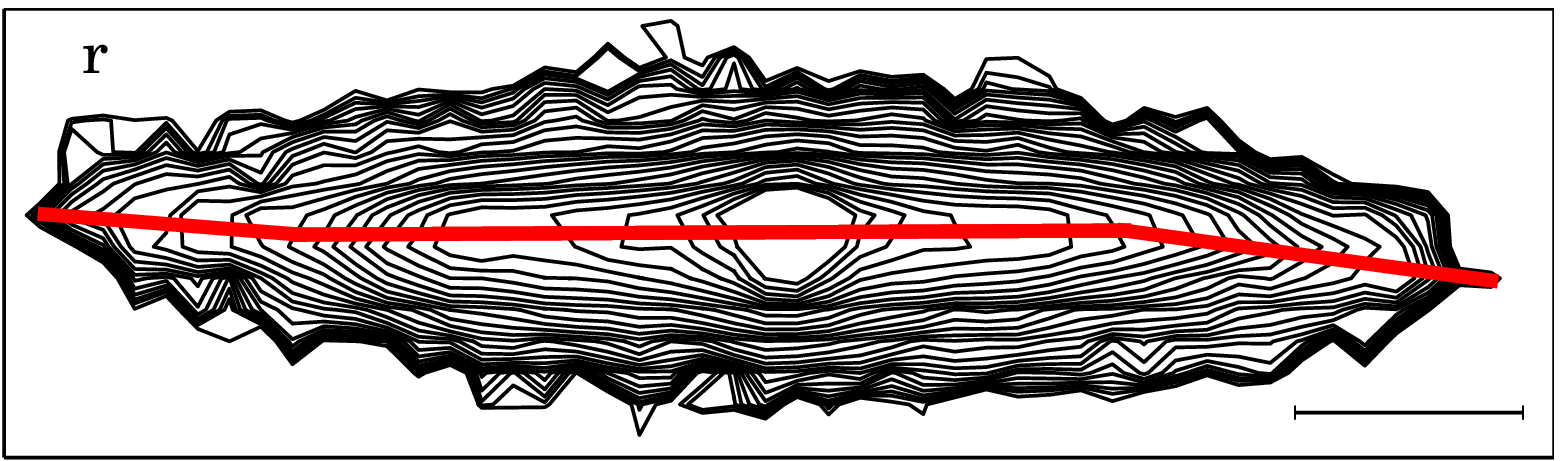}
\includegraphics[width=4.4cm, angle=0, clip=]{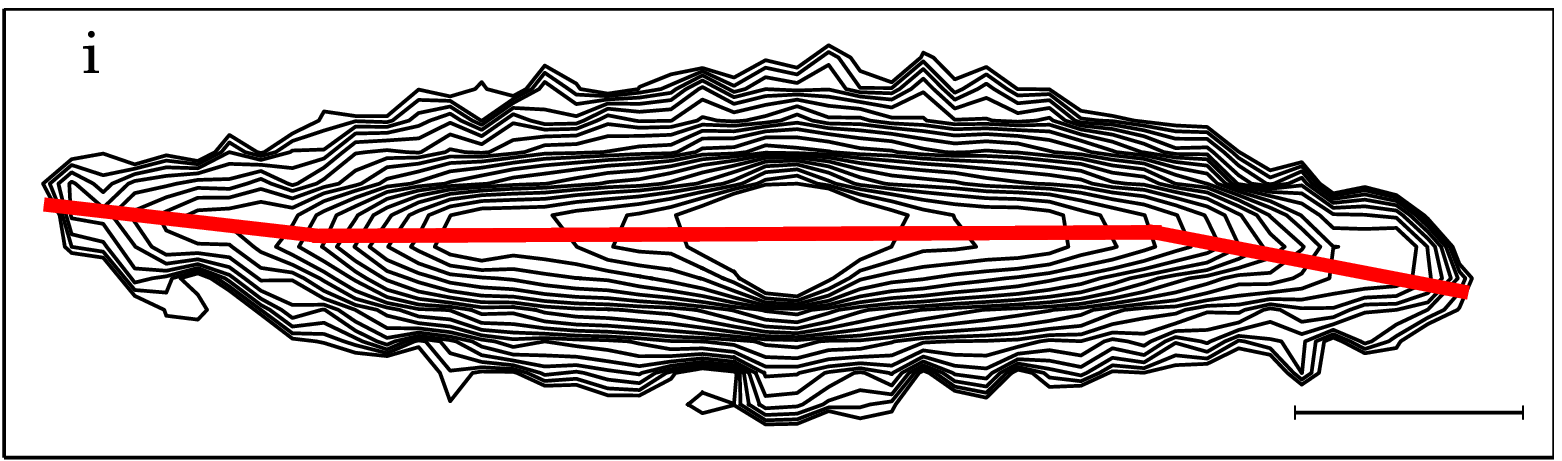}
\includegraphics[width=4.4cm, angle=0, clip=]{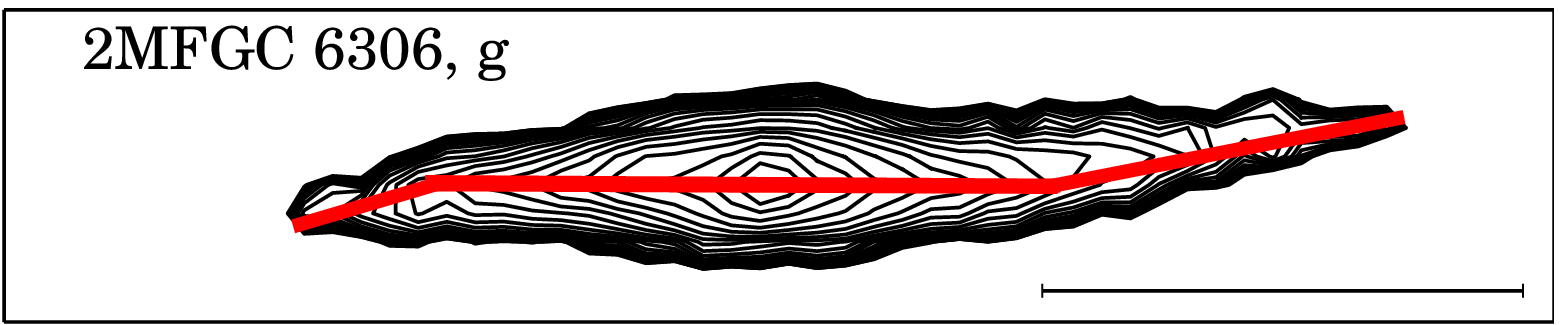}
\includegraphics[width=4.4cm, angle=0, clip=]{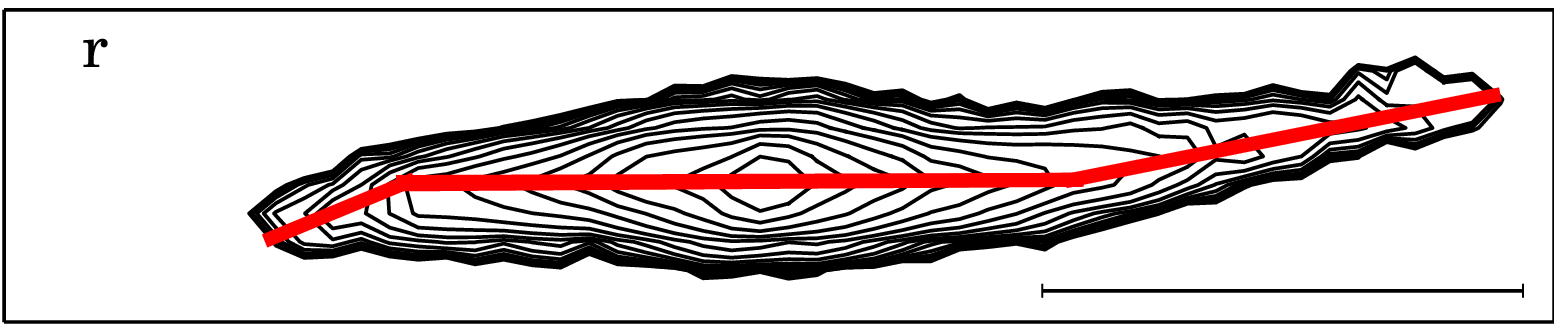}
\includegraphics[width=4.4cm, angle=0, clip=]{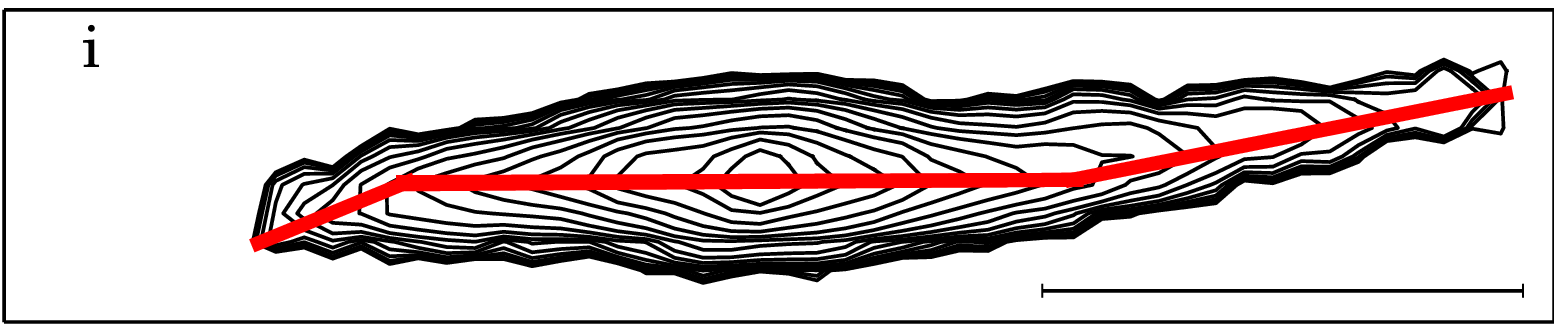}
\includegraphics[width=4.4cm, angle=0, clip=]{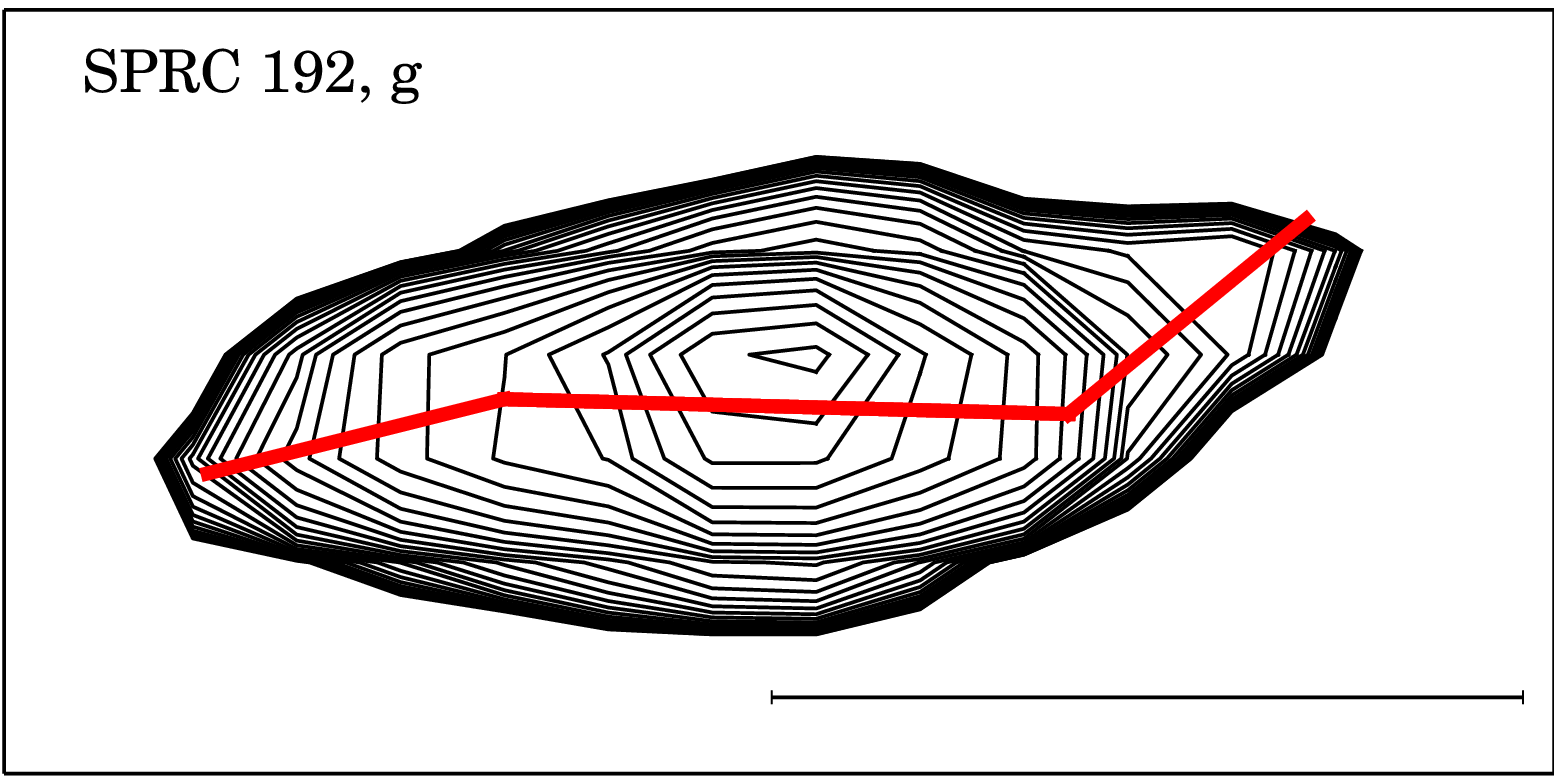}
\includegraphics[width=4.4cm, angle=0, clip=]{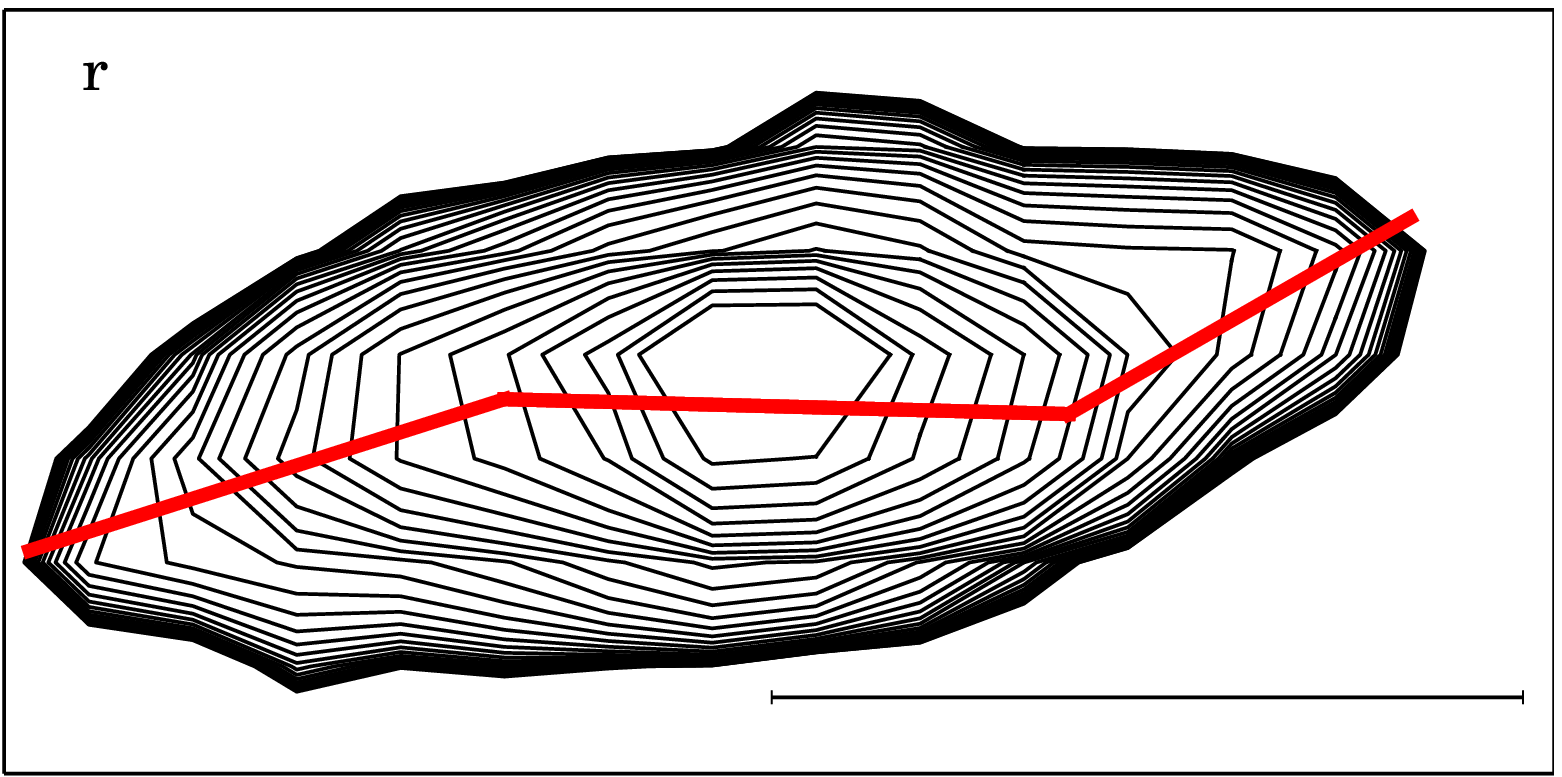}
\includegraphics[width=4.4cm, angle=0, clip=]{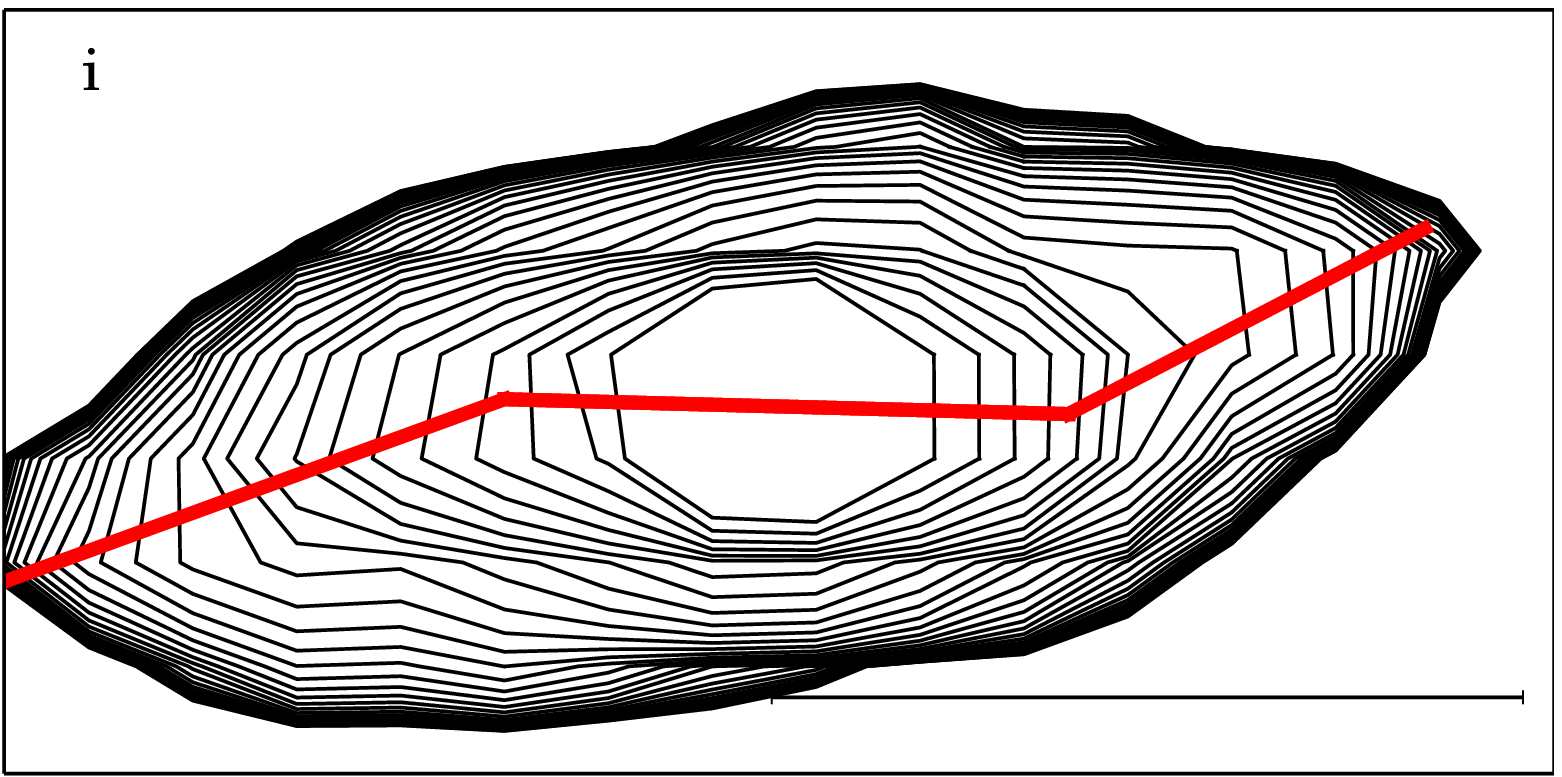}
\includegraphics[width=4.4cm, angle=0, clip=]{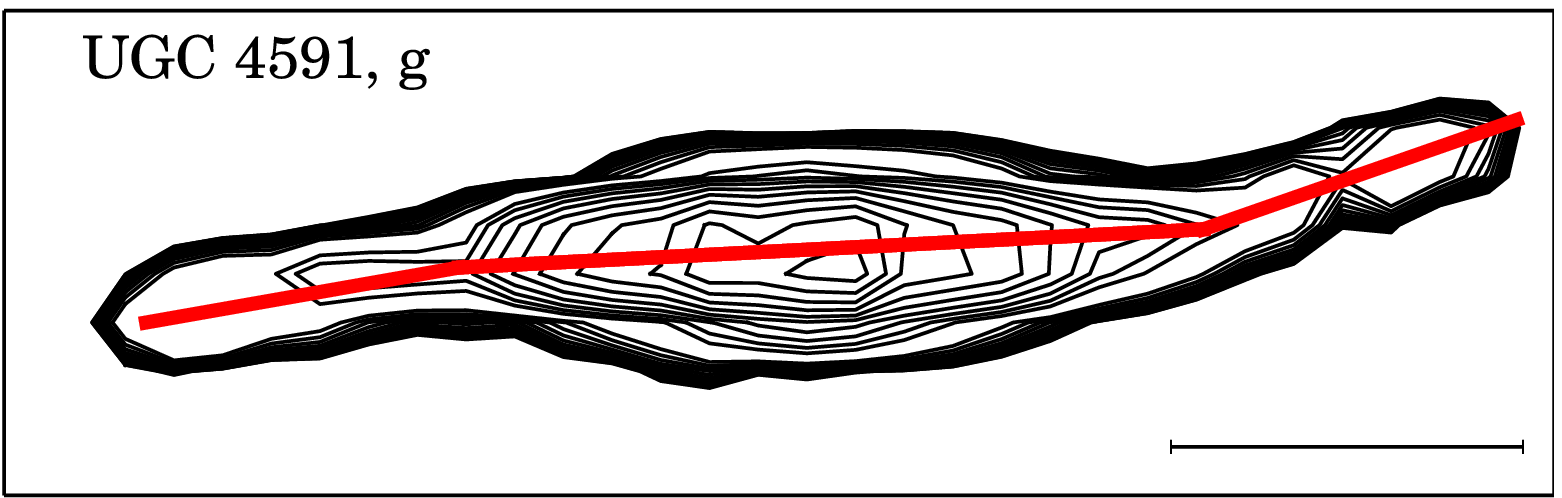}
\includegraphics[width=4.4cm, angle=0, clip=]{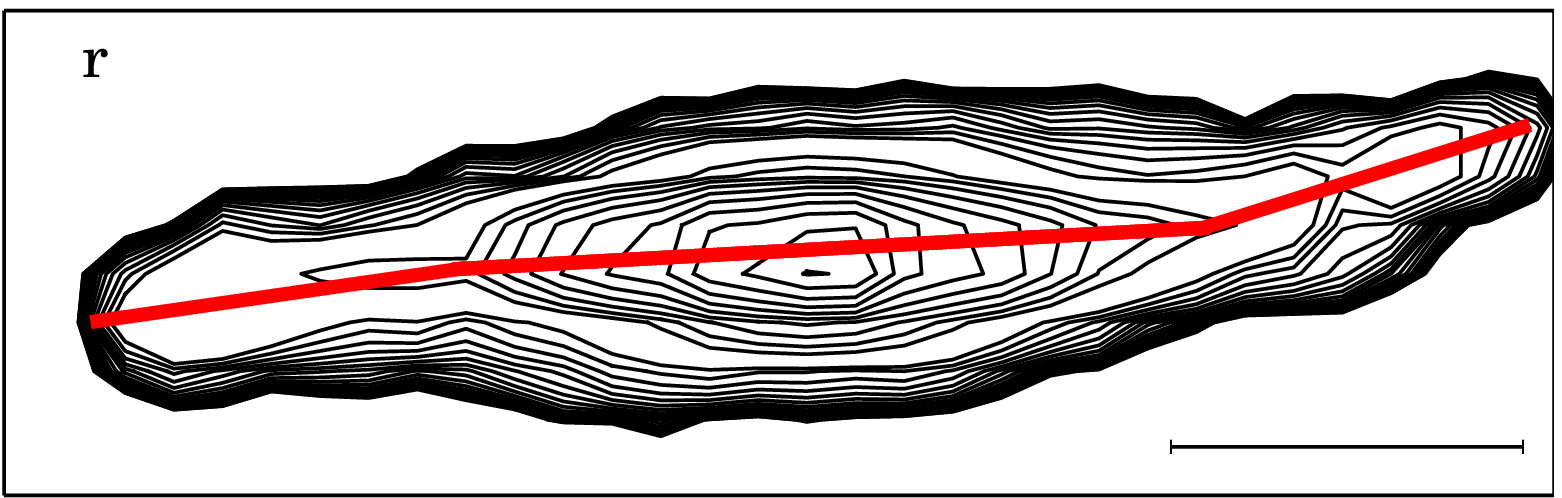}
\includegraphics[width=4.4cm, angle=0, clip=]{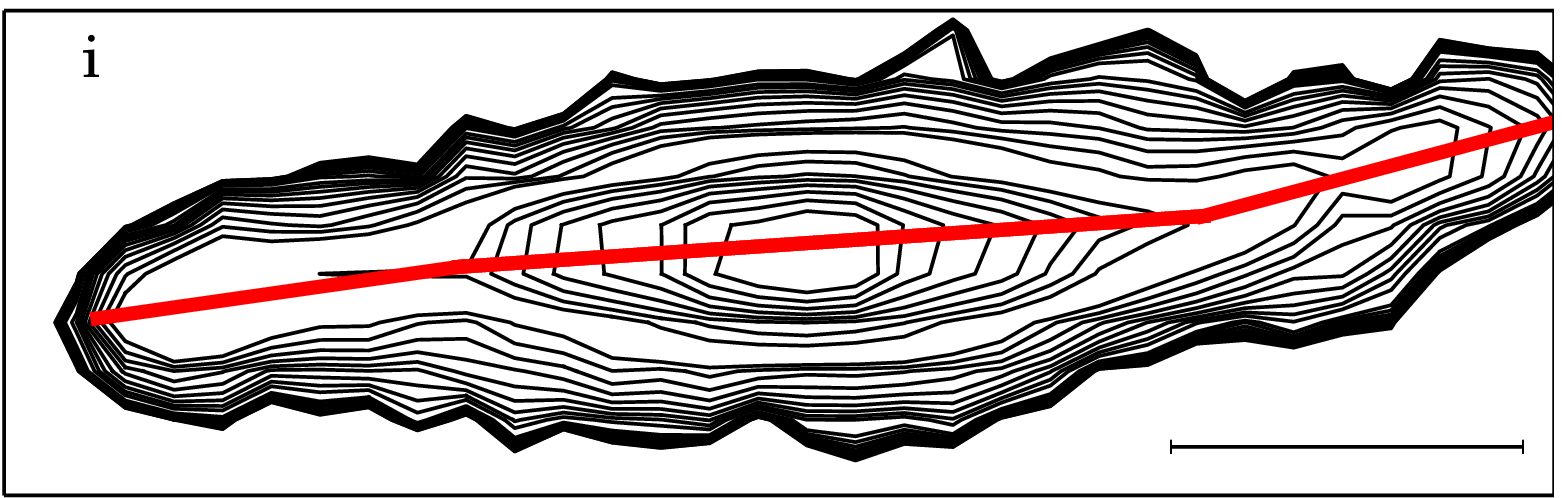}
\includegraphics[width=4.4cm, angle=0, clip=]{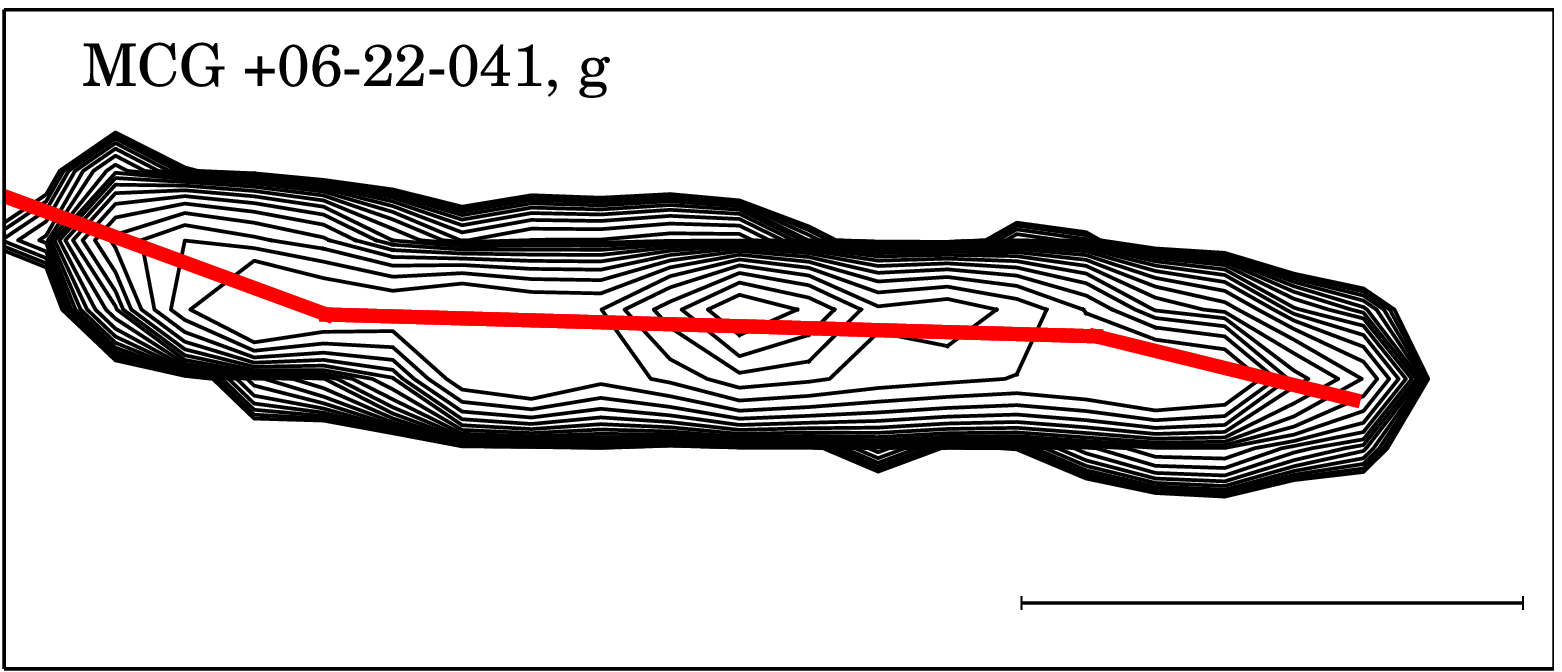}
\includegraphics[width=4.4cm, angle=0, clip=]{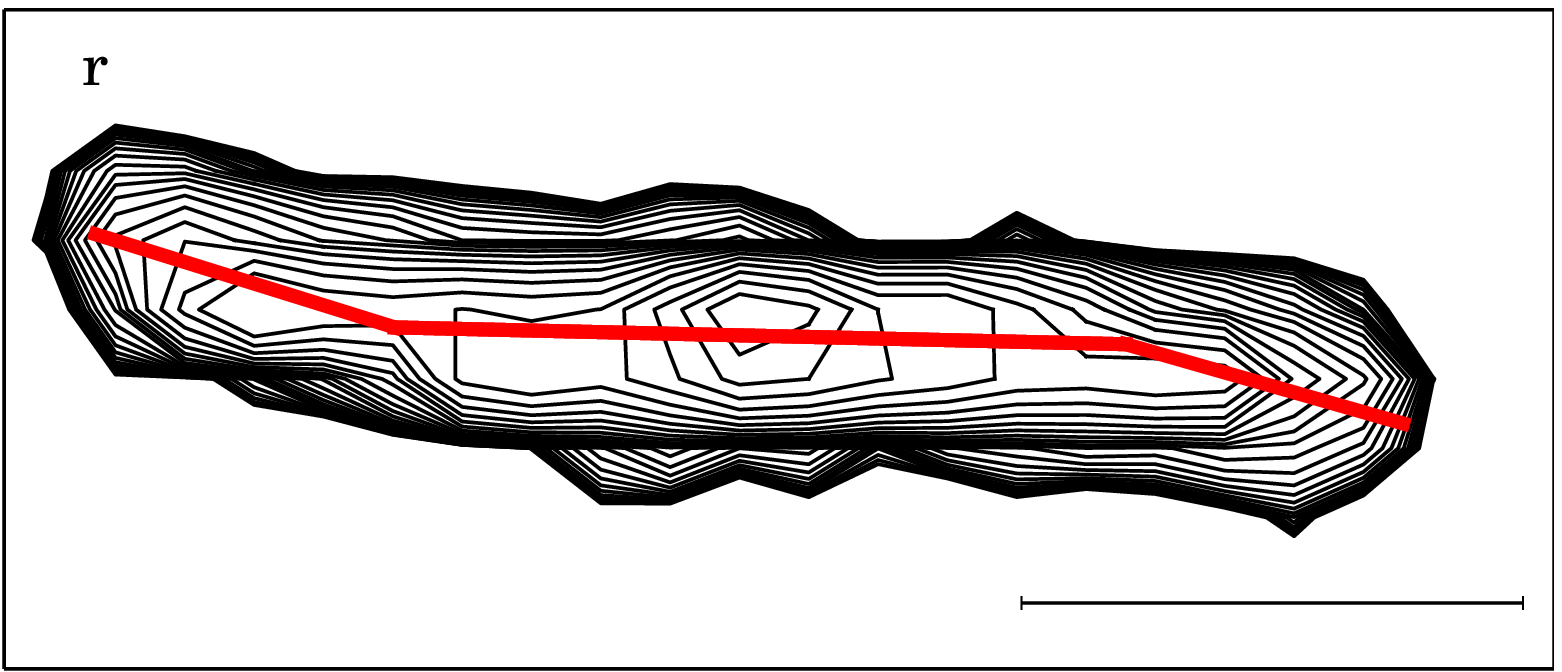}
\includegraphics[width=4.4cm, angle=0, clip=]{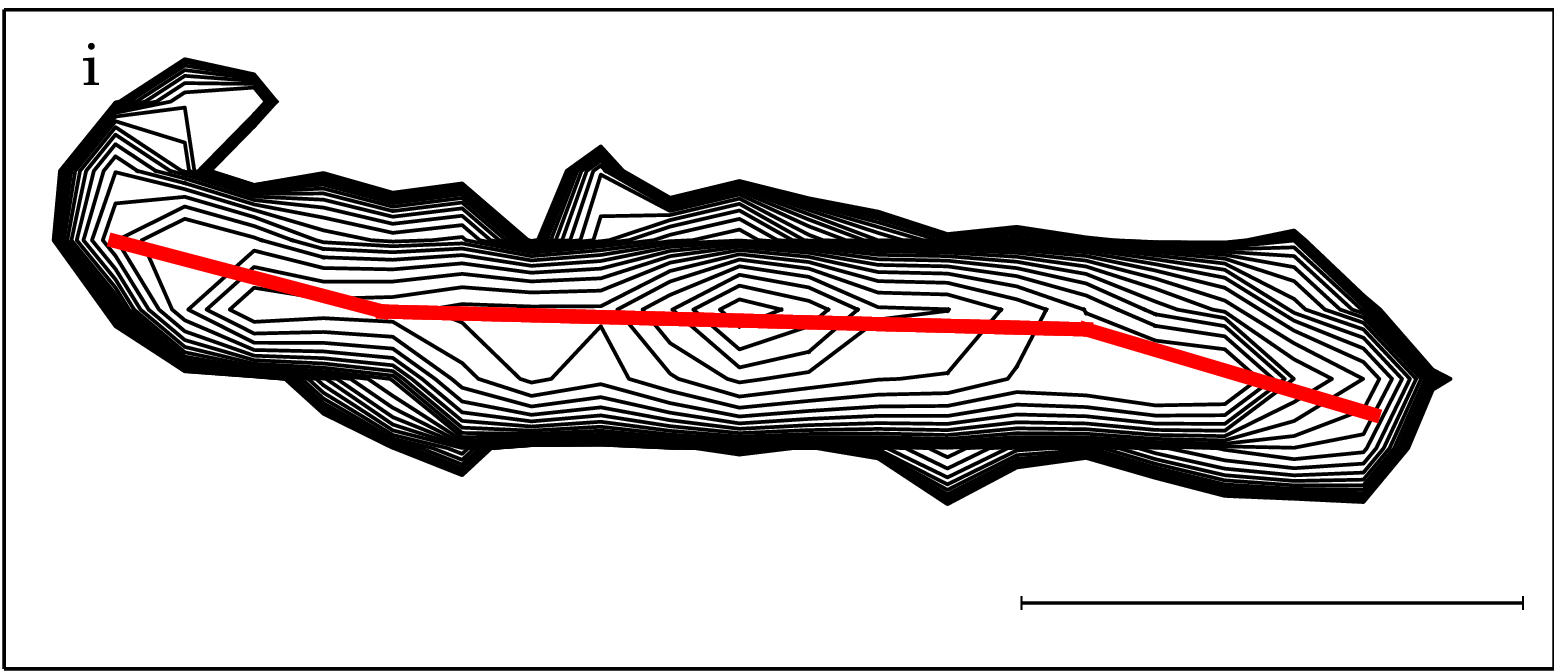}
\includegraphics[width=4.4cm, angle=0, clip=]{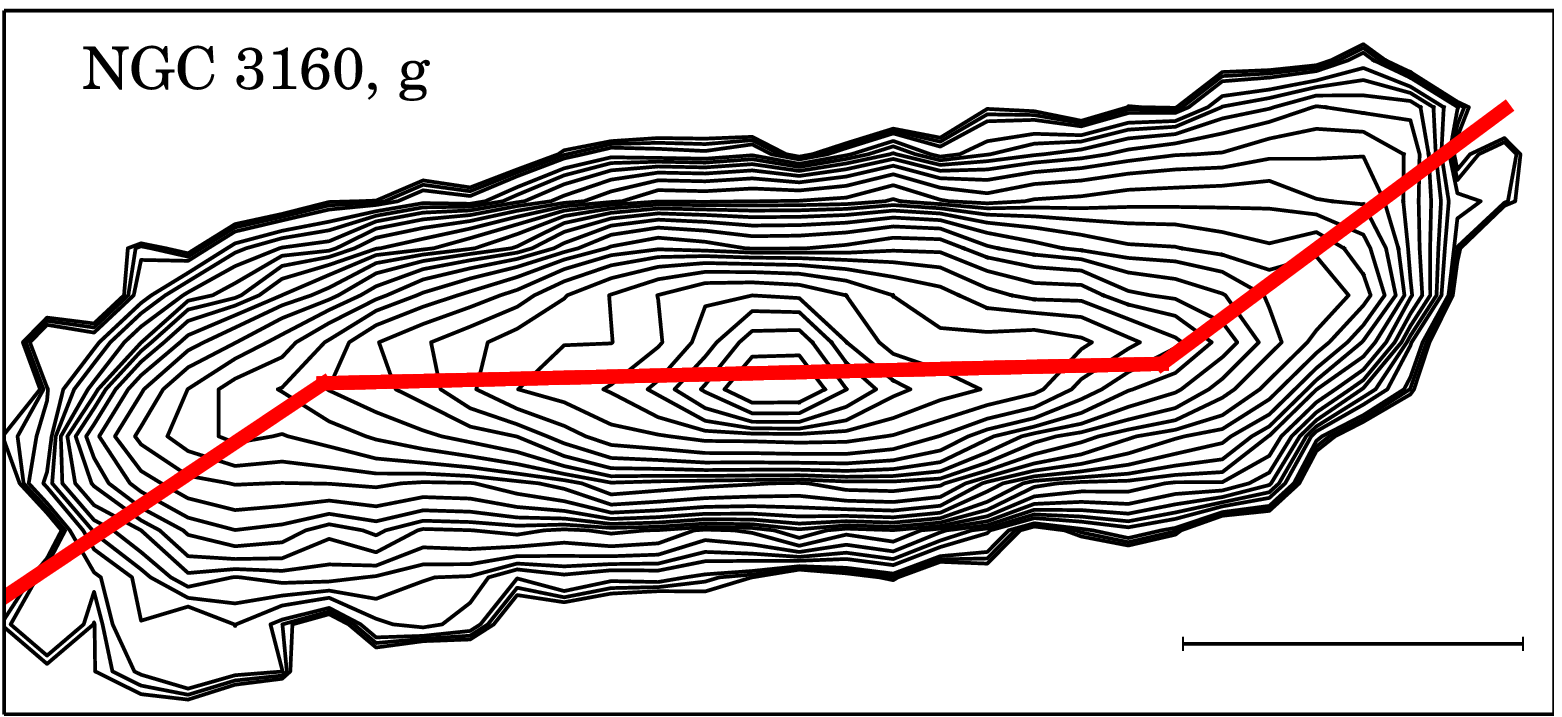}
\includegraphics[width=4.4cm, angle=0, clip=]{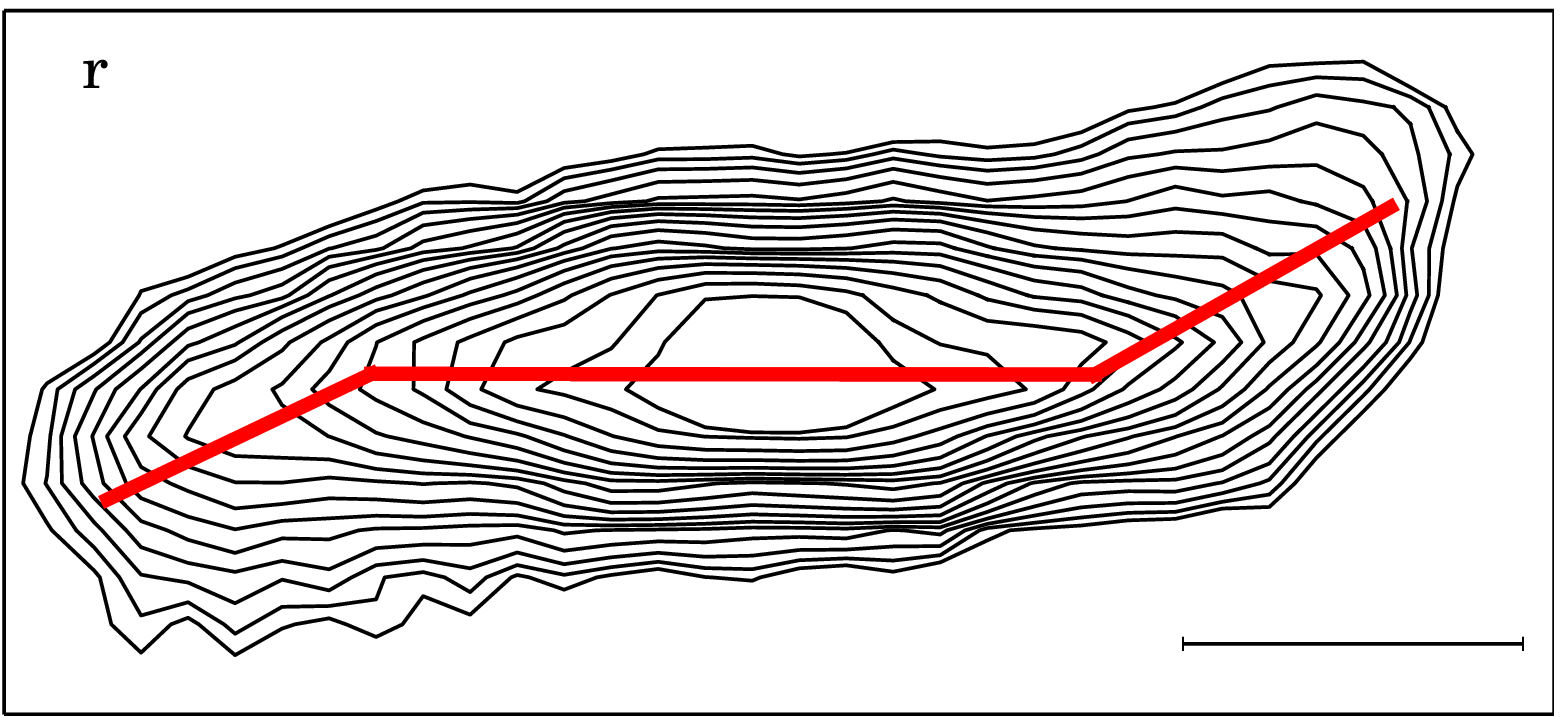}
\includegraphics[width=4.4cm, angle=0, clip=]{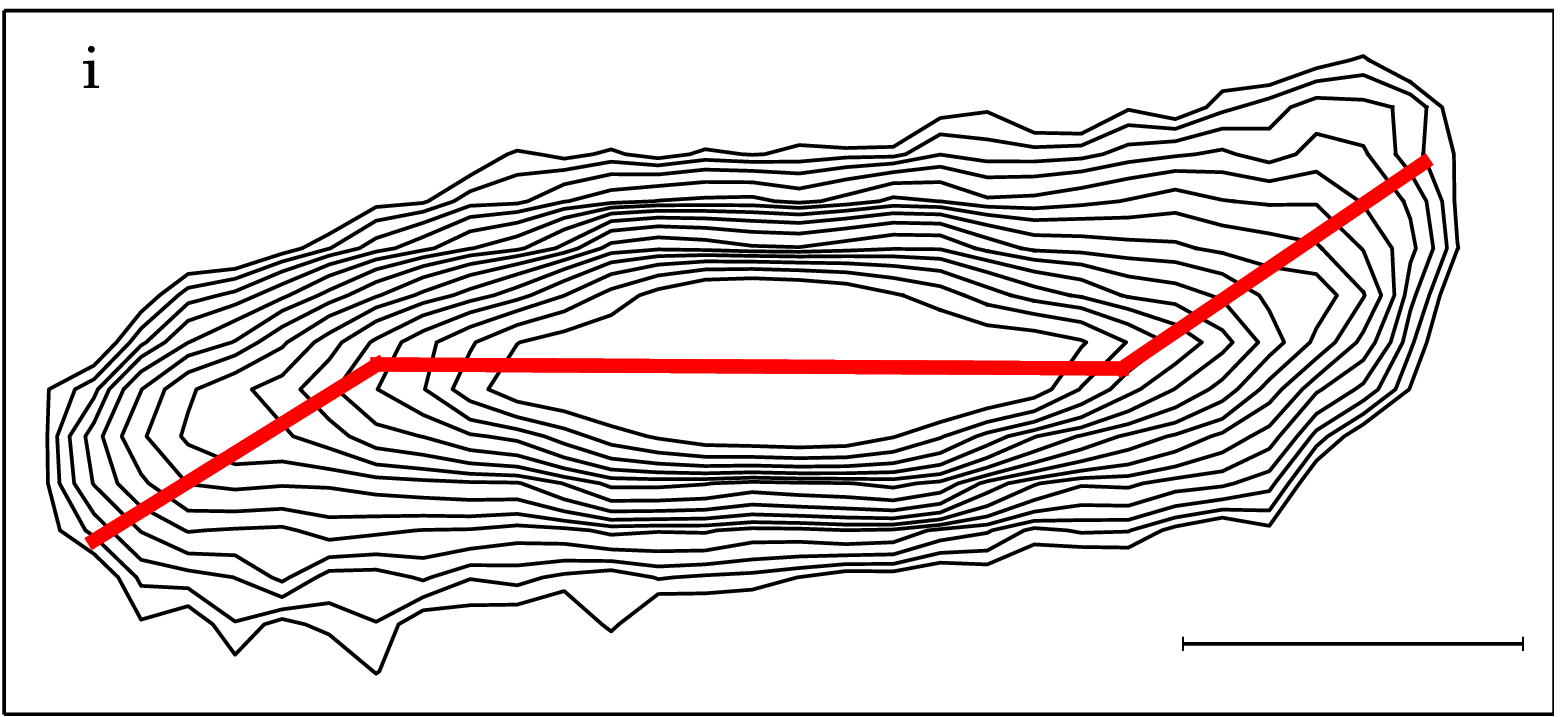}
\includegraphics[width=4.4cm, angle=0, clip=]{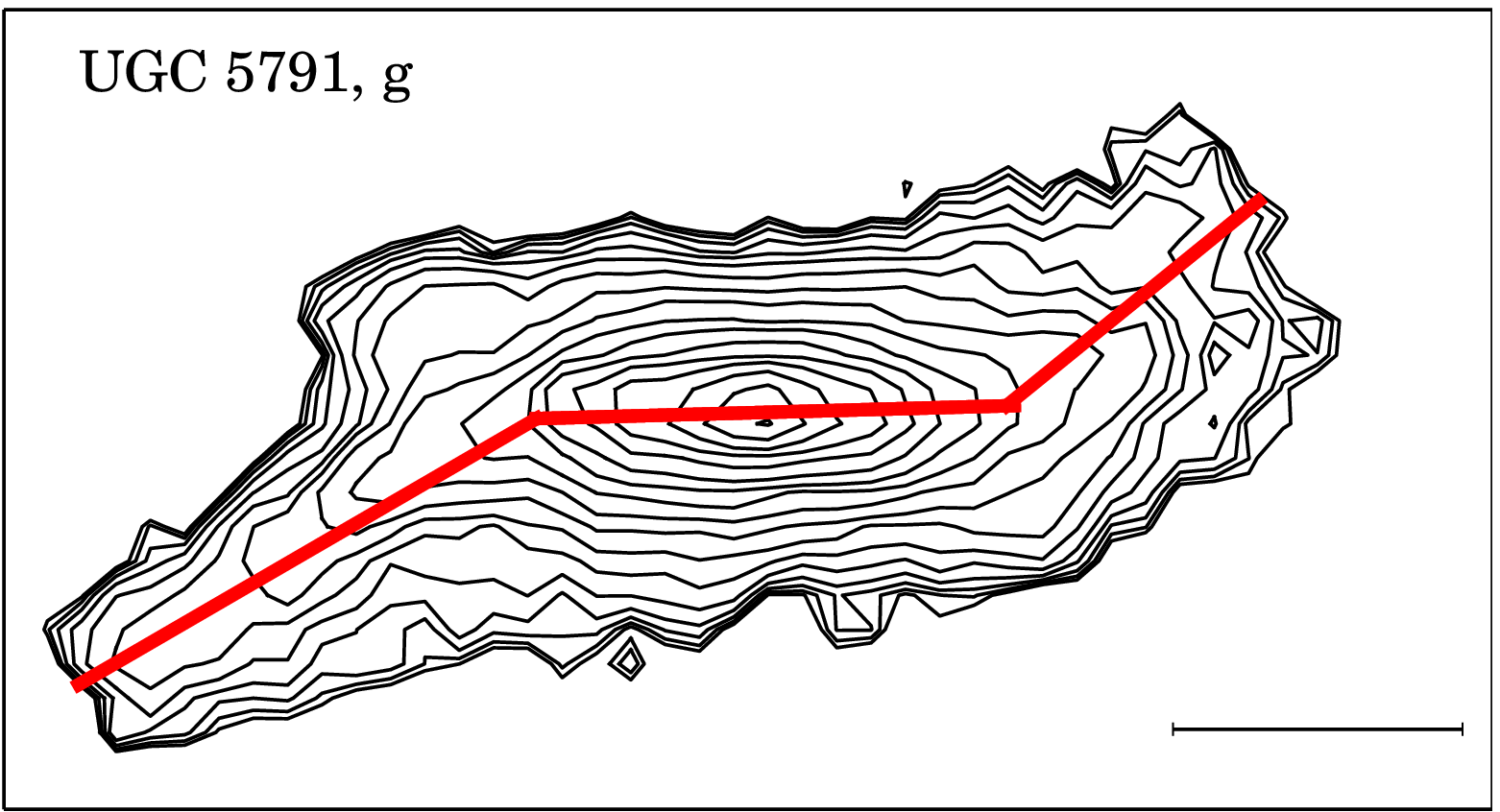}
\includegraphics[width=4.4cm, angle=0, clip=]{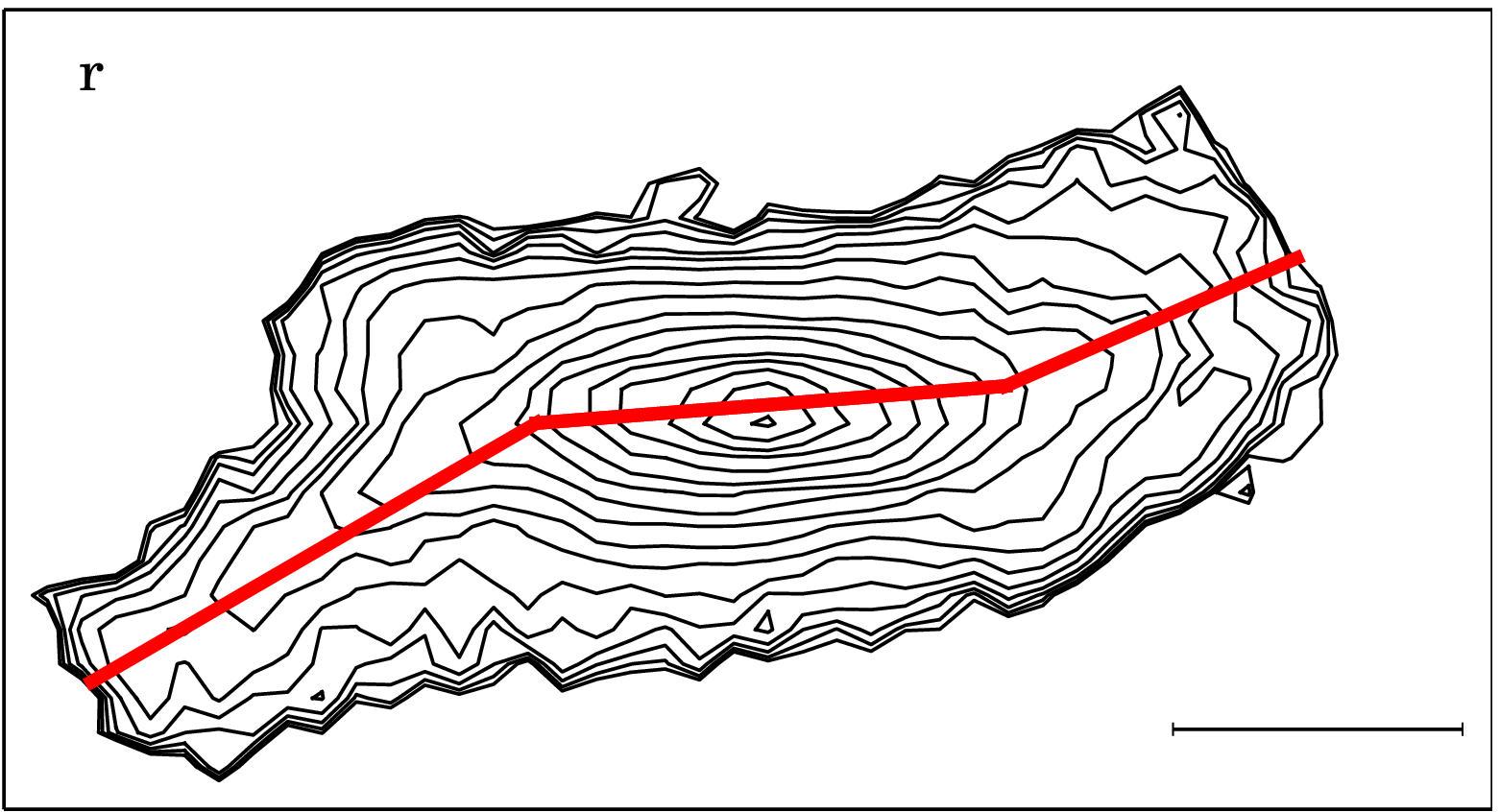}
\includegraphics[width=4.4cm, angle=0, clip=]{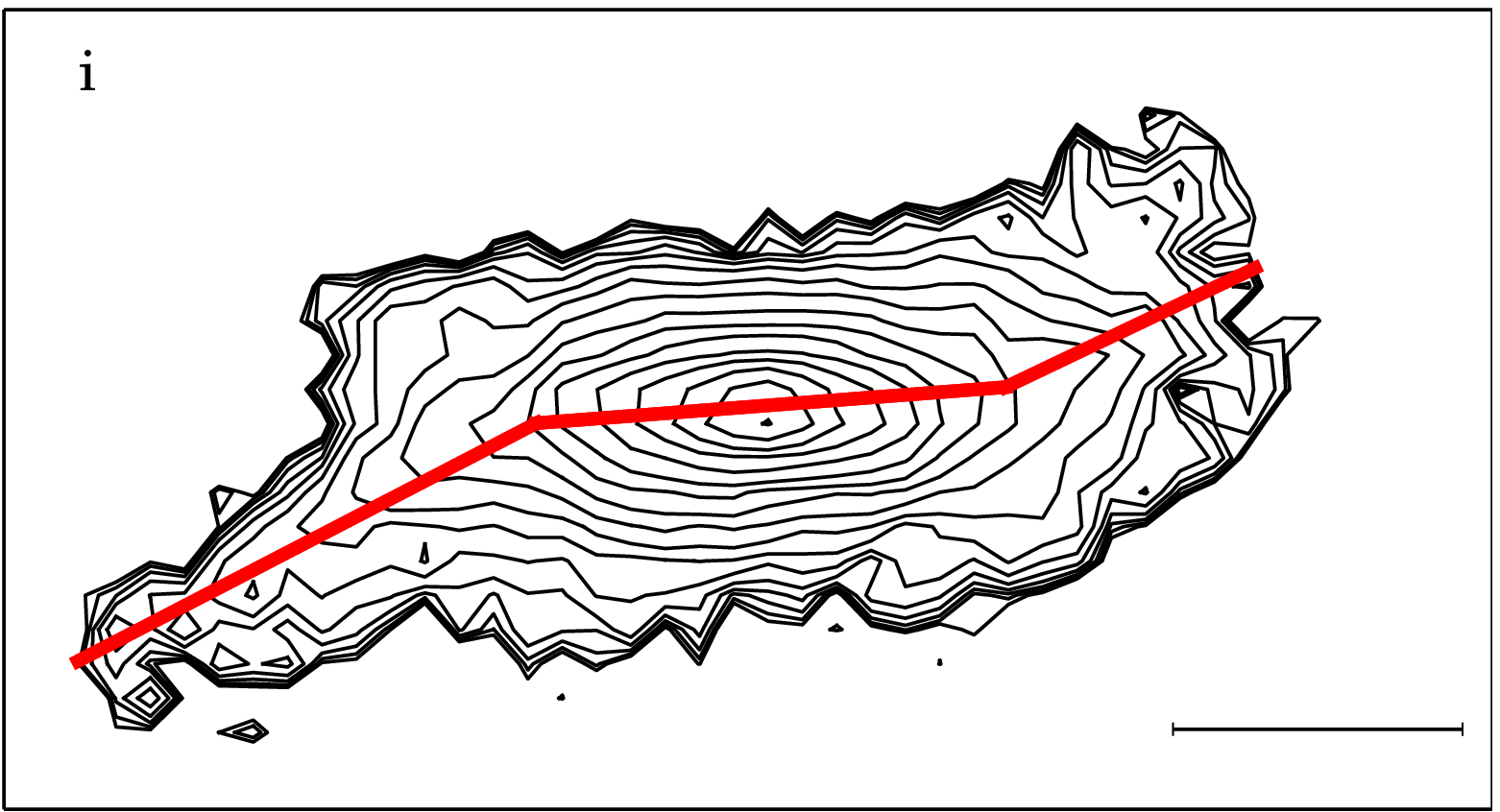}
\includegraphics[width=4.4cm, angle=0, clip=]{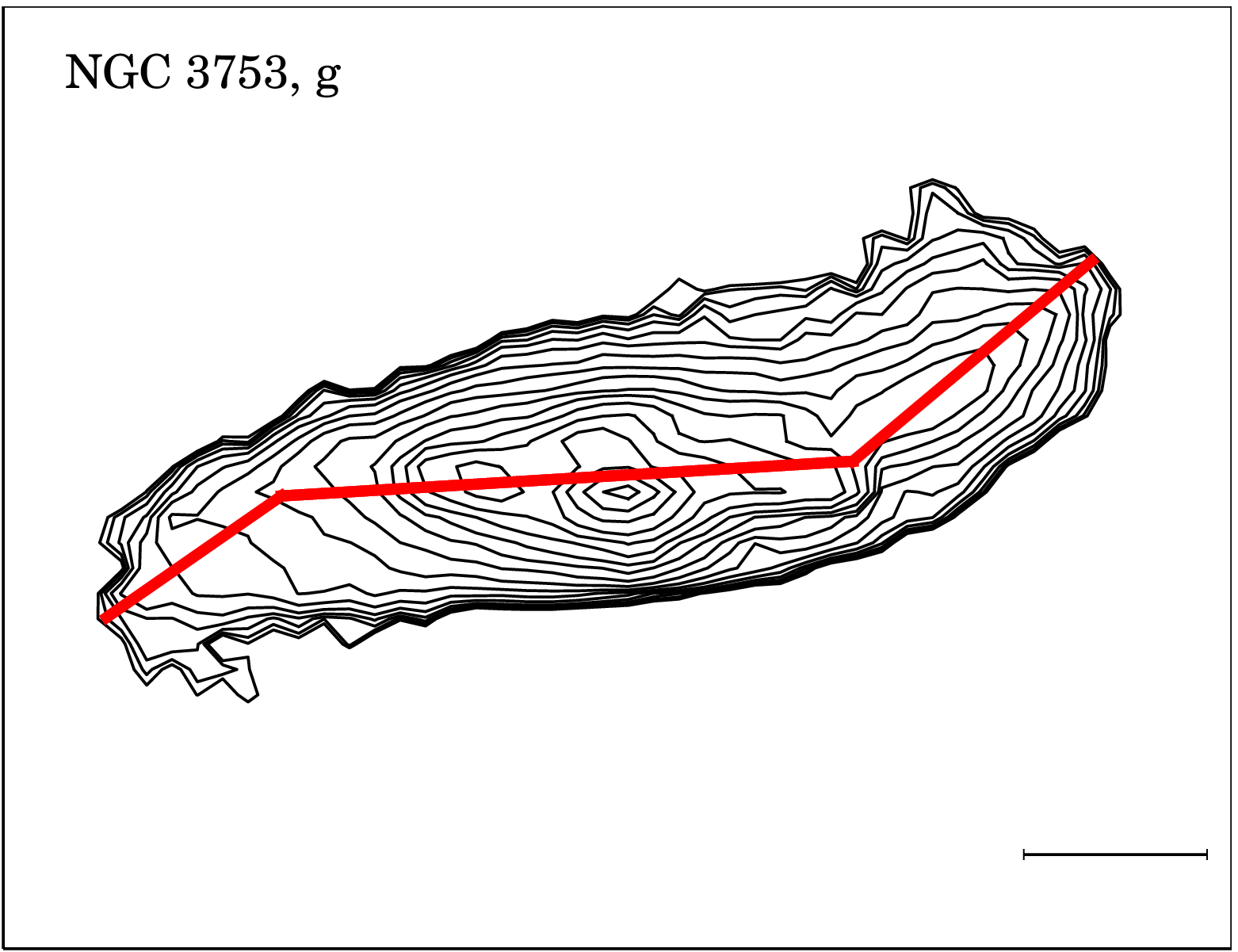}
\includegraphics[width=4.4cm, angle=0, clip=]{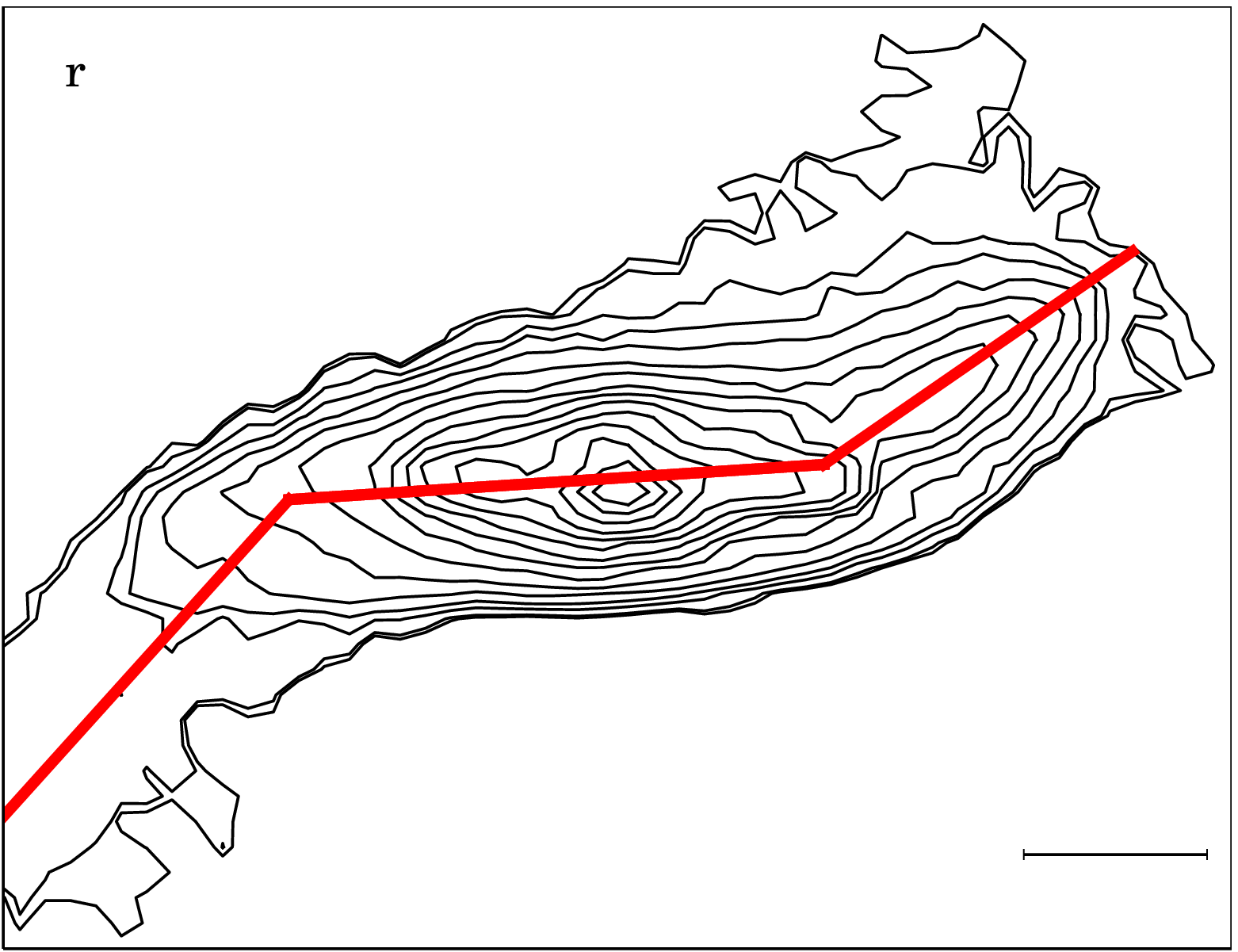}
\includegraphics[width=4.4cm, angle=0, clip=]{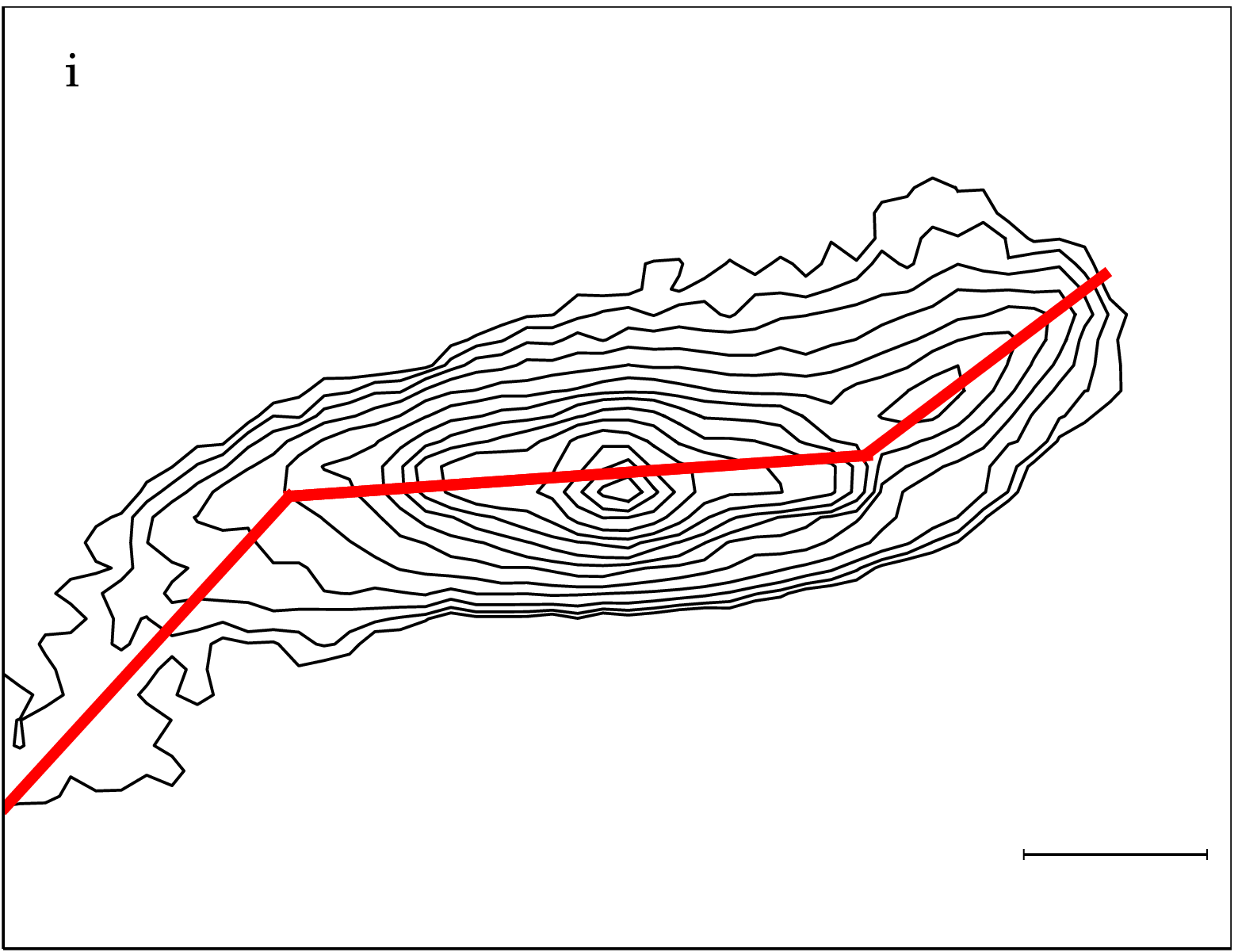}
\includegraphics[width=4.4cm, angle=0, clip=]{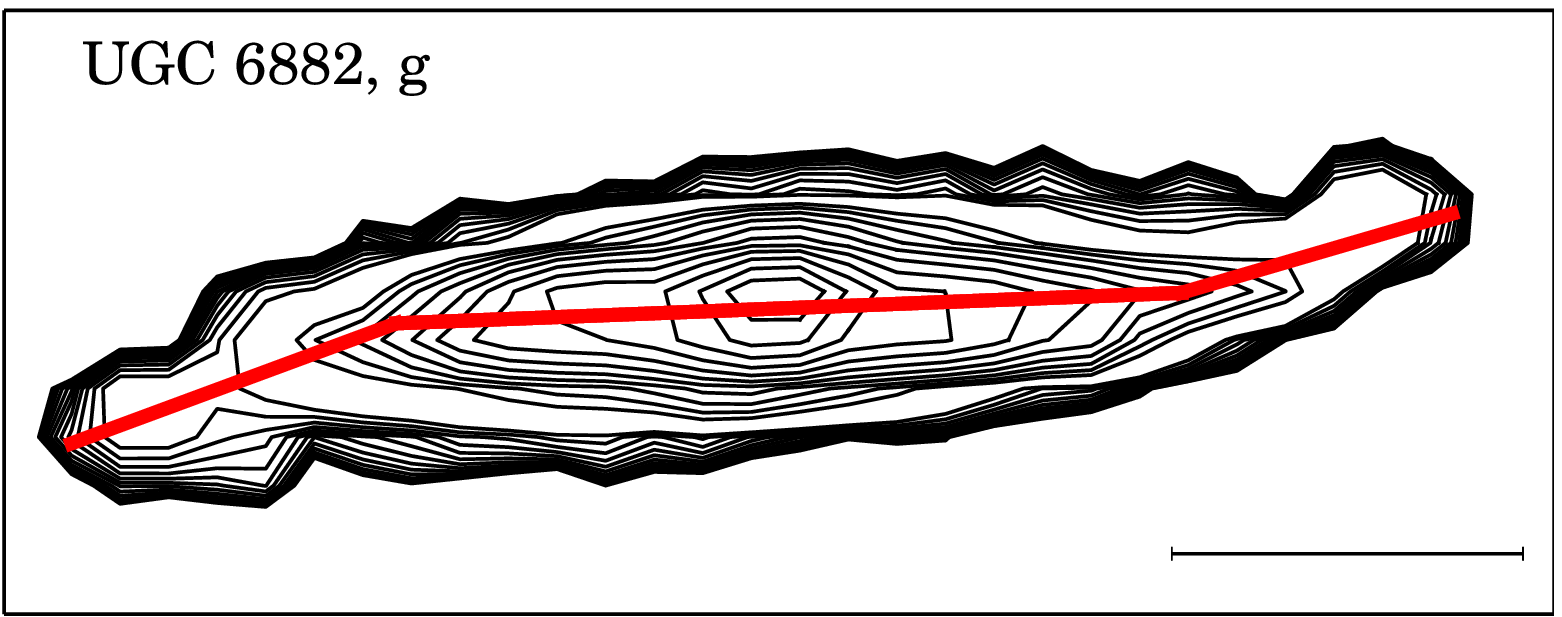}
\includegraphics[width=4.4cm, angle=0, clip=]{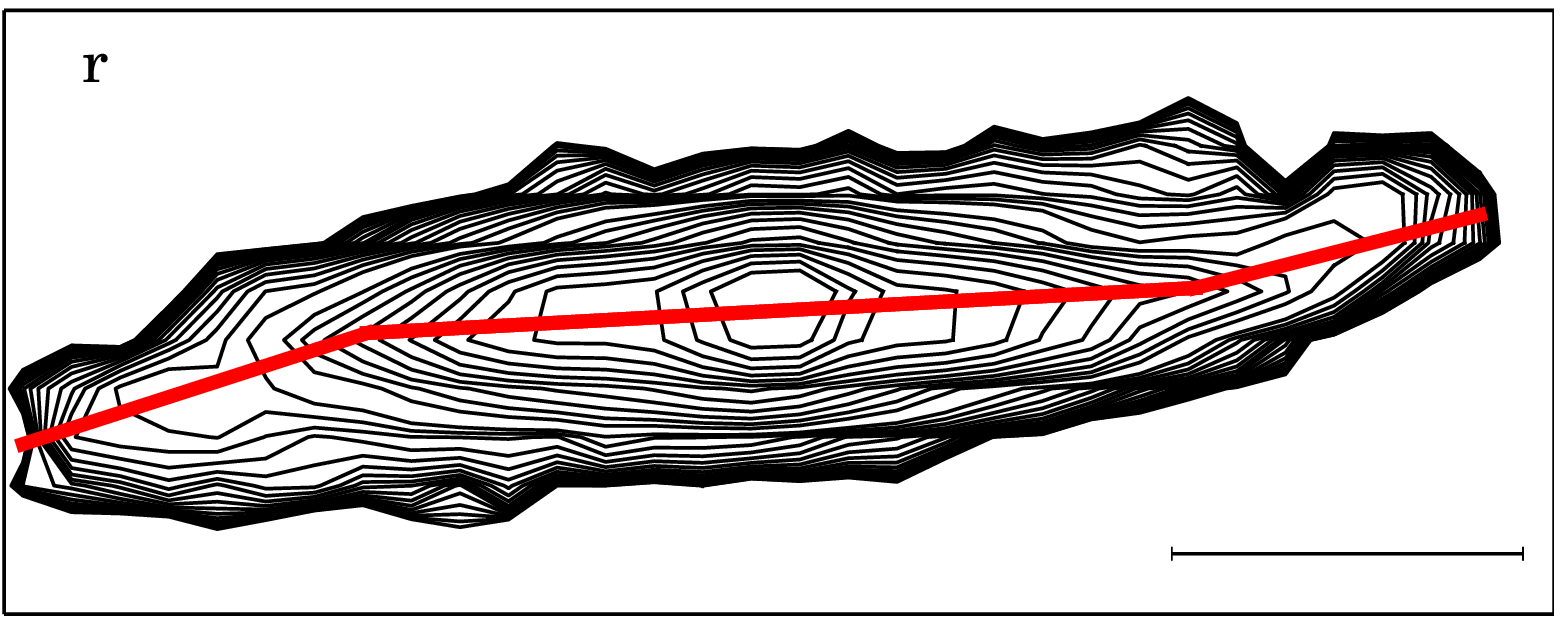}
\includegraphics[width=4.4cm, angle=0, clip=]{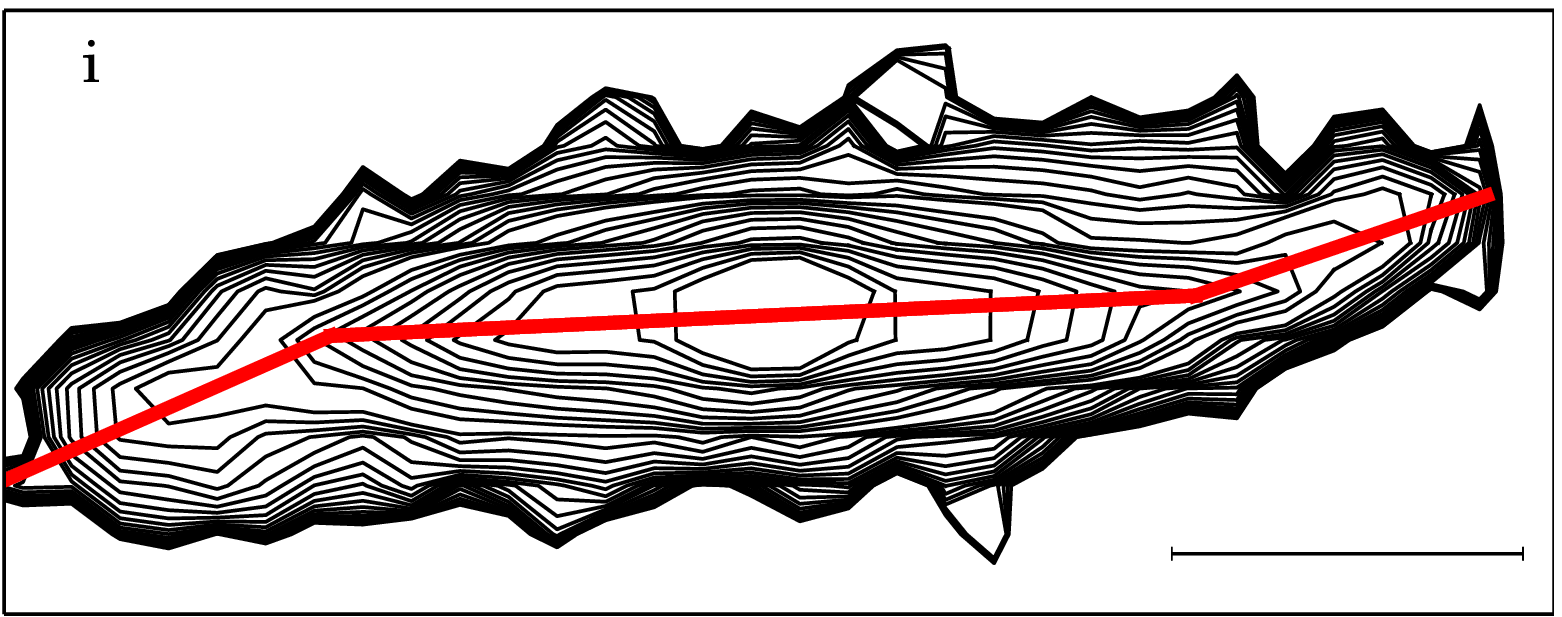}
\includegraphics[width=4.4cm, angle=0, clip=]{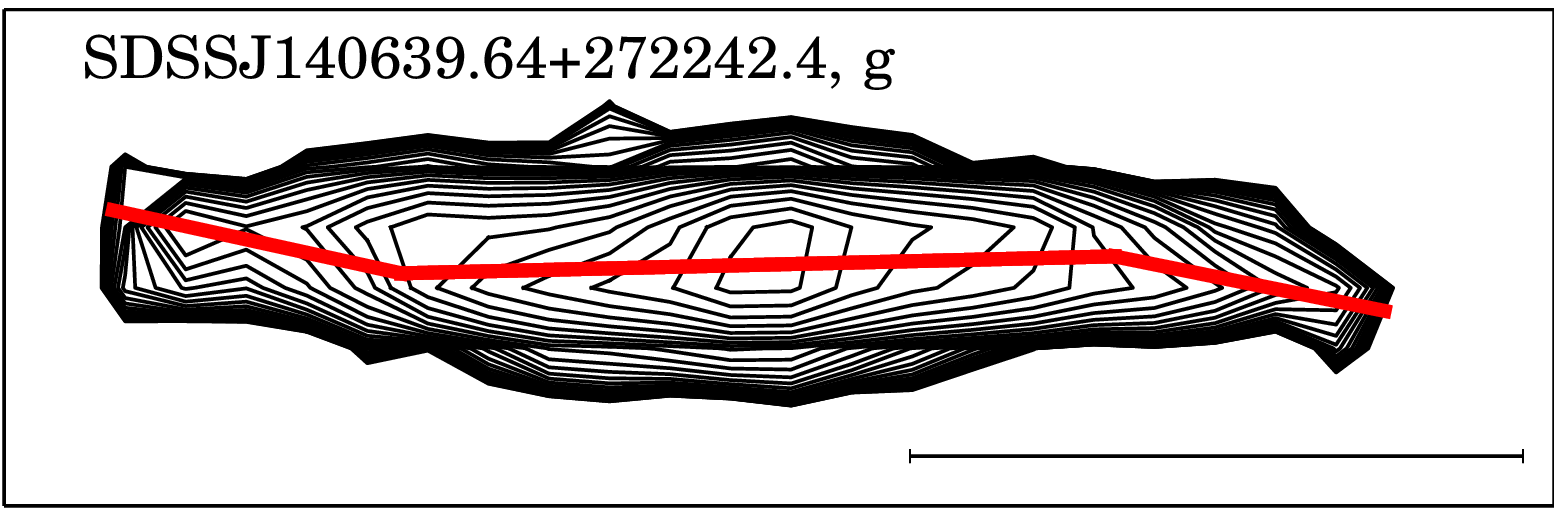}
\includegraphics[width=4.4cm, angle=0, clip=]{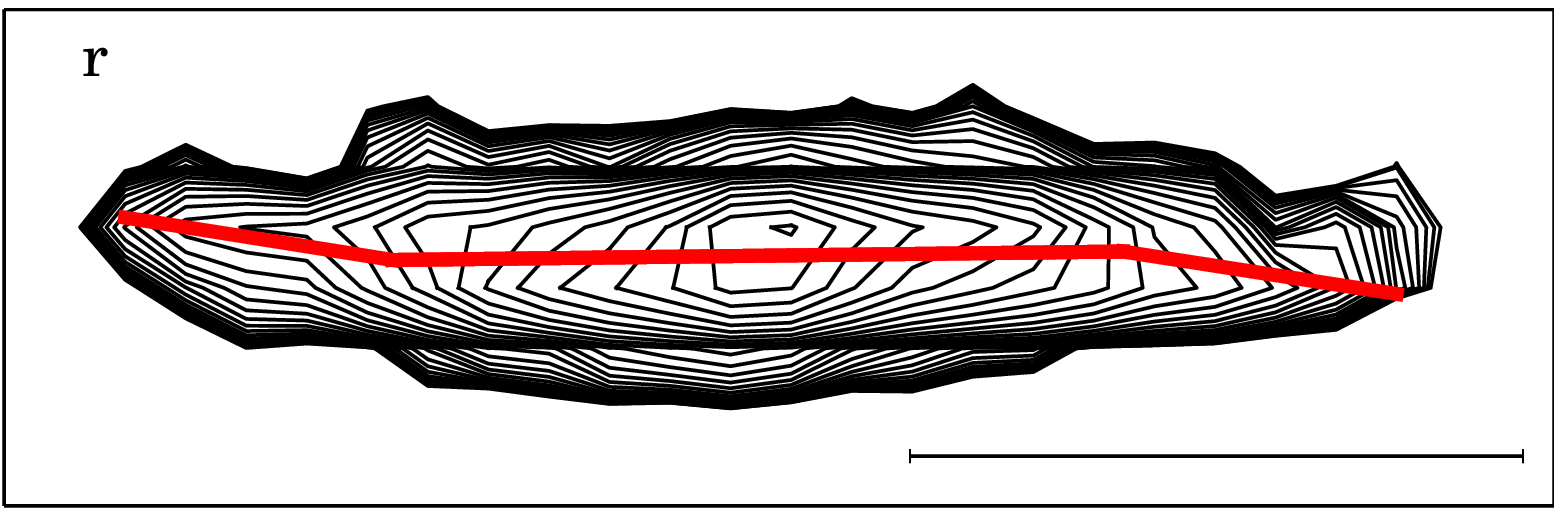}
\includegraphics[width=4.4cm, angle=0, clip=]{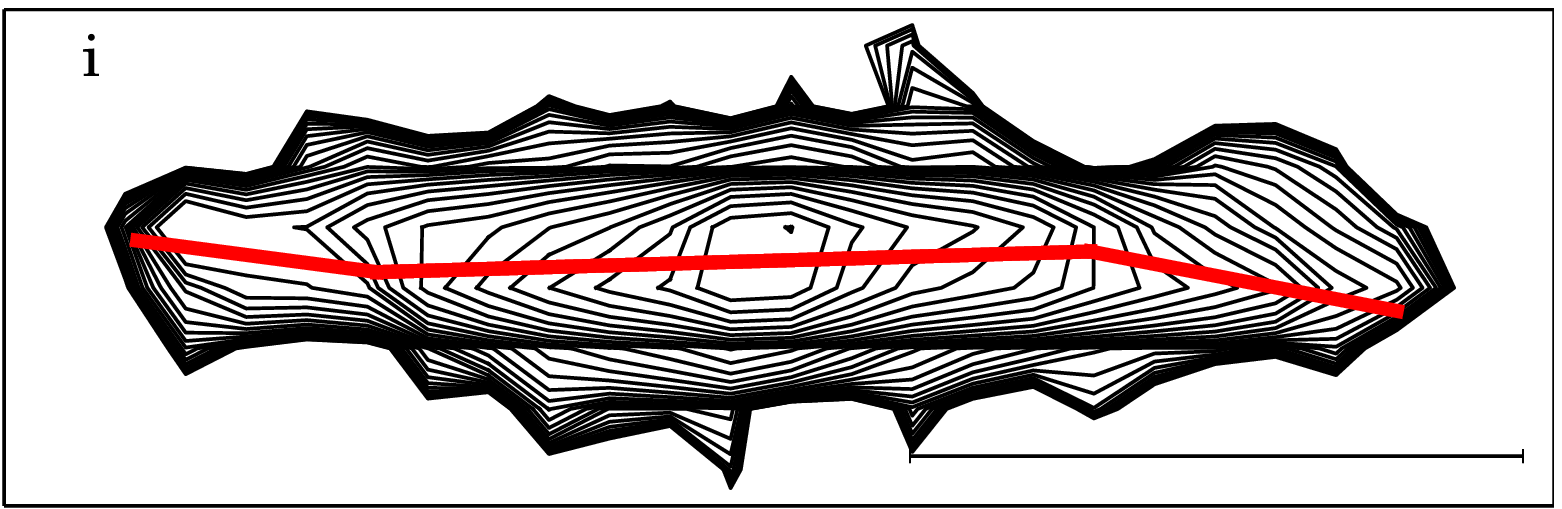}
\includegraphics[width=4.4cm, angle=0, clip=]{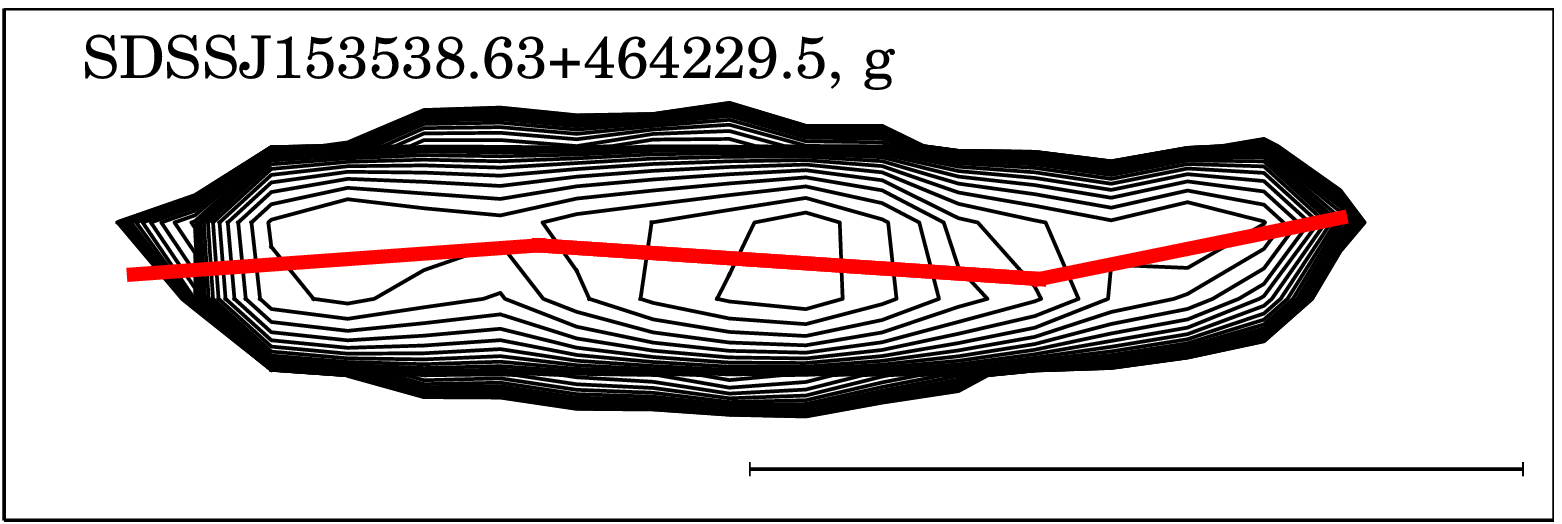}
\includegraphics[width=4.4cm, angle=0, clip=]{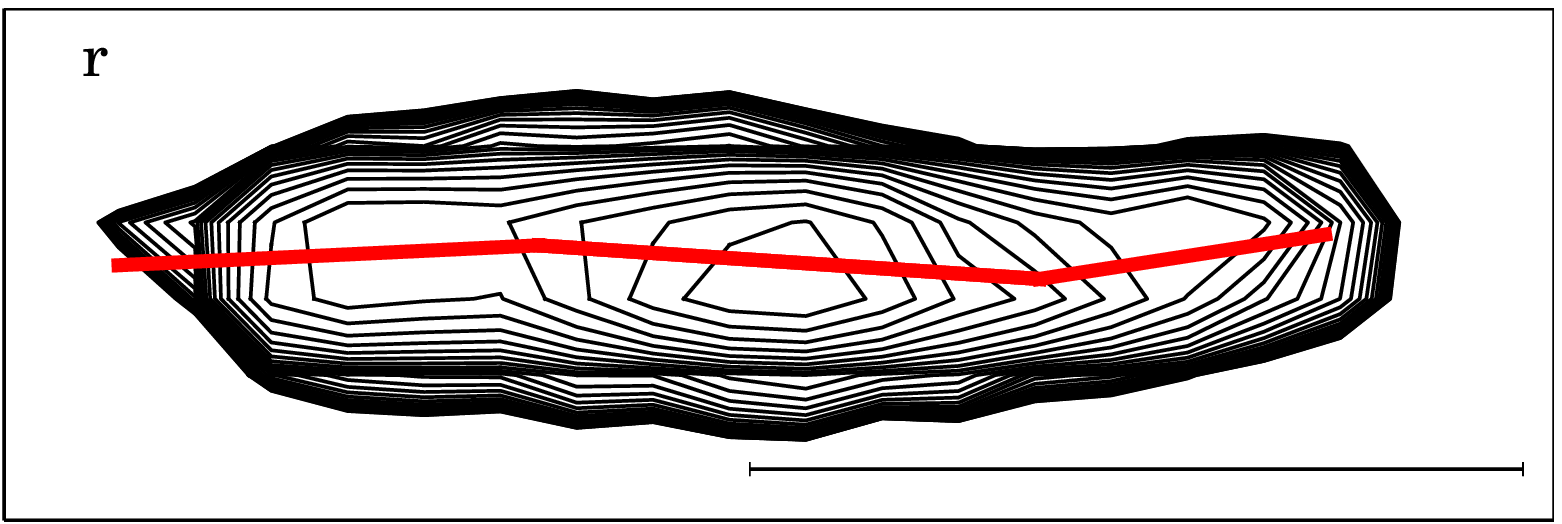}
\includegraphics[width=4.4cm, angle=0, clip=]{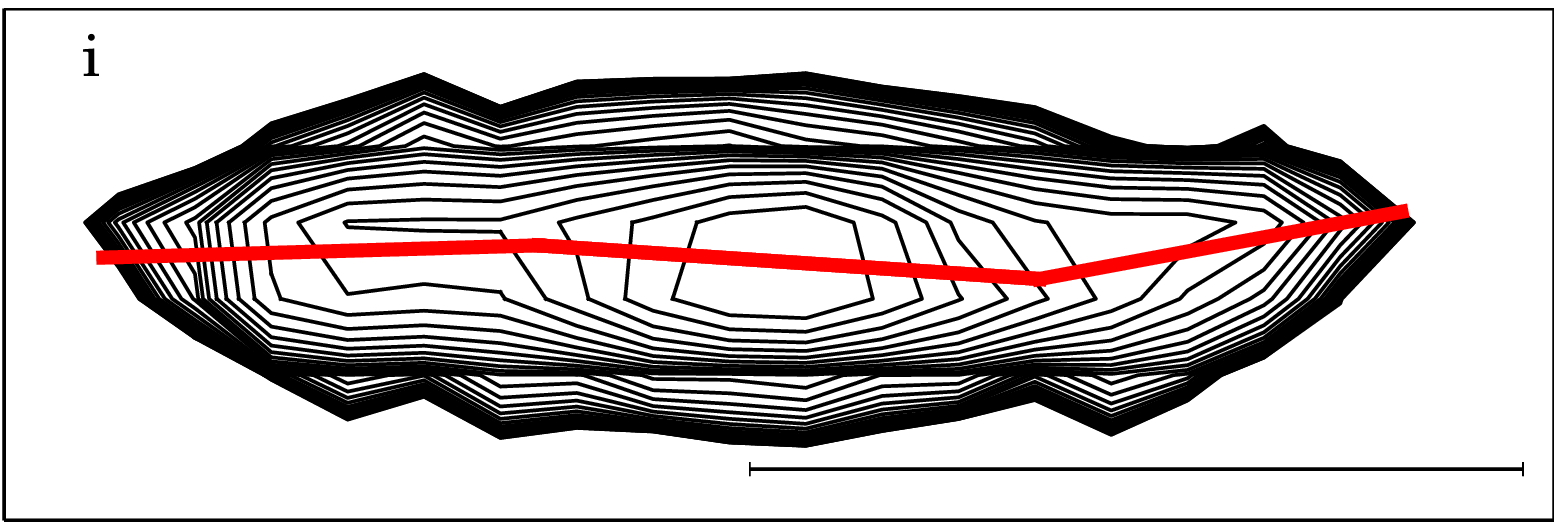}
\includegraphics[width=4.4cm, angle=0, clip=]{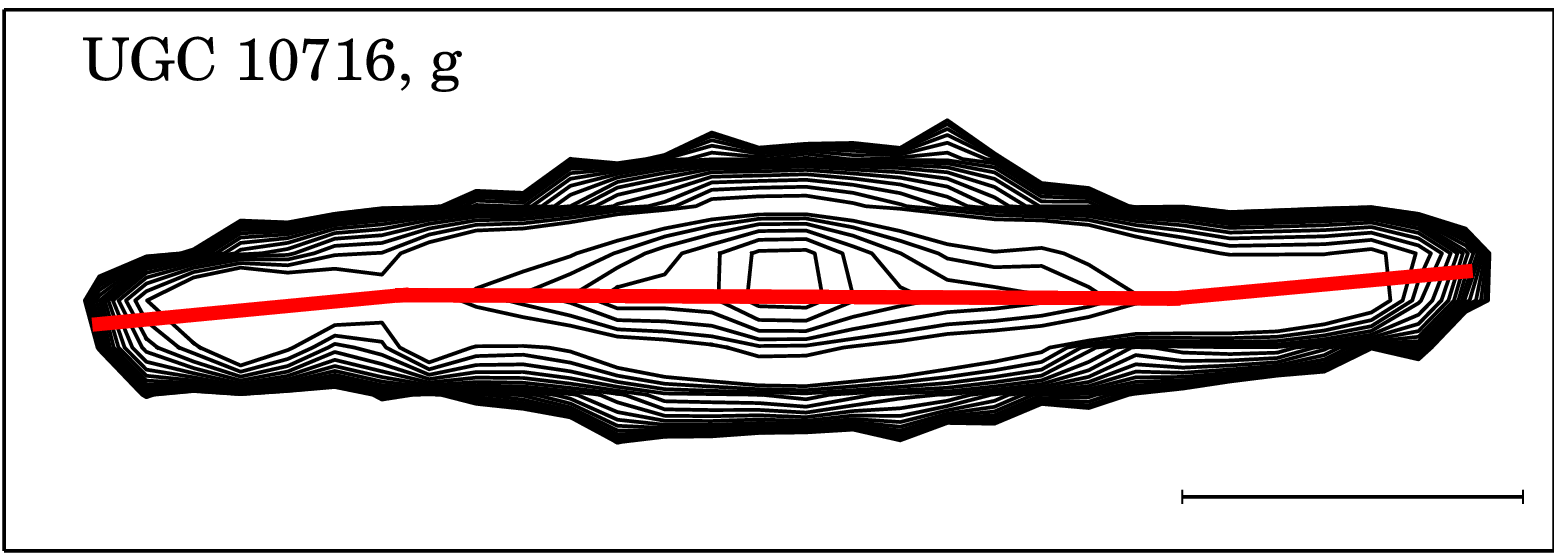}
\includegraphics[width=4.4cm, angle=0, clip=]{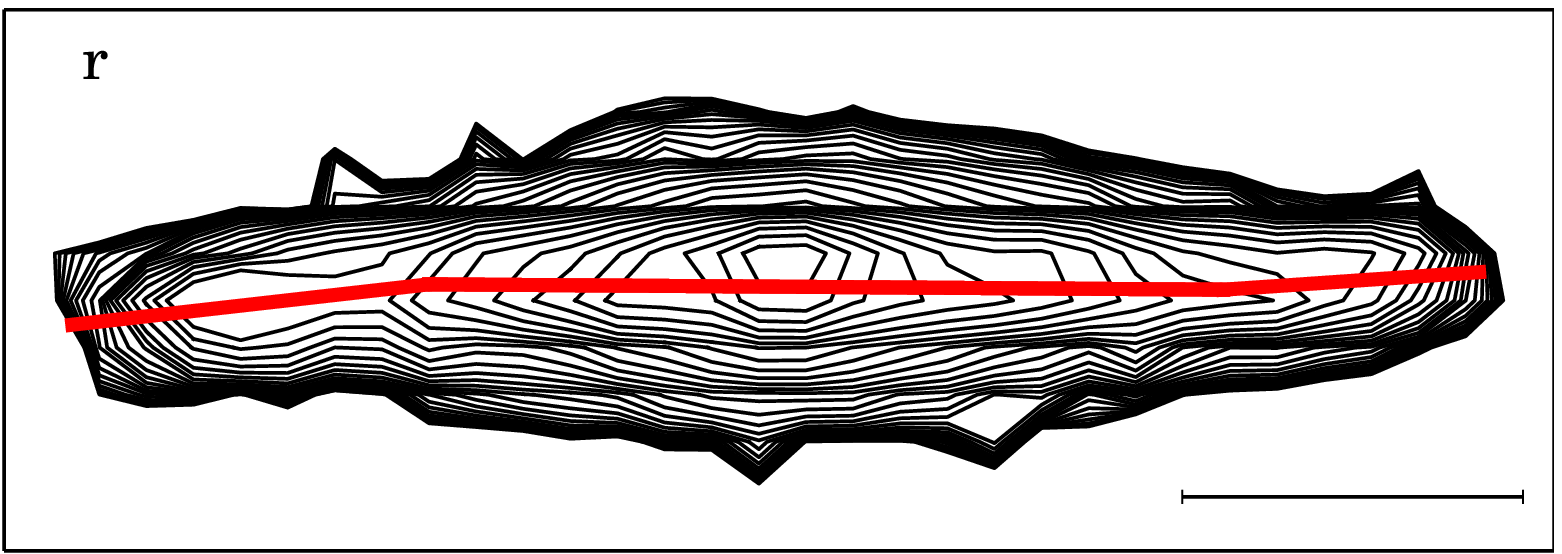}
\includegraphics[width=4.4cm, angle=0, clip=]{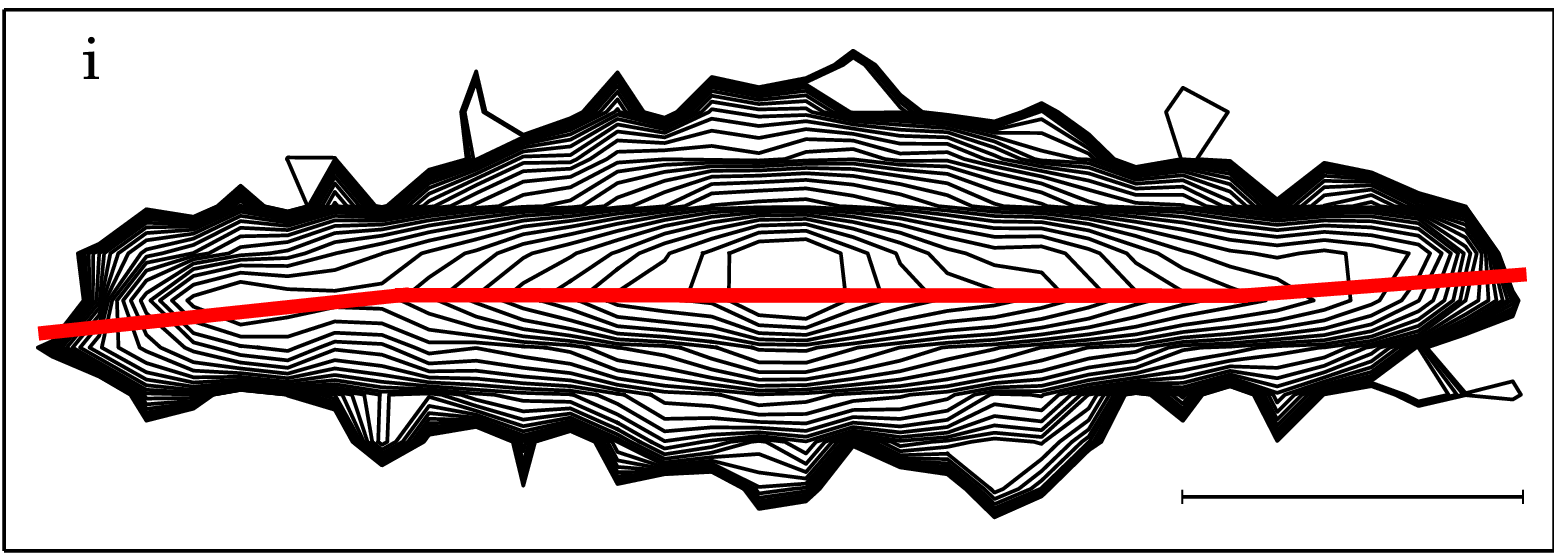}
\includegraphics[width=4.4cm, angle=0, clip=]{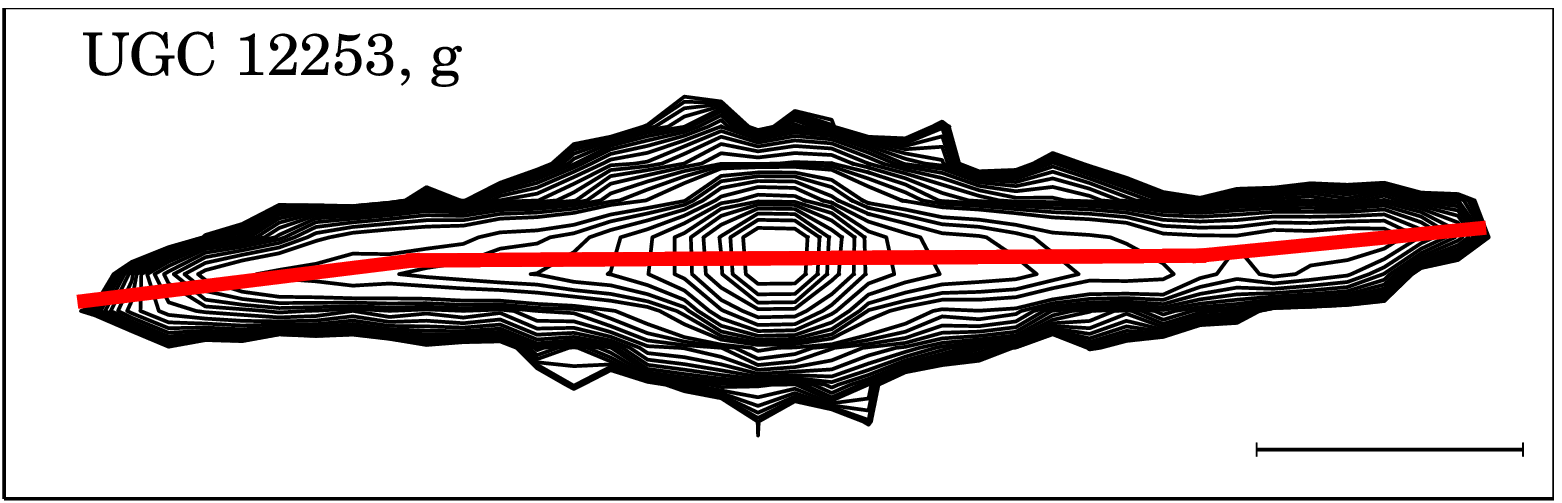}
\includegraphics[width=4.4cm, angle=0, clip=]{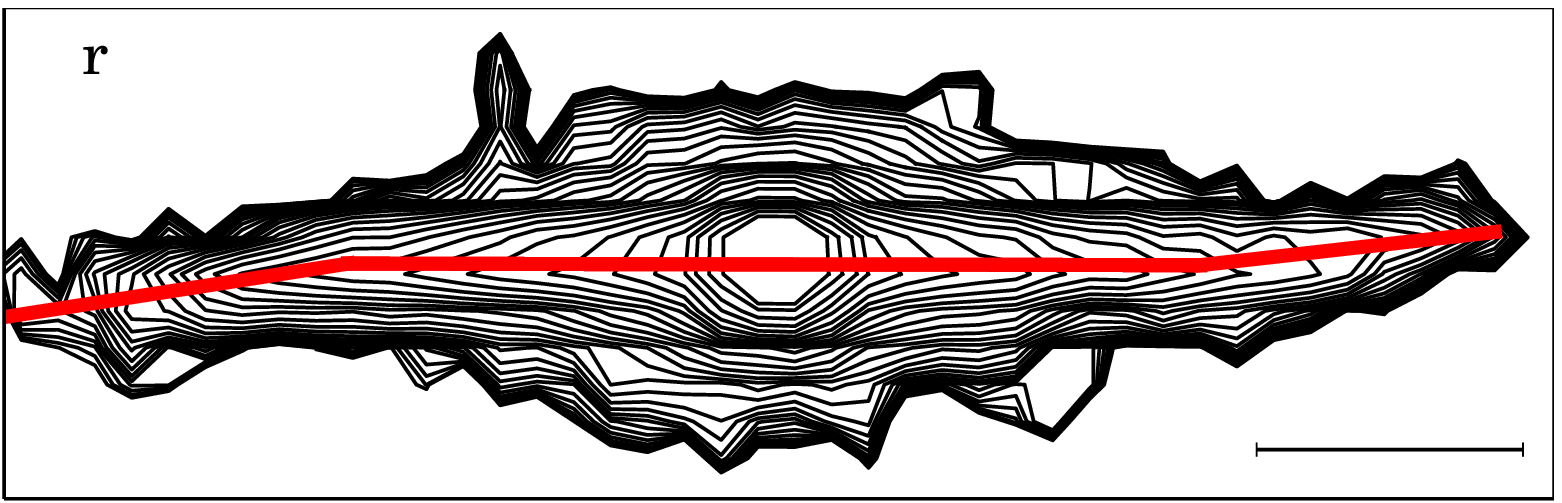}
\includegraphics[width=4.4cm, angle=0, clip=]{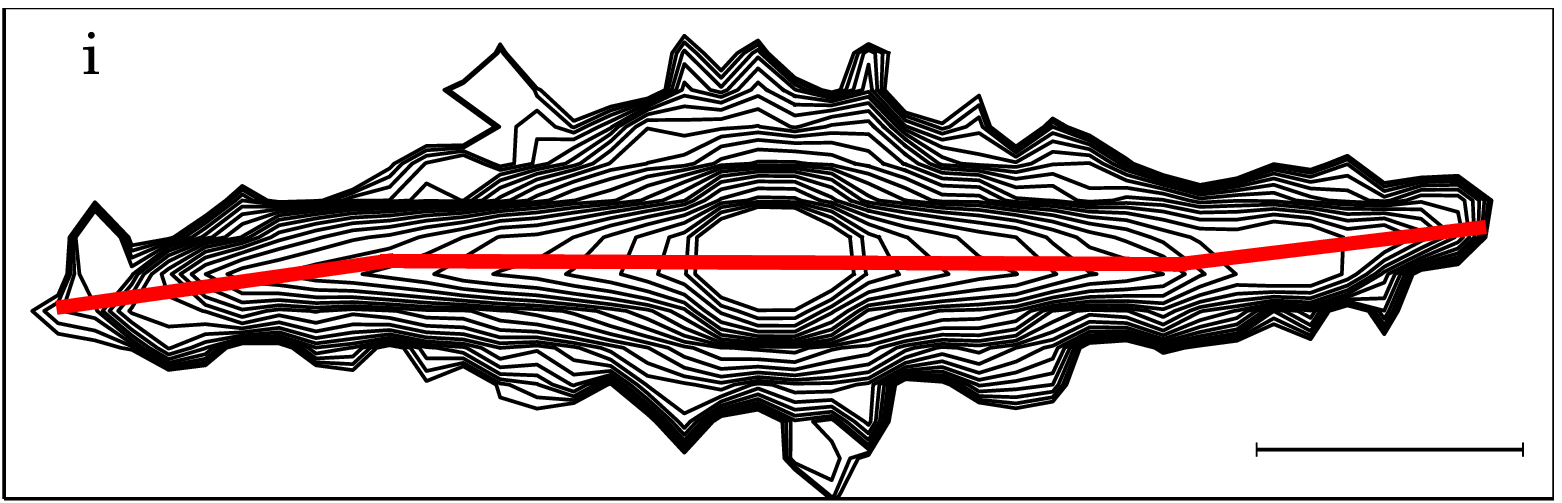}
\caption{The $g$-,$r$- and $i$-band contour maps of the sample galaxies.
The red line corresponds to the best fit of the center-line with the 
piecewise linear function. The outer isophote is of 
25.5 mag/$\Box''$ (24.5 mag/$\Box''$ for 2MFGC~6306). 
The length of the bar in the bottom right corner is 20$\arcsec$.}
\label{warps_pics}
\end{figure*}

The results of the fitting are presented in Table~\ref{Table3} and in 
Fig.~\ref{warps_pics}. In Fig.~\ref{warps_pics}, the galaxies are oriented in such a way that their major axes are horizontal and the north part of the galaxies is up.
Signs of warp angles in Table~\ref{Table3} are given as follows. For left (east) halves 
of the galaxy images, up-bending warps have positive angles, down-bending warps 
have negative values. For right (west) halves, on the contrary, 
up-bending has negative value and down-bending is described by positive warp 
angles.

\begin{table*}
 \centering
\begin{minipage}{150mm}
\parbox[t]{150mm} {\caption{Warp parameters for the sample galaxies in the
$g$, $r$ and $i$ bands. $\mu_{w}$ is the isophote level where the warp of the 
disc begins to appear (corrected for Galactic extinction and K-correction) which 
corresponds to the radius of the warp $R_\mathrm{w}$ in units of the disc scale 
length. $\psi$ is the warp angle. ``North'' and ``South'' mean the
north and south halfs of galaxies.}
\label{Table3}}
\begin{tabular}{|ccc|ccc|ccc|}
\hline
\hline
\# & Galaxy & Band &  \multicolumn{1}{c}{$\mu_\mathrm{w}$} & \multicolumn{1}{c}{$R_\mathrm{w}$} & \multicolumn{1}{c|}{$\psi$} &  \multicolumn{1}{c}{$\mu_\mathrm{w}$} & \multicolumn{1}{c}{$R_\mathrm{w}$} & \multicolumn{1}{c|}{$\psi$} \\ 
   &        &      &  \multicolumn{1}{c}{($\mathrm{mag}/\square''$)}& \multicolumn{1}{c}{($h$)}   & 
   \multicolumn{1}{c|}{(deg)} &  \multicolumn{1}{c}{($\mathrm{mag}/\square''$)}& 
   \multicolumn{1}{c}{($h$)}   & \multicolumn{1}{c|}{(deg)} \\
\hline
   &        &      & \multicolumn{3}{c|}{North} & \multicolumn{3}{c|}{South}     \\ 
\hline
1&IC 194&g&23.9 $\pm$ 0.3&2.35 $\pm$ 0.04&-1.8 $\pm$ 0.2&22.3 $\pm$ 0.1&1.59 $\pm$ 0.43&4.0 $\pm$ 0.5  \tabularnewline
&&r&23.2 $\pm$ 0.3&2.6 $\pm$ 0.05&-1.8 $\pm$ 0.2&21.7 $\pm$ 0.1&1.88 $\pm$ 0.49&4.1 $\pm$ 0.5  \tabularnewline
&&i&22.8 $\pm$ 0.4&2.59 $\pm$ 0.05&-2.5 $\pm$ 0.3&21.4 $\pm$ 0.1&2.11 $\pm$ 0.52&5.3 $\pm$ 0.7  \tabularnewline
2&2MFGC 6306&g&22.7 $\pm$ 0.2&2.91 $\pm$ 0.12&6.2 $\pm$ 0.6&22.6 $\pm$ 0.2&2.35 $\pm$ 0.24&-4.6 $\pm$ 0.5  \tabularnewline
&&r&21.9 $\pm$ 0.1&3.4 $\pm$ 0.14&7.5 $\pm$ 0.5&21.9 $\pm$ 0.1&2.98 $\pm$ 0.28&-6.3 $\pm$ 0.4  \tabularnewline
&&i&21.7 $\pm$ 0.2&3.84 $\pm$ 0.14&7.2 $\pm$ 0.5&21.3 $\pm$ 0.1&2.68 $\pm$ 0.29&-6.2 $\pm$ 0.5  \tabularnewline
3&SPRC 192&g&21.7 $\pm$ 0.2&1.65 $\pm$ 0.36&5.8 $\pm$ 2.7&21.9 $\pm$ 0.4&1.75 $\pm$ 0.02&-15.9 $\pm$ 1.7  \tabularnewline
&&r&21.0 $\pm$ 0.2&1.95 $\pm$ 0.41&9.8 $\pm$ 2.4&21.4 $\pm$ 0.4&2.06 $\pm$ 0.01&-13.2 $\pm$ 1.8  \tabularnewline
&&i&20.7 $\pm$ 0.2&2.04 $\pm$ 0.42&13.9 $\pm$ 2.7&20.9 $\pm$ 0.3&2.15 $\pm$ 0.0&-14.2 $\pm$ 1.6  \tabularnewline
4&UGC 4591&g&22.7 $\pm$ 0.3&2.03 $\pm$ 0.09&2.1 $\pm$ 0.4&22.9 $\pm$ 0.1&2.7 $\pm$ 0.56&-6.1 $\pm$ 0.6  \tabularnewline
&&r&22.3 $\pm$ 0.2&2.48 $\pm$ 0.14&1.5 $\pm$ 0.5&22.4 $\pm$ 0.1&3.29 $\pm$ 0.72&-5.4 $\pm$ 1.0  \tabularnewline
&&i&21.9 $\pm$ 0.1&2.71 $\pm$ 0.16&1.4 $\pm$ 0.6&22.2 $\pm$ 0.1&3.61 $\pm$ 0.79&-5.5 $\pm$ 1.1  \tabularnewline
5&MCG +06-22-041&g&22.6 $\pm$ 0.2&3.03 $\pm$ 0.57&-9.0 $\pm$ 2.4&22.8 $\pm$ 0.4&2.12 $\pm$ 0.06&4.3 $\pm$ 0.4  \tabularnewline
&&r&22.6 $\pm$ 0.2&2.94 $\pm$ 0.63&-7.4 $\pm$ 2.8&22.8 $\pm$ 0.4&2.65 $\pm$ 0.07&6.3 $\pm$ 0.6  \tabularnewline
&&i&22.8 $\pm$ 0.3&3.18 $\pm$ 0.64&-6.9 $\pm$ 2.1&22.8 $\pm$ 0.5&2.47 $\pm$ 0.08&5.8 $\pm$ 0.7  \tabularnewline
6&NGC 3160&g&22.6 $\pm$ 0.2&2.4 $\pm$ 0.5&15.3 $\pm$ 3.3&22.2 $\pm$ 0.2&2.02 $\pm$ 0.06&-17.7 $\pm$ 1.7  \tabularnewline
&&r&21.5 $\pm$ 0.1&2.39 $\pm$ 0.48&11.7 $\pm$ 2.7&20.9 $\pm$ 0.2&1.85 $\pm$ 0.06&-14.4 $\pm$ 1.5  \tabularnewline
&&i&21.2 $\pm$ 0.1&2.19 $\pm$ 0.46&14.7 $\pm$ 2.8&20.8 $\pm$ 0.2&1.87 $\pm$ 0.06&-17.4 $\pm$ 1.7  \tabularnewline
7&UGC 5791&g&22.2 $\pm$ 0.1&1.57 $\pm$ 0.19&16.0 $\pm$ 5.8&22.6 $\pm$ 0.2&1.92 $\pm$ 0.54&-12.8 $\pm$ 3.3  \tabularnewline
&&r&21.9 $\pm$ 0.1&1.58 $\pm$ 0.16&17.3 $\pm$ 4.6&22.3 $\pm$ 0.2&1.93 $\pm$ 0.51&-10.8 $\pm$ 2.3  \tabularnewline
&&i&21.9 $\pm$ 0.1&1.51 $\pm$ 0.12&16.4 $\pm$ 0.7&22.3 $\pm$ 0.2&1.84 $\pm$ 0.45&-11.7 $\pm$ 1.9  \tabularnewline
8&NGC 3753&g&23.6 $\pm$ 0.3&3.0 $\pm$ 0.59&10.0 $\pm$ 2.8&24.7 $\pm$ 0.4&2.1 $\pm$ 0.08&-15.2 $\pm$ 2.1  \tabularnewline
&&r&22.9 $\pm$ 0.2&3.08 $\pm$ 0.58&8.8 $\pm$ 1.4&24.1 $\pm$ 0.4&1.93 $\pm$ 0.07&-30.0 $\pm$ 1.7  \tabularnewline
&&i&22.5 $\pm$ 0.3&3.17 $\pm$ 0.64&12.3 $\pm$ 2.1&22.8 $\pm$ 0.2&2.42 $\pm$ 0.08&-28.7 $\pm$ 1.7  \tabularnewline
9&UGC 6882&g&22.7 $\pm$ 0.3&1.79 $\pm$ 0.41&8.3 $\pm$ 5.7&22.6 $\pm$ 0.3&1.88 $\pm$ 0.04&-6.1 $\pm$ 0.5  \tabularnewline
&&r&22.3 $\pm$ 0.2&2.15 $\pm$ 0.48&6.7 $\pm$ 4.5&22.2 $\pm$ 0.3&2.17 $\pm$ 0.05&-4.8 $\pm$ 0.5  \tabularnewline
&&i&22.3 $\pm$ 0.2&2.46 $\pm$ 0.53&10.2 $\pm$ 6.4&21.9 $\pm$ 0.2&2.29 $\pm$ 0.05&-6.7 $\pm$ 0.6  \tabularnewline
10&SDSS J140639.64+272242.4&g&23.3 $\pm$ 0.1&2.44 $\pm$ 0.04&-5.6 $\pm$ 0.3&23.0 $\pm$ 0.1&2.21 $\pm$ 0.04&6.0 $\pm$ 0.3  \tabularnewline
&&r&22.0 $\pm$ 0.1&3.07 $\pm$ 0.05&-4.5 $\pm$ 0.2&21.8 $\pm$ 0.1&2.80 $\pm$ 0.05&3.8 $\pm$ 0.2  \tabularnewline
&&i&22.0 $\pm$ 0.1&3.44 $\pm$ 0.04&-3.4 $\pm$ 0.4&21.3 $\pm$ 0.1&2.72 $\pm$ 0.04&6.2 $\pm$ 0.2  \tabularnewline
11&SDSS J153538.63+464229.5&g&22.5 $\pm$ 0.1&0.92 $\pm$ 0.09&3.5 $\pm$ 0.1&22.7 $\pm$ 0.1&1.52 $\pm$ 0.59&-5.9 $\pm$ 0.1  \tabularnewline
&&r&22.0 $\pm$ 0.1&1.05 $\pm$ 0.10&0.2 $\pm$ 0.1&22.2 $\pm$ 0.1&1.70 $\pm$ 0.08&-1.6 $\pm$ 0.4  \tabularnewline
&&i&21.9 $\pm$ 0.1&1.19 $\pm$ 0.12&2.2 $\pm$ 0.1&22.0 $\pm$ 0.1&1.93 $\pm$ 0.10&-6.4 $\pm$ 0.1  \tabularnewline
12&UGC 10716&g&23.5 $\pm$ 0.3&1.51 $\pm$ 0.04&3.9 $\pm$ 0.3&22.9 $\pm$ 0.1&1.58 $\pm$ 0.35&-0.9 $\pm$ 0.2  \tabularnewline
&&r&22.8 $\pm$ 0.2&1.76 $\pm$ 0.06&4.0 $\pm$ 0.4&22.5 $\pm$ 0.1&2.24 $\pm$ 0.46&-1.0 $\pm$ 0.1  \tabularnewline
&&i&22.5 $\pm$ 0.3&2.12 $\pm$ 0.07&4.1 $\pm$ 0.3&22.4 $\pm$ 0.3&2.59 $\pm$ 0.54&-0.4 $\pm$ 0.1  \tabularnewline
13&UGC 12253&g&22.9 $\pm$ 0.1&1.69 $\pm$ 0.41&3.6 $\pm$ 2.2&23.2 $\pm$ 0.2&1.94 $\pm$ 0.04&-1.7 $\pm$ 0.2  \tabularnewline
&&r&22.3 $\pm$ 0.2&2.51 $\pm$ 0.55&4.9 $\pm$ 2.5&22.7 $\pm$ 0.3&2.45 $\pm$ 0.06&-1.7 $\pm$ 0.2  \tabularnewline
&&i&21.9 $\pm$ 0.2&2.47 $\pm$ 0.56&4.4 $\pm$ 3.3&22.2 $\pm$ 0.3&2.52 $\pm$ 0.06&-2.2 $\pm$ 0.3  \tabularnewline

\hline
\end{tabular}
\end{minipage}

\end{table*} 

\section{Spectroscopic observations and data reduction}
\label{spectro}

The spectroscopic observations of our sample galaxies were carried out at 
the prime focus of the SAO RAS 6-m telescope with the multi-mode focal 
reducer SCORPIO-2 \citep{AfansievMoiseev2011}, UGC~10716 was observed 
with the previous version of the focal 
reducer \citep[SCORPIO,][]{AfansievMoiseev2005}. In the long-slit mode 
both devices have the same slit size (6 arcmin $\times$ 1 arcsec) with a 
scale of 0.36 arcsec per pixel. The exception was UGC 5791 observed 
with 0.7 arcsec slit width.

The spectral resolution was similar in all cases (FWHM about  5 \AA), 
while SCORPIO-2 provides a twice broader spectral range than old SCORPIO. 
The log of spectral observation is given in Table~\ref{tab4}. The slit 
position  was chosen along the major axis of each galaxy. For several objects 
we used two slit positions -- along the major axis 
and along the tips of apparent warp.

Data reduction was performed in a standard way using the IDL software package developed at the SAO RAS for reducing the data obtained with SCORPIO/SCORPIO-2. The brightest emission lines H$\alpha$ and [NII]$\lambda\lambda6548,6583$ were fitted by Gaussian profiles in order to calculate the ionized gas velocities along the slit. The line-of-sight velocity distributions were then converted to the galaxy rotation curves (RCs) assuming a constant disc inclination $i=90^{\rm o}$.

\begin{table*}
 \centering
\begin{minipage}{150mm}
 \centering
\parbox[t]{150mm} {\caption{Log of spectral observations.}
}
\begin{tabular}{ccccccc}
\hline 
\hline
\# & Galaxy       	& Slit PA& Date & Spectral range & Exp. time & Seeing     \\ 
   &                          	& (deg)&      & (\AA\AA)    & (s)       & ($''$)  \\  \hline
1  & IC 194             	& 193  & 20.11.2014   & 3650--7250& 2$\times$900  & 1.7    \tabularnewline
2  & 2MFGC 6306   	& 164  & 10.10.2012   &3650--7250& 5$\times$1200 & 2.1    \tabularnewline
3  & SPRC 192       	&   98  &  10.10.2012  &3650--7250& 5$\times$1200 & 2.3    \tabularnewline
    &         			& 110  &  08.12.2012  &3650--7250& 7$\times$900  & 2.9    \tabularnewline
4  & UGC 4591     	& 14    &  14.12.2014  &3650--7250& 6$\times$900  & 1.0    \tabularnewline
5  & MCG +06-22-041& 132&  15.12.2014  &3650--7250& 7$\times$900  & 1.1    \tabularnewline
6  & NGC 3160     	& 140  &  24.02.14     &3650--7250& 6$\times$900  & 1.3    \tabularnewline
   &              		& 150  &  15.12.2014 &3650--7250& 4$\times$900  & 1.1    \tabularnewline
7 & UGC 5791     	& 60    &  10.12.2012 & 3600--8500& 3$\times$900  & 2.9    \tabularnewline
8 & NGC 3753     	& 107  &  06.03.2013 &3650--7250& 4$\times$1200 & 1.8    \tabularnewline
   &              		& 125  &  06.03.2013 &3650--7250& 4$\times$1200 & 3      \tabularnewline
9 & UGC 6882     	& 120  &  12.12.2012 &3650--7250& 4$\times$1200 & 1.1    \tabularnewline
   &              		& 130  &  12.12.2012 &3650--7250& 4$\times$1200 & 1.2    \tabularnewline
10& SDSS~J140639.64+272242.4 	& 140   &  07.03.2013  &3650--7250 & 5$\times$1200 & 1.9     \tabularnewline
11& SDSS~J153538.63+464229.5   	& 74    &   07.03.2013 &3650--7250 & 3$\times$1200 & 1.5  \tabularnewline   
12& UGC 10716   	& 12    &  12.07.2013  & 5700--7500 & 2$\times$1200 & 1.4    \tabularnewline
13& UGC 12253   	& 144  &  15.12.2014 &3650--7250& 8$\times$900  & 1.7    \tabularnewline

\hline\\
\end{tabular}
\end{minipage}
\label{tab4}
\end{table*} 


\section[]{Results and discussion}
\label{Results}

\subsection{Photometric characteristics of galaxies}
\label{phot_char}
The mean characteristics of the sample galaxies -- $\langle M_r \rangle = -20.51 \pm 1.43$,
$\langle g - r \rangle = +0.64 \pm 0.25$, $\langle r - i \rangle = +0.36 \pm 0.18$, 
$\langle B/T \rangle = 0.16 \pm 0.18$ -- 
are typical for bright edge-on disc-dominated spirals (e.g. \citealp{biz2014}).
As expected, bulges of our galaxies are redder than their discs:
$\langle g - r \rangle_{\rm B} = +0.91 \pm 0.27$ vs. $\langle g - r \rangle_{\rm D} = +0.68 \pm 0.26$ 
and $\langle r - i \rangle_{\rm B} = +0.60 \pm 0.15$ vs. $\langle r - i \rangle_{\rm D} = +0.42 \pm 0.14$.

The galaxies show notable radial colour gradients: 
$\langle h_g/h_r \rangle = 1.15 \pm 0.08$
and $\langle h_g/h_i \rangle = 1.20 \pm 0.13$, where $h_g$, $h_r$, $h_i$ -- 
scale lengths of the discs in the $g$, $r$ and $i$ passbands respectively. 
Such gradients are usual for the discs of late-type spirals 
(e.g. \citealp{ee1984}, \citealp{degr1998}).

The average disc flattening for the sample is $\langle z_0/h_r \rangle = 0.34\pm0.15$ 
or, excluding four most inclined galaxies (SPRC-192, NGC~3753, UGC~6882 
and SDSS~J153538.63+464229.5 -- see Table~\ref{Table1}),
$\langle z_0/h_r \rangle = 0.30\pm0.11$ ($r$ passband). Both values are somewhat 
higher as compared to normal spirals of the same morphological types, luminosities 
and colours \citep{2015MNRAS.451.2376M} but, within the quoted scatter, the
difference is not significant.

\subsection{Kinematical characteristics of galaxies}
\label{kin_char}

\begin{figure*}
\includegraphics[width=4cm, angle=-90, clip=]{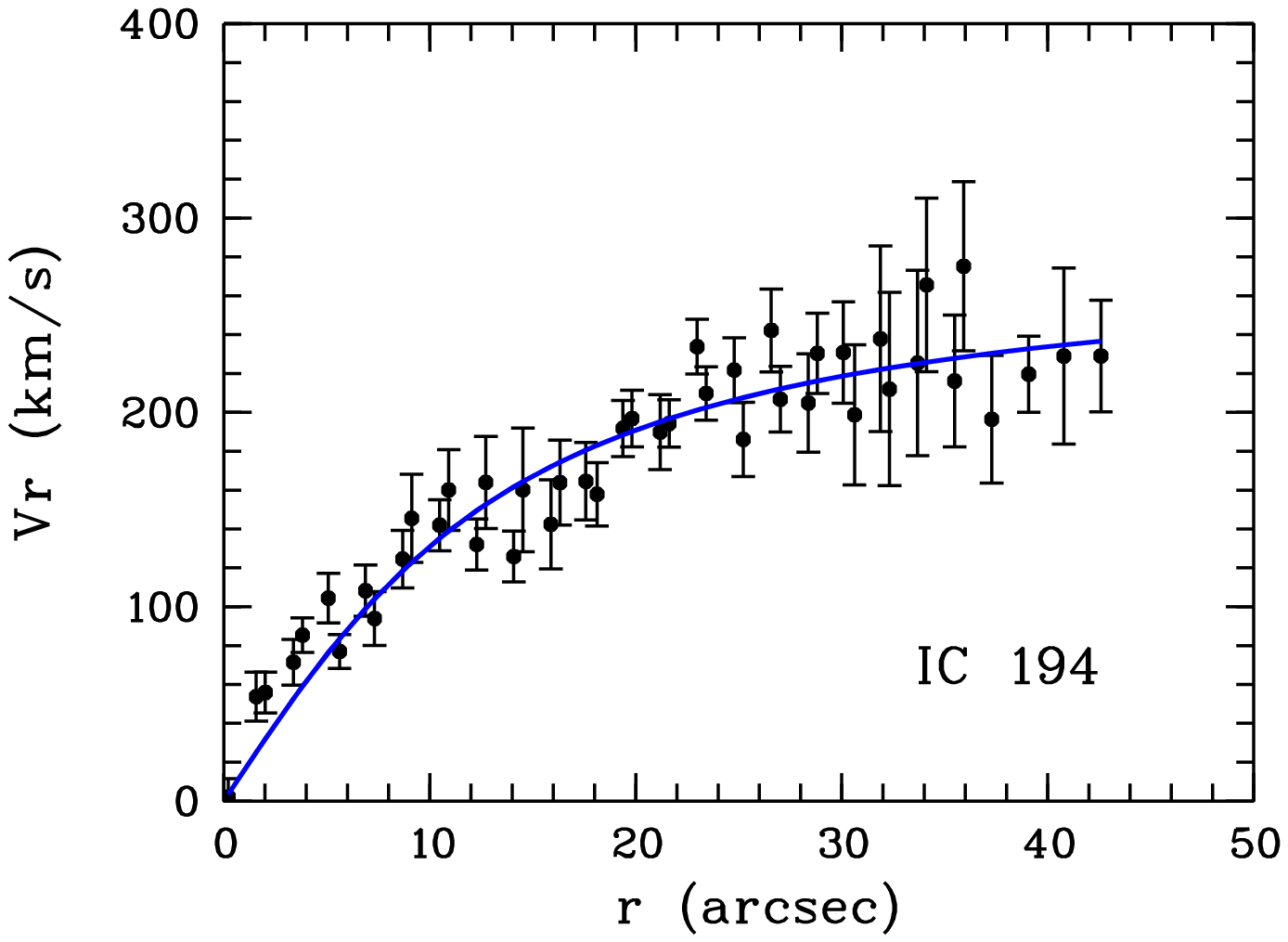}
\includegraphics[width=4cm, angle=-90, clip=]{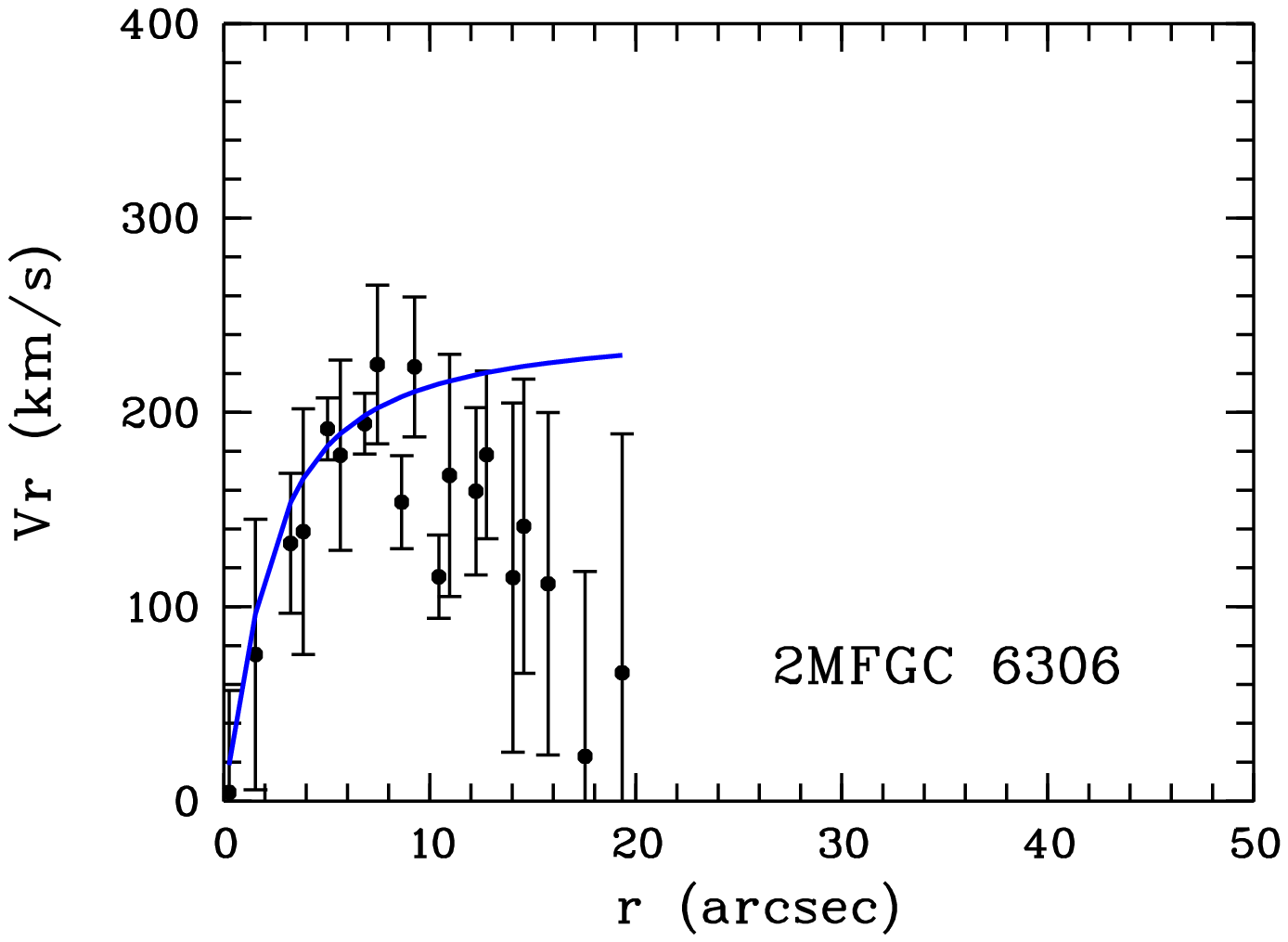}
\includegraphics[width=4cm, angle=-90, clip=]{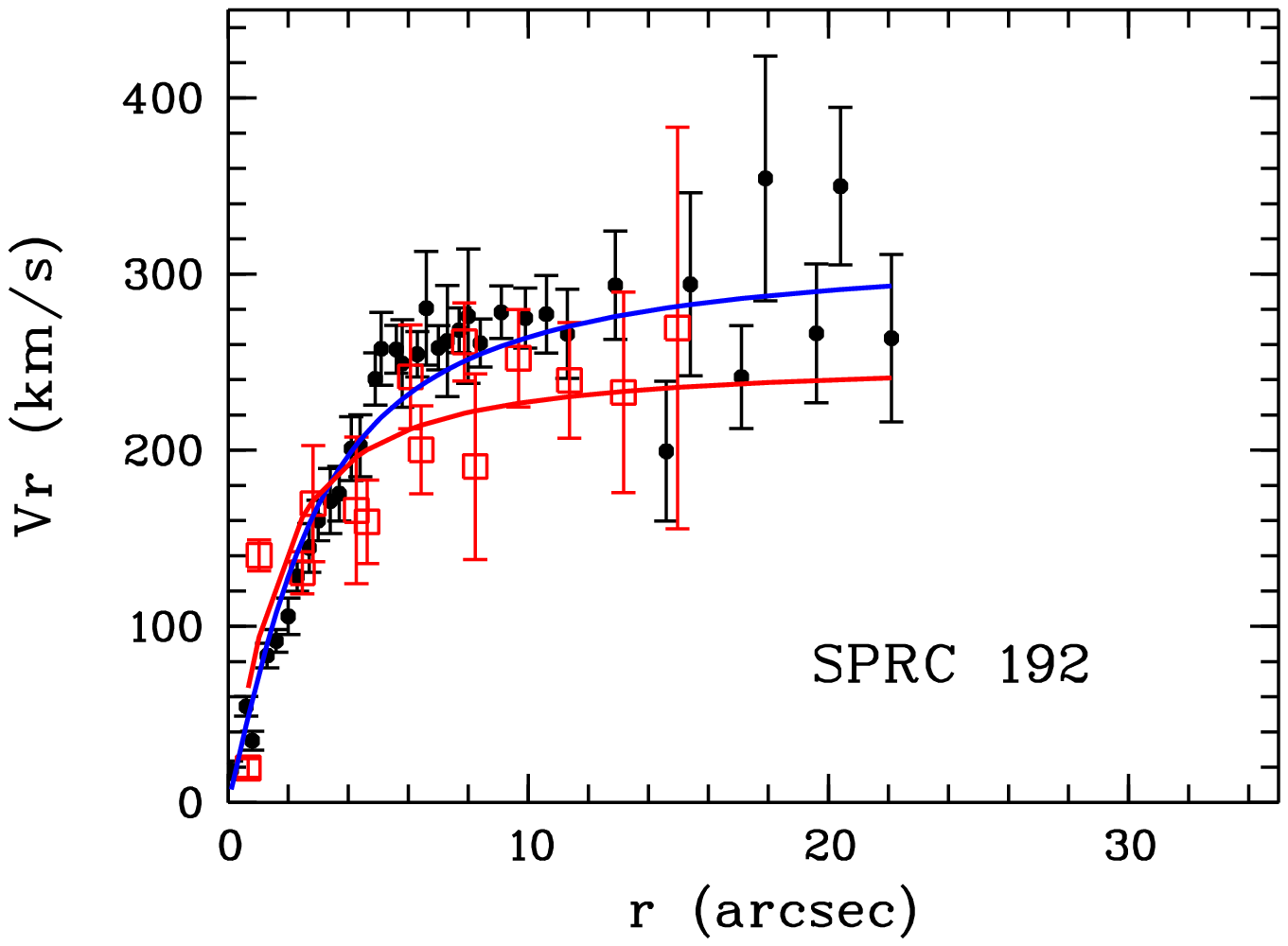}
\includegraphics[width=4cm, angle=-90, clip=]{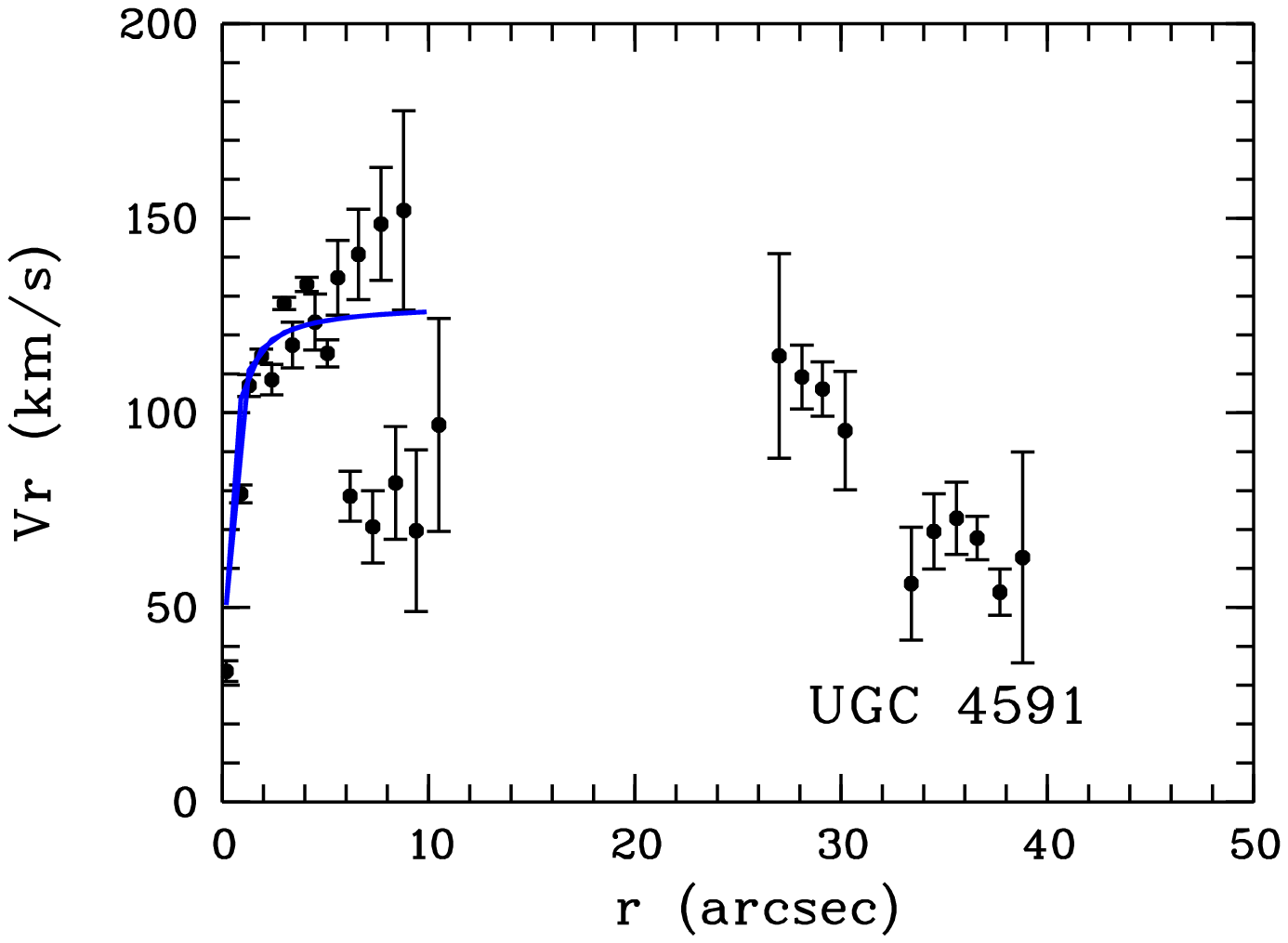}
\includegraphics[width=4cm, angle=-90, clip=]{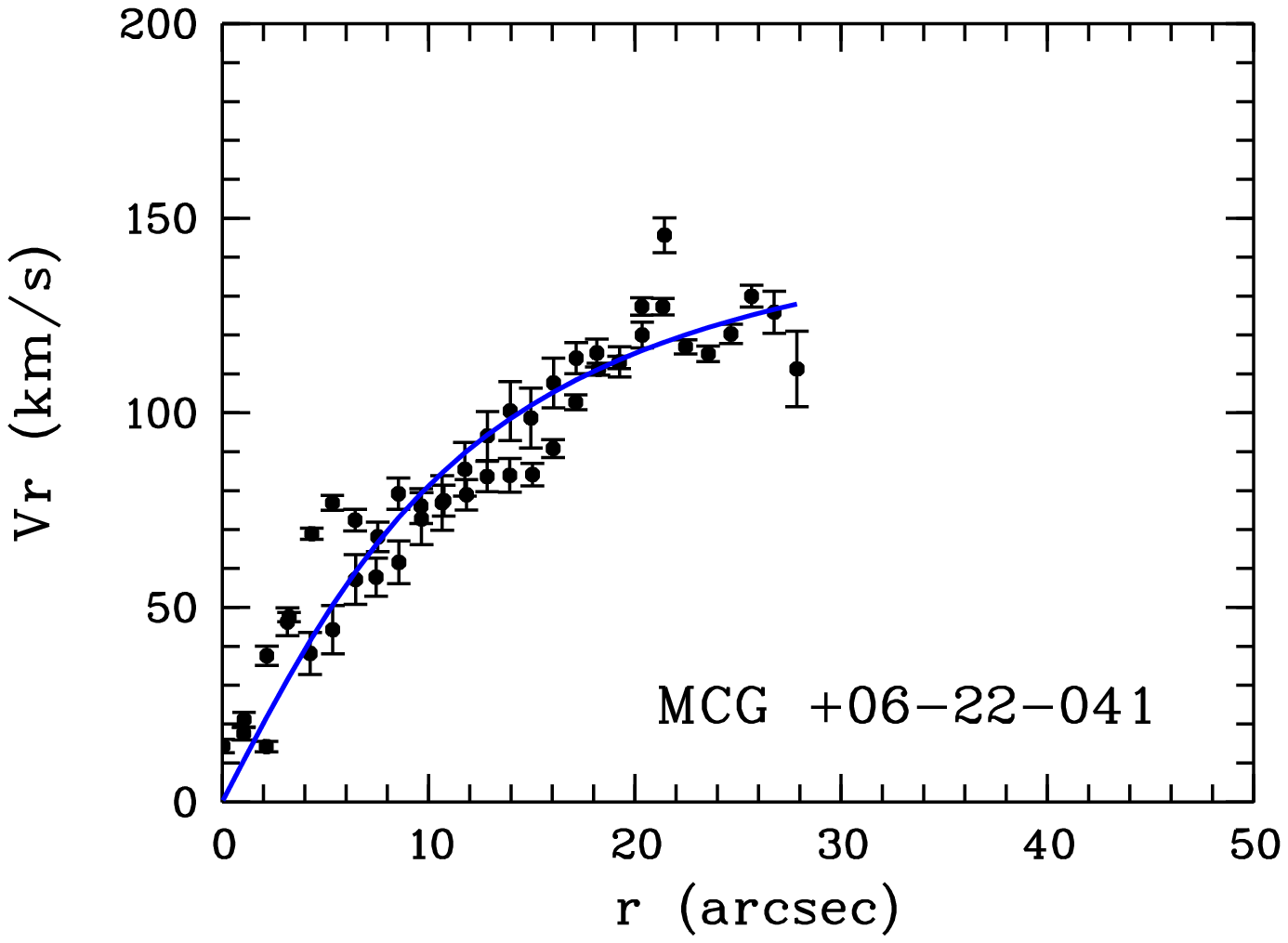}
\includegraphics[width=4cm, angle=-90, clip=]{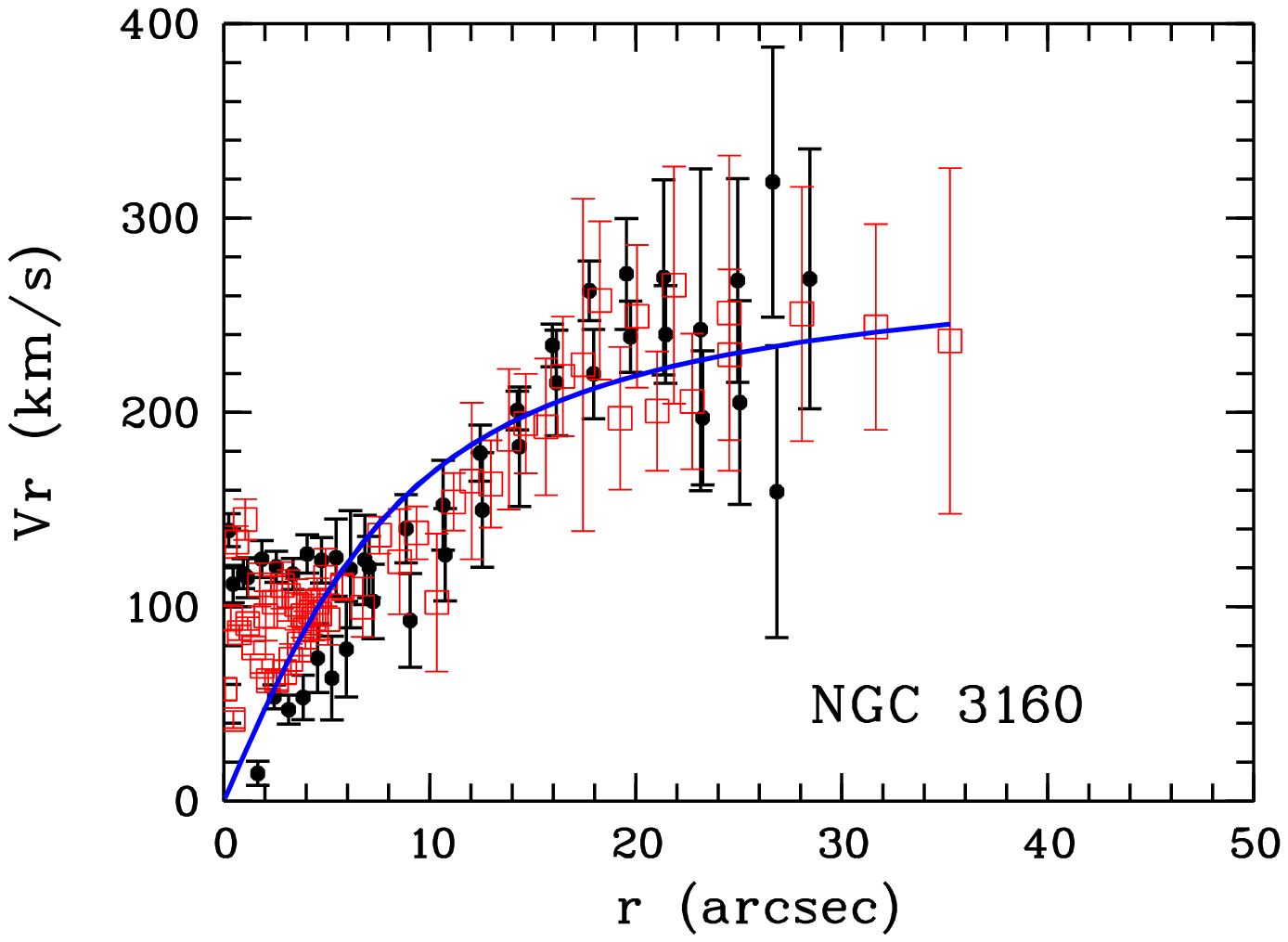}
\includegraphics[width=4cm, angle=-90, clip=]{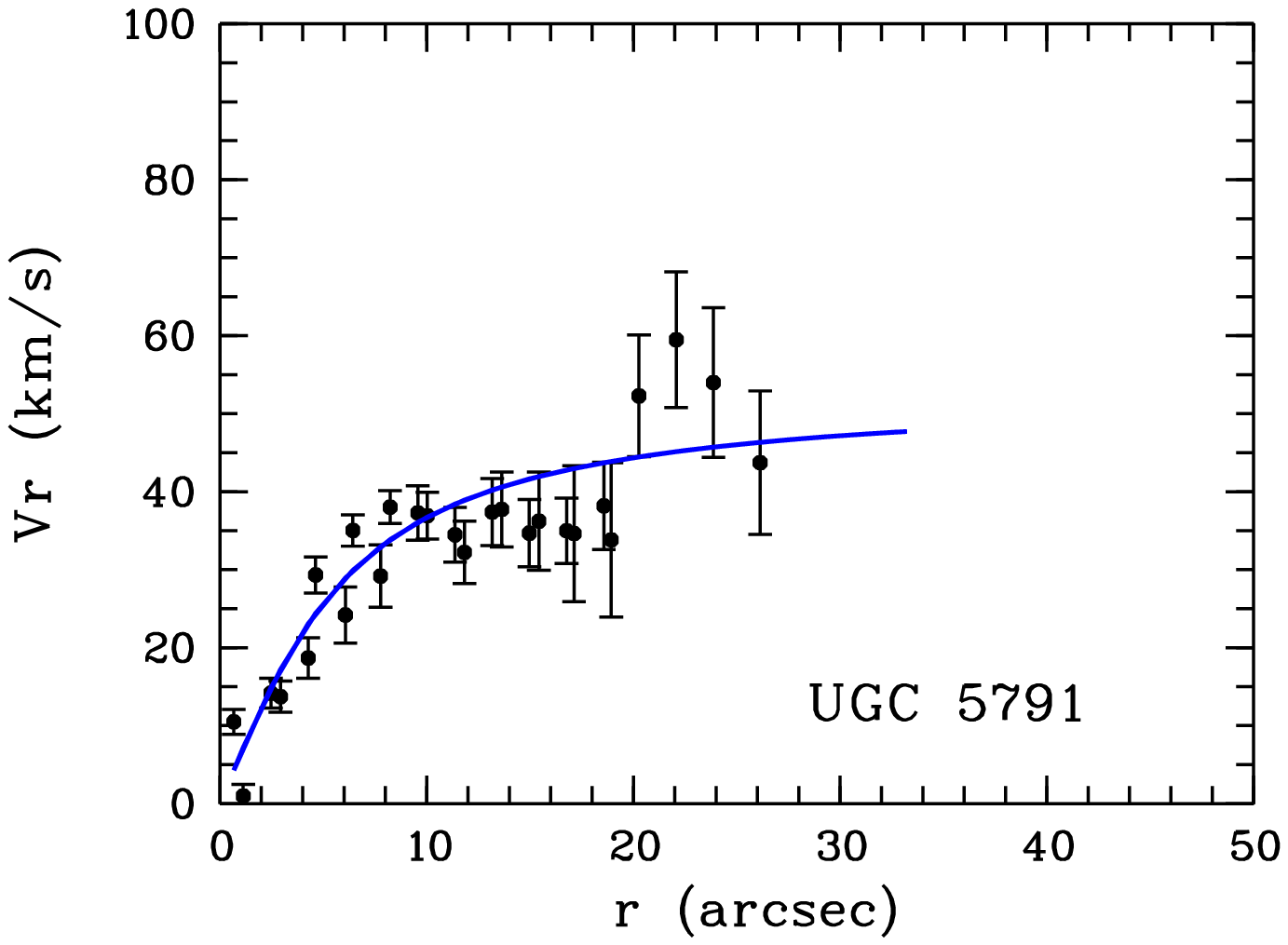}
\includegraphics[width=4cm, angle=-90, clip=]{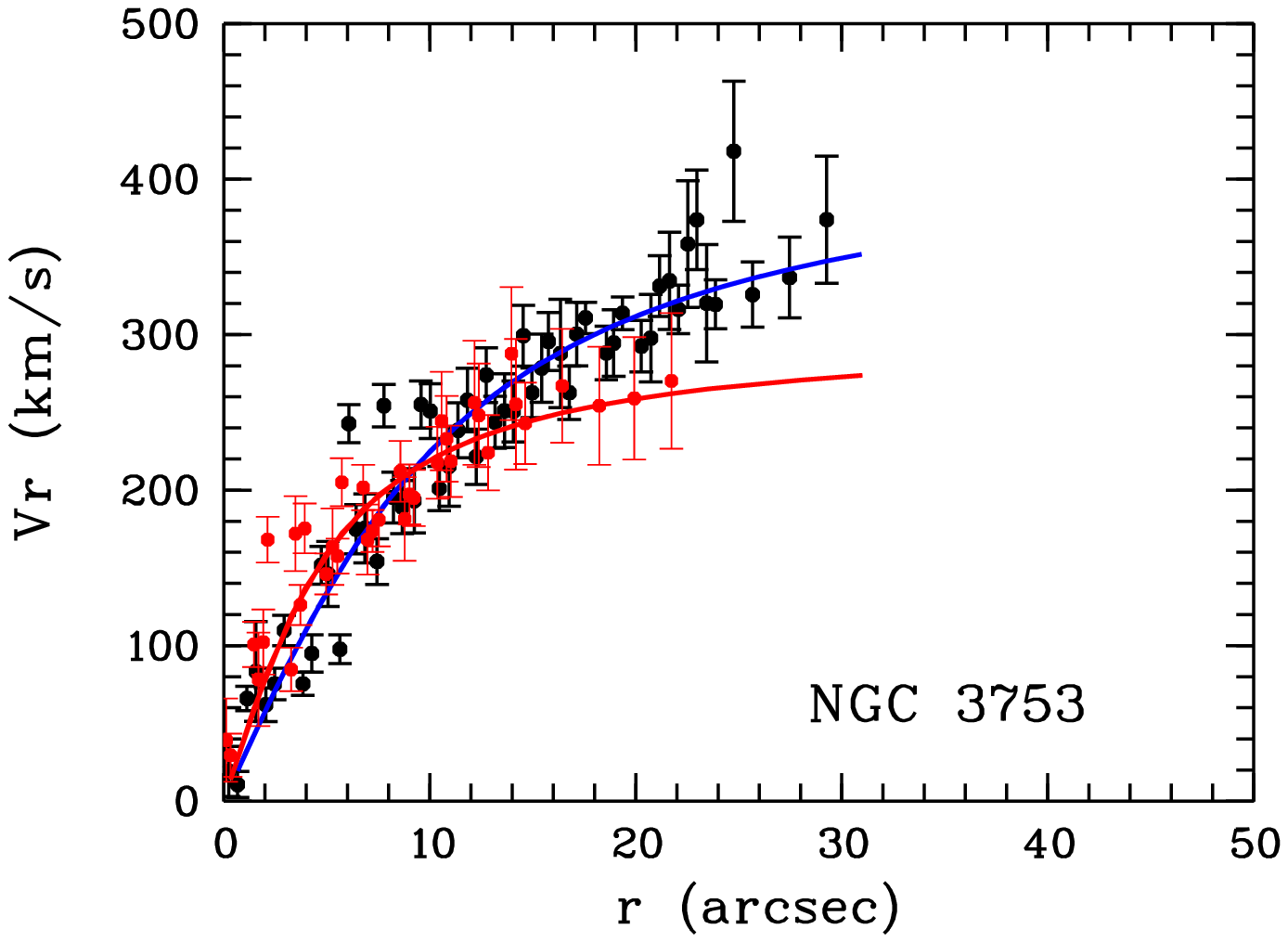}
\includegraphics[width=4cm, angle=-90, clip=]{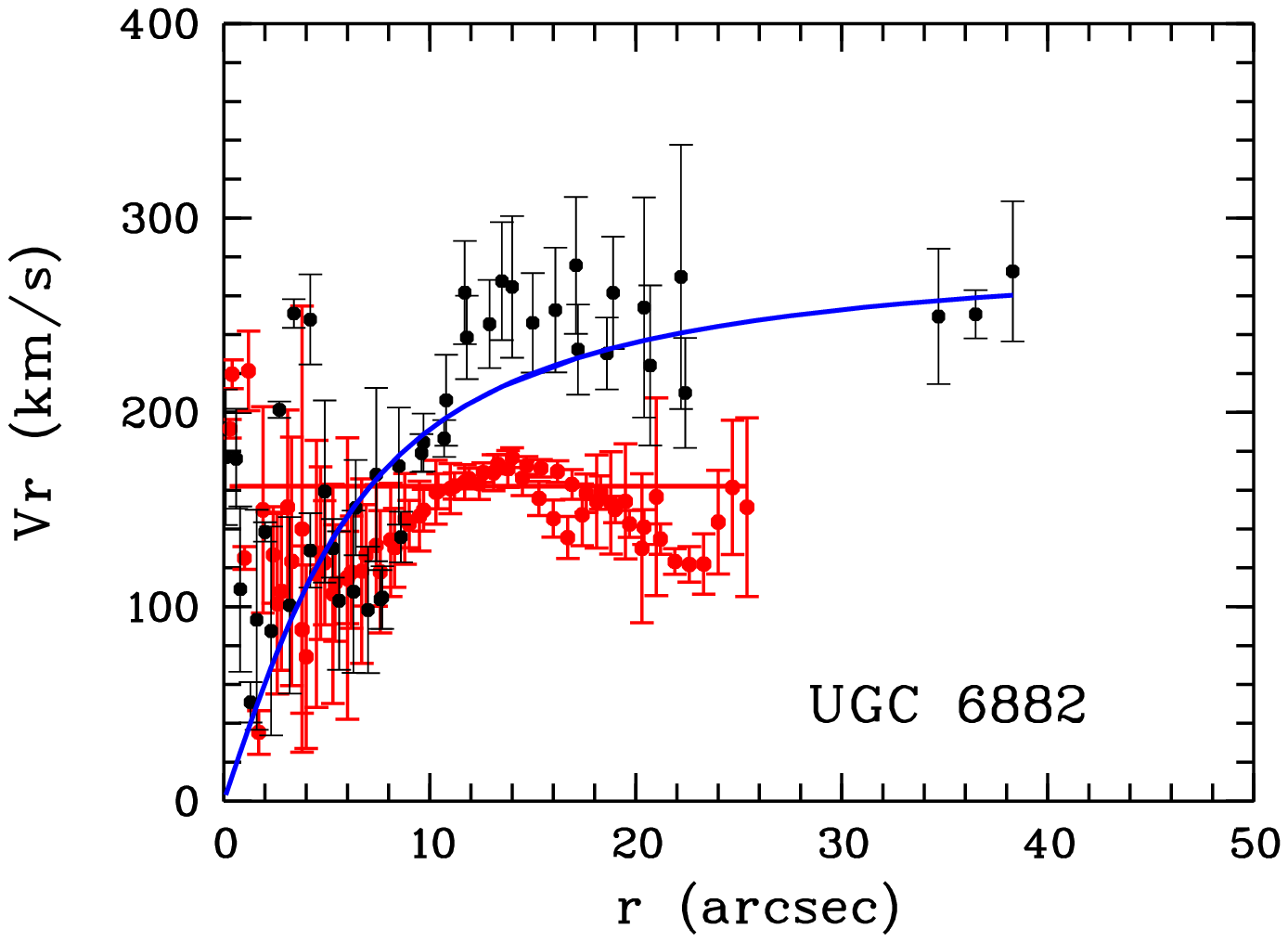}
\includegraphics[width=4cm, angle=-90, clip=]{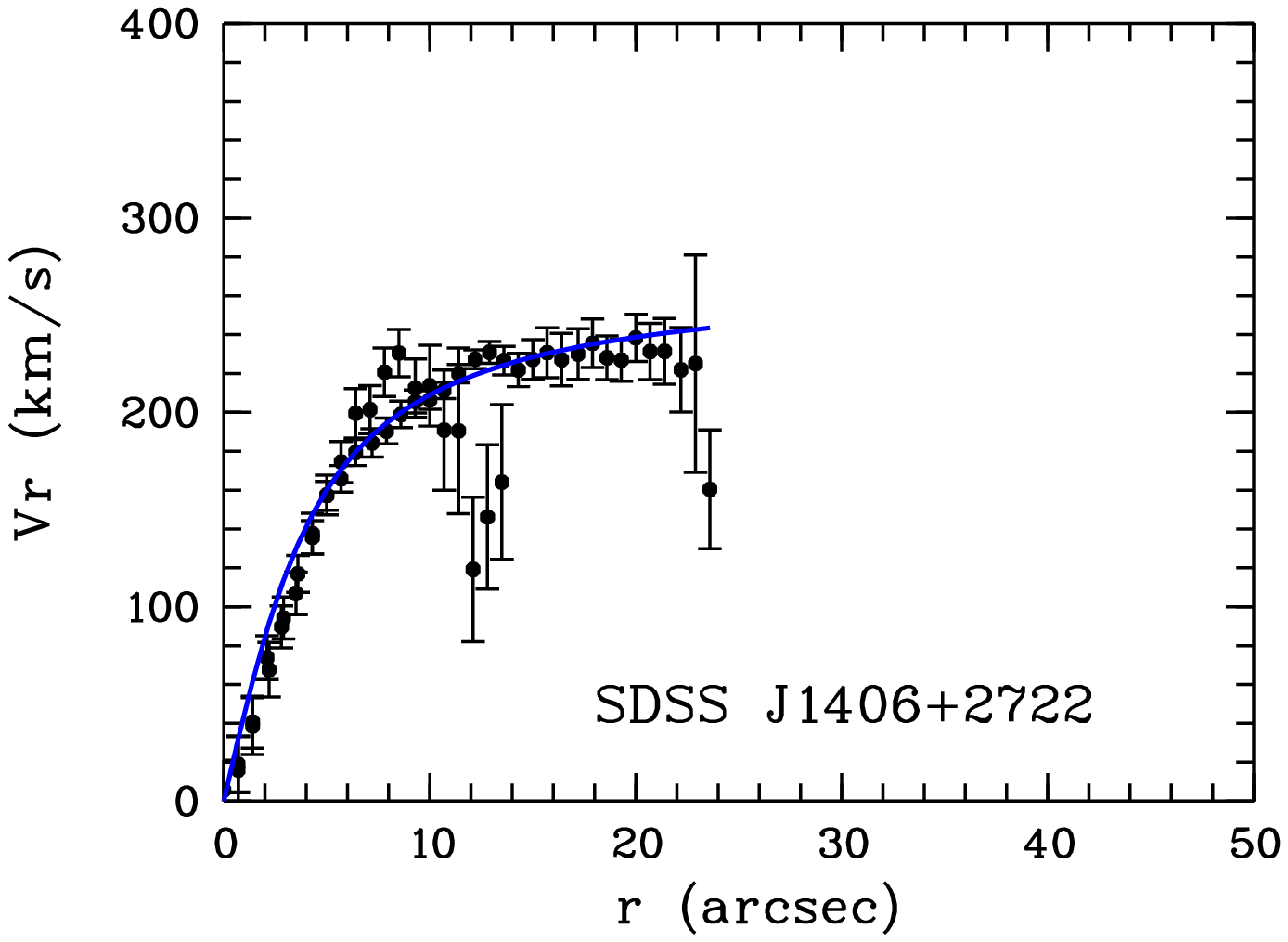}
\includegraphics[width=4cm, angle=-90, clip=]{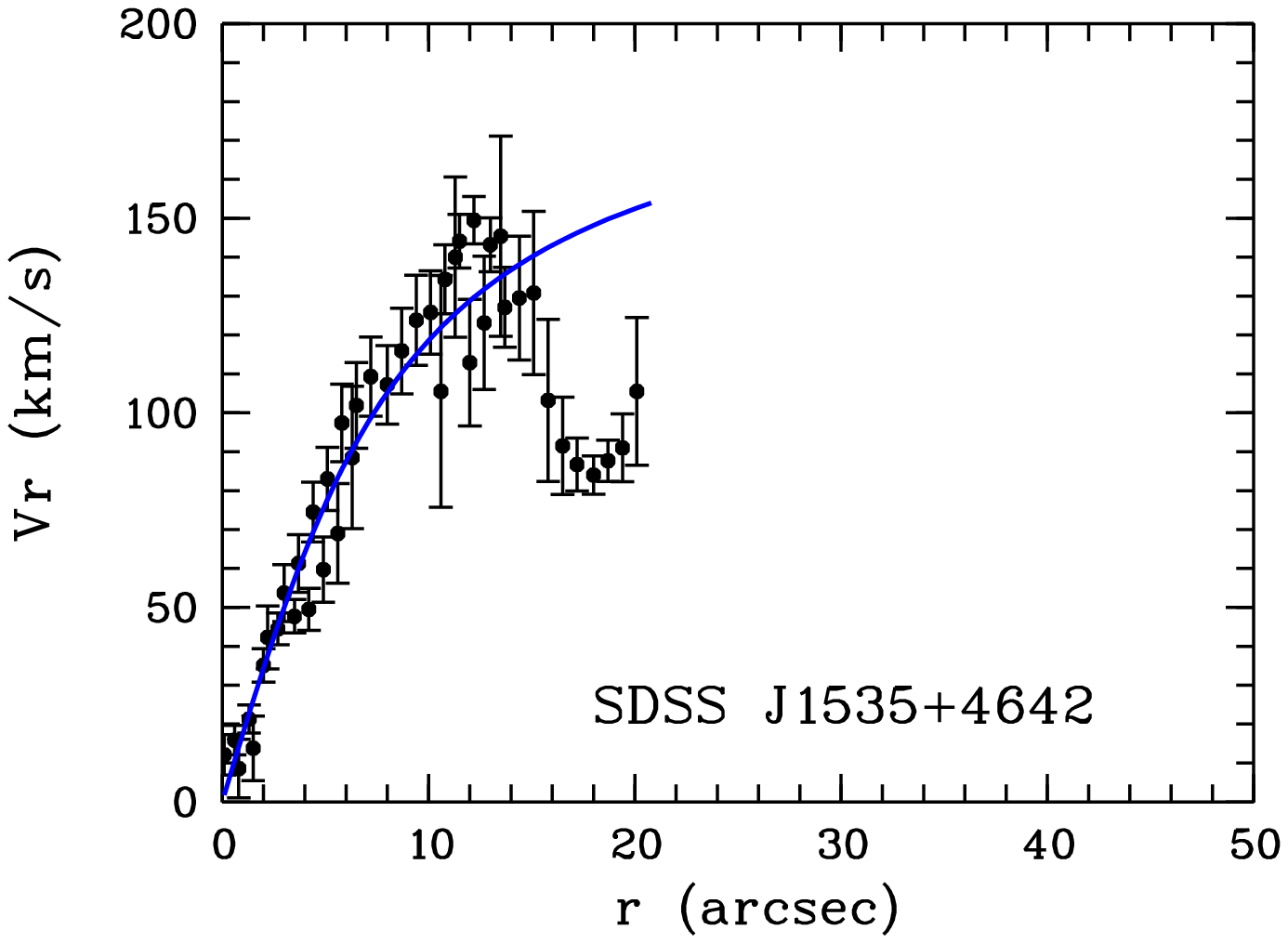}
\includegraphics[width=4cm, angle=-90, clip=]{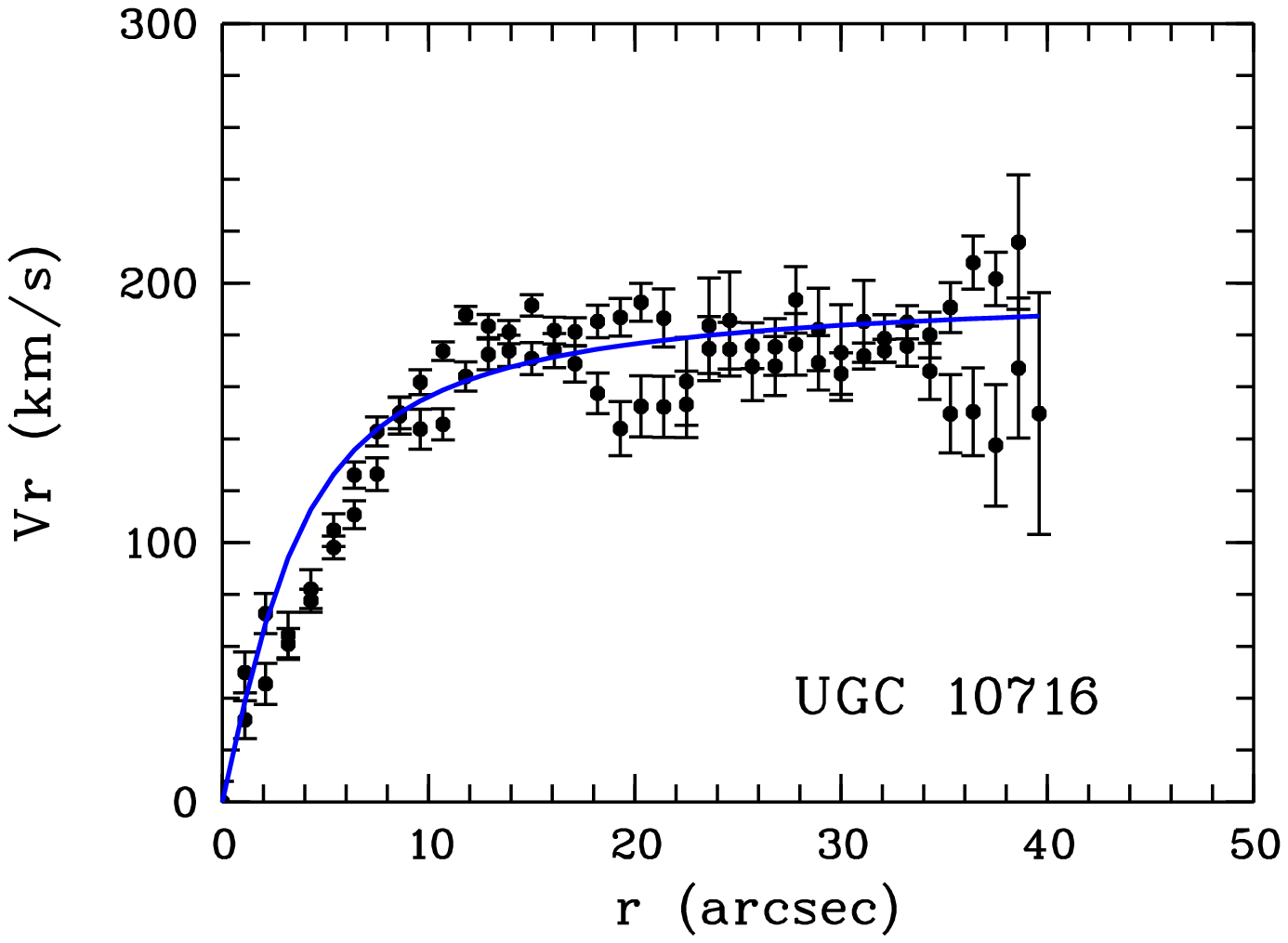}
\includegraphics[width=4cm, angle=-90, clip=]{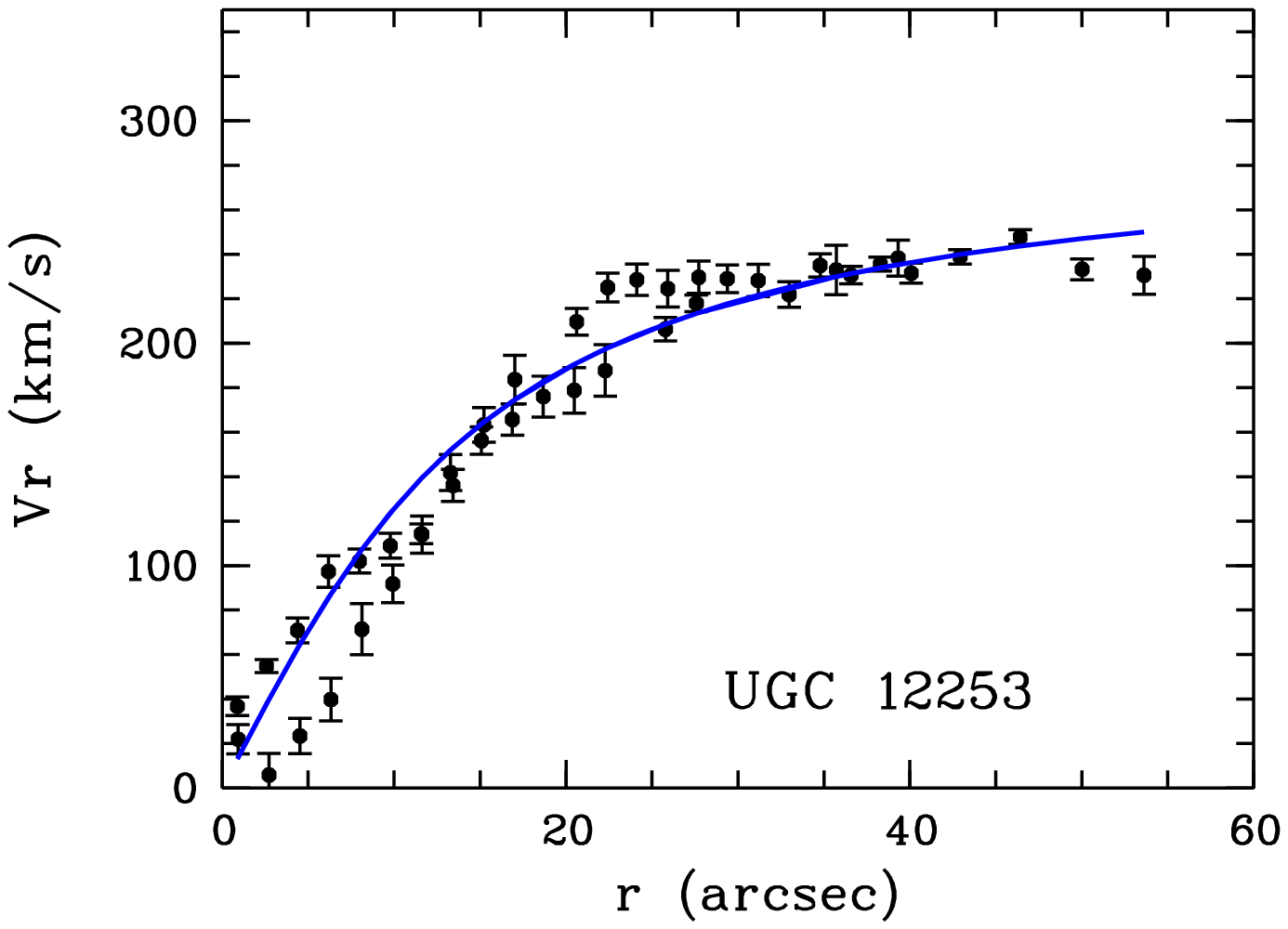}
\caption{Rotation curves of 13 sample galaxies. Solid curves show
the best-fit arctangent functions. {\bf SPRC-192}: black symbols and blue curve
correspond to position angle of slit P.A. = 110$^{\rm o}$, red symbols and red
curve -- P.A. = 98$^{\rm o}$; {\bf NGC~3160}: black symbols -- P.A. = 140$^{\rm o}$,
red symbols -- P.A. = 149.5$^{\rm o}$, blue curve gives arctangent fit for
joint rotation curve; {\bf NGC~3753}: black symbols and blue curve -- 
P.A. = 107$^{\rm o}$, red symbols and curve -- P.A. = 125$^{\rm o}$;
{\bf UGC~6882}: black symbols and blue curve -- P.A. = 130$^{\rm o}$, 
red symbols and curve -- P.A. = 120$^{\rm o}$.}
\label{rc}
\end{figure*}

Fig.~\ref{rc} presents rotation curves (RCs) of the galaxies averaged with respect to their
dynamic centres. Optical RCs of edge-on galaxies are strongly suffered by 
internal absorption which changes their shapes (e.g. \citealp{bosma1992}, 
\citealp{zasov2003}, \citealp{step2013}). However, our RCs are sufficiently
extended (mean extension of the curves is (0.80 $\pm$ 0.19) $a$, where $a$ is 
a semi-major axis of the galaxy, see Table~\ref{Table1}) to estimate rotational velocities and to check 
the Tully-Fisher (TF) relation for the sample galaxies.

Following the standard practice, we have fitted RCs by the arctangent function 
(\citealp{cour1997}, \citealp{wil1999})
\begin{equation}
V(r) = \frac{2}{\pi}~ V_a~ \mathrm{arctan}\left(\frac{r}{r_t}\right), 
\end{equation}
where $V_a$ is an asymptotic velocity and  $r_t$ is a turnover radius where the 
rotation curve goes from rising to flat. (For two galaxies with
apparently peculiar RCs -- 2MFGC~6306 and UGC~4591 -- we have used the observed 
velocities within $r \le 10''$ in our calculations, for SDSS~J153538.63+464229.5
we have took the data within $r \le 15''$. At larger radii, the spectrograph 
slit, probably, shifted away from the warped edge-on disc for these galaxies.) 

The fitted value $V_a$ can
significantly overestimate the true rotation velocity when the observed points do
not reach the flat part of the rotation curve. In order to avoid this, 
the rotation speed at 2.2 disc scale lengths ($V_{2.2}$) is used in our work. 
To obtain $V_{2.2}$, we find the value of the arctangent function at a 
distance of 2.2$h_r$.

The results of our analysis are summarized in Table~\ref{Table5}. For each galaxy
the table gives fitted $r_t$ and $V_a$ values, as well as their uncertainties.
The last column of Table~\ref{Table5} presents $V_{2.2}$ values corrected for the
cosmological broadening. 

\begin{table*}
 \centering
 \centering
\parbox[t]{150mm} {\caption{Kinematical characteristics of the sample galaxies.}
\label{Table5}}
\begin{tabular}{ccccccc}
\hline 
\hline
\# & Galaxy       &  $r_t$    & $\sigma_{r_t}$ &  $V_a$  & $\sigma_{V_a}$ & $V_{2.2}$      \\ 
   &              &  ($''$) & ($''$)       &  (km/s) & (km/s) & (km/s) \\  \hline
1  & IC 194       &  11.2    & 1.3  &  283   & 13             & 223     \tabularnewline
2  & 2MFGC 6306   &  2.2     & 1.1  &  247:   & 30            & 202:     \tabularnewline
3  & SPRC 192 (PA=110)  &  2.7    & 0.3   &  318   & 10       & 239          \tabularnewline
   & SPRC 192 (PA=278)  &  1.6    & 0.3   &  252     & 13     & 210            \tabularnewline
4  & UGC 4591     &  0.3    & 0.3   &  128:   & 10            & 124:    \tabularnewline
5  & MCG +06-22-041  &  10.2   & 1.6   &  165   & 10          & 87       \tabularnewline
6  & NGC 3160     & 7.4    & 0.8    &  282   & 14             & 220     \tabularnewline
7  & UGC 5791     &  5.3    & 1.3   &  53    & 5              & 45     \tabularnewline
8  & NGC 3753 (PA=107)  &  9.4   & 0.8   &  433   & 16        & 326            \tabularnewline
   & NGC 3753 (PA=125)  &  4.6   & 0.6   &  302   & 14        & 260           \tabularnewline
9  & UGC 6882 (PA=300)& -  & -     &  162:   & 4              & 157:    \tabularnewline
   & UGC 6882 (PA=310)& 5.8& 0.9   &  288   & 13              & 236   \tabularnewline
10 & SDSS~J140639.64+272242.4  & 3.7 & 0.3   &  270   & 8     & 188    \tabularnewline
11 & SDSS~J153538.63+464229.5  & 7.0 & 0.9   &  194:  & 15    & 113:    \tabularnewline
12 & UGC 10716  & 3.5 & 0.4   &  198   & 4                    & 175    \tabularnewline
13 & UGC 12253     &  12.5    & 1.1   &  293    & 6            & 210        \tabularnewline
\hline\\
\end{tabular}
\label{tab5}
\end{table*} 

\begin{figure}
\includegraphics[width=7.5cm, angle=-90, clip=]{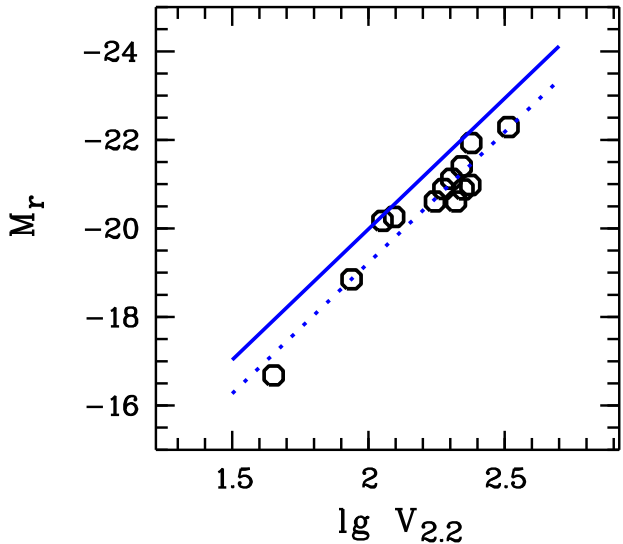}
\caption{The $r$ band Tully-Fisher relation for edge-on spiral galaxies
with warped discs (circles). Solid line represents the TF relation according 
to \citet{piz2007}, dotted line -- the same relation shifted by 
$\Delta M_r$ = 0$\fm$76.}
\label{tf}
\end{figure}

The Tully-Fisher relation for the sample galaxies is presented in Fig.~\ref{tf}.
(For the galaxies with rotation curves obtained at two position angles, we
use the larger value of $V_{2.2}$.)
As one can see, warped galaxies follow the relation for normal spirals
in a wide range of luminosities. The slope of the TF for warped galaxies
(--5.99$\pm$0.50) is consistent with the slope for usual spirals
(--5.91$\pm$0.20 -- see \citealp{piz2007}).
Some systematic shift of warped spirals towards fainter absolute magnitudes 
($\Delta M_r$ = 0$\fm$76) can be explained by the internal extinction taking place in 
edge-on spiral galaxies.

\subsection{Characteristics of warps}
\label{warp_char}
Table~\ref{Table6} presents average characteristics of stellar disc warps 
in the three filters. 

\begin{table}
 \centering
 \centering
\parbox[t]{150mm} {\caption{Characteristics of warps in the $g$, $r$, $i$ filters}
\label{Table6}}
\begin{tabular}{cccc}
\hline 
\hline
 & North  & South &  All \\  \hline
$\langle \mu_g \rangle$ & 22.84$\pm$0.60 & 22.80$\pm$0.67 & 22.82$\pm$0.62 \\
$\langle \mu_r \rangle$ & 22.21$\pm$0.59 & 22.22$\pm$0.77 & 22.22$\pm$0.68 \\
$\langle \mu_i \rangle$ & 22.01$\pm$0.60 & 21.87$\pm$0.67 & 21.94$\pm$0.63 \\
 &    &   &      \\
$\langle R_{w,g} \rangle$ & 2.10$\pm$0.65 & 1.98$\pm$0.33 & 2.04$\pm$0.51 \\
$\langle R_{w,r} \rangle$ & 2.38$\pm$0.67 & 2.30$\pm$0.49 & 2.34$\pm$0.58 \\
$\langle R_{w,i} \rangle$ & 2.53$\pm$0.75 & 2.40$\pm$0.47 & 2.47$\pm$0.62 \\
 &    &   &      \\
$\langle \psi_g \rangle$ & 7.0$\pm$4.6 & 7.8$\pm$5.6 & 7.4$\pm$5.0 \\
$\langle \psi_r \rangle$ & 6.6$\pm$4.7 & 8.0$\pm$7.9 & 7.3$\pm$6.4 \\
$\langle \psi_i \rangle$ & 7.7$\pm$5.2 & 9.0$\pm$7.5 & 8.3$\pm$6.4 \\
\hline\\
\end{tabular}

\parbox[t]{150mm}{Rows: \\
(1)--(3)  average surface brightness where warp starts, \\
(4)--(6)  average projected distance where warp begins (in units of scale length $h$), \\
(7)--(9)  average warp angle in degrees}
\end{table} 

On average, warps start at a projected distance of $\approx$2.5\,$h$. 
There is a clear dependence of $R_w$ on a passband (Table~\ref{Table6}), but
this dependence just reflects the fact that the scale length depends on the colour. The scale lengths
are larger in the $g$ filter (Sect.~\ref{phot_char}), therefore, $R_w$ values in units
of blue scale lengths are smaller. As for $R_w$ values in absolute units
(in arcseconds or in kpc), warps start from about the same projected distance 
in all three filters: average ratio of the projected distances in the $g$ and $r$ 
passbands is 1.00$\pm$0.09, in the $r$ and $i$ passbands -- 0.99$\pm$0.07.

Projected starting points of stellar warps vary among different
sources. For example, \cite{grijs1997} gives 
$\langle R_w \rangle = (2.1 \pm 1.0)\,h$ in the $I$ filter.
\cite{ap2006} studied 325 edge-on galaxies from the Digitized Sky Survey (DSS) and found
that $\langle R_w \rangle = (0.9 \pm 0.3)\,R_{25}$, where $R_{25}$ is the
semi-major axis of the $\mu_B = 25$ mag/$\Box''$ isophote. For the standard
exponential disc (\citealp{Freeman1970}) with the central surface brightness 
in the $B$ filter of $\mu_0 \approx (21 - 22)$ mag/$\Box''$ this result transforms 
into $\langle R_w \rangle = (2.5 - 3.3)\,h$. For ten galaxies observed
in the {\it Spitzer}/IRAC 4.5-$\mu$m band, the mean warp radius is (4.1$\pm$1.0)\,$h$
(\citealp{saha2009}). Since the disc scale length decreases with 
wavelength \citep[see e.g.][]{degr1998, 1999A&A.344.868X}, 
the last estimate of $R_w$ is in general agreement with the values obtained 
in optical filters.

\begin{figure}
\includegraphics[width=9cm, angle=-90, clip=]{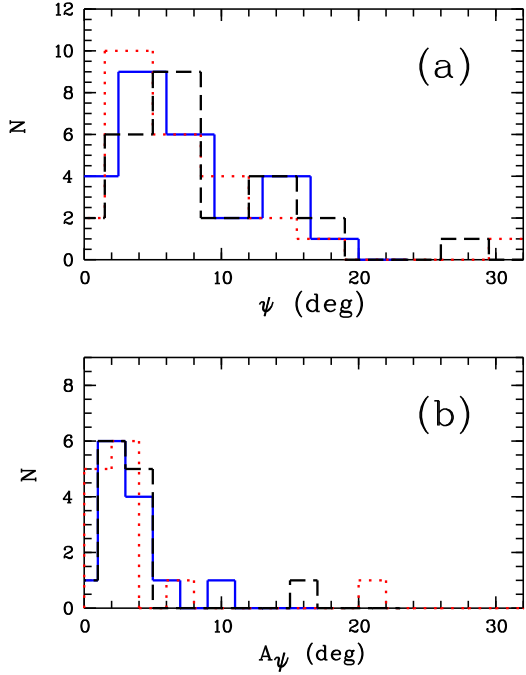}
\caption{Distributions of the sample galaxies over the warp angle $\psi$ (a) 
and over the warp asymmetry $A_{\psi}$ (b). Solid (blue) lines show the 
distributions in the $g$ passband, dotted (red) -- in the $r$, and dashed (black) 
-- in the $i$ bands.}
\label{hist}
\end{figure}

Fig.~\ref{hist}a shows the observed distribution of warp angles $\psi$ in the north
and south sides of galaxies ($\psi_N$ and $\psi_S$ together). As one can see, 
the warp angles have a relatively wide distribution which extends to $\psi \sim 20^{\rm o}$.
Average values of $\psi$ (see Table~\ref{Table6}) are large in comparison with the typical 
values for spiral galaxies (e.g. \citealp{ss1990}, \citealp{rc1998}, \citealp{ap2006}).
Apparently, this can be regarded a consequence of the selection effect, since our sample includes 
galaxies with perceptible warps only.

There is a tendency for stronger warps to start from a closer distance
to the center. This tendency is almost absent in the $g$ band but 
quite pronounced in the $r$ and $i$ filters. For instance, the mean projected
starting point of warp $\langle R_{w,r} \rangle$ for the galaxies with the mean
warp angle $\langle \psi_r \rangle < 7^{\rm o}$ is (2.45$\pm$0.57)\,$h_r$, but
for $\langle \psi_r \rangle > 7^{\rm o}$ this value is (2.10$\pm$0.31)\,$h_r$.
The same values in the $i$-band: $\langle R_{w,i} \rangle = (2.64\pm0.57)\,h_i$
for $\langle \psi_i \rangle < 7^{\rm o}$ and 
$\langle R_{w,i} \rangle = (2.19\pm0.42)\,h_i$ for 
$\langle \psi_i \rangle > 7^{\rm o}$. Probably, the dependence in the $g$ filter 
does not appear due to the influence of the dust extinction on the observable 
structure of edge-on galaxies.

If we exclude two galaxies in which spiral arms can mimick warps (UGC~6882
and SDSS~J153538.63+464229.5), the effect becomes stronger: 

$\langle R_{w,r} \rangle = (2.65 \pm 0.42)\,h_r$ vs. 
$\langle R_{w,r} \rangle = (2.10 \pm 0.31)\,h_r$ \\
($\langle \psi_r \rangle < 7^{\rm o}$ and $\langle \psi_r \rangle > 7^{\rm o}$,
respectively)

and

$\langle R_{w,i} \rangle = (2.79 \pm 0.39)\,h_i$ vs.
$\langle R_{w,i} \rangle = (2.15 \pm 0.47)\,h_i$ \\
($\langle \psi_i \rangle < 7^{\rm o}$ and  $\langle \psi_i \rangle > 7^{\rm o}$,
respectively).

The data presented in Table~\ref{Table6} clearly show that warps 
measured within the certain isophotes are asymmetric for all galaxies in the sample
(partially this may be explained again by the presence of dust lanes).
To quantify the observed asymmetry, we used different asymmetry indexes
(see \citealp{gsk2002}, \citealp{clsb2002}, \citealp{saha2009}).
First two indexes provide asymmetry as the
difference of $\psi$ and $R_w$ values in two sides of galaxies (both
values of $\psi$ must be positive, regardless of the sign in Table~\ref{Table3}):

$A_\psi = |\psi_N - \psi_S|$ (in degrees),

$A_R = |R_{w,N} - R_{w,S}|$ (in units of scale length $h$).\\
Two other indexes give dimensionless measure of warp (both
values of $\psi$ are taken positive):

$\delta_\psi = \frac{|\psi_N - \psi_S|}{\psi_N + \psi_S}$    and
$\delta_R = \frac{|R_{w,N} - R_{w,S}|}{R_{w,N} + R_{w,S}}$.

\begin{table}
 \centering
 \centering
\parbox[t]{150mm} {\caption{Average warps asymmetries in the $g$, $r$, $i$ filters.}
\label{Table7}}
\begin{tabular}{cccc}
\hline 
\hline
 & $g$  & $r$ &  $i$ \\  \hline
$A_\psi$ &  3.3$\pm$2.4 & 4.0$\pm$5.4 & 3.8$\pm$4.0 \\
$A_R$    &  0.45$\pm$0.30 & 0.45$\pm$0.32 & 0.53$\pm$0.33 \\
 &      &     &      \\
$\delta_\psi$ & 0.28$\pm$0.18 & 0.33$\pm$0.24 & 0.30$\pm$0.29 \\
$\delta_R$ & 0.11$\pm$0.07 & 0.10$\pm$0.08 & 0.11$\pm$0.06 \\
\hline\\
\end{tabular}

\parbox[t]{150mm}{Rows: \\
(1)--(2)  absolute asymmetries of warp ($A_\psi$ in degrees, 
$A_R$ in units \\ of scale length $h$), \\
(3)--(4)  relative warp asymmetries.}
\end{table} 

Table~\ref{Table7} presents average values of the warp asymmetries. As can be
seen in the table, typical asymmetries reach several degrees 
in $\psi$ (see also Fig.~\ref{hist}) and about 50 per cent of the exponential scale length in $R_w$.

\begin{figure}
\includegraphics[width=4.7cm, angle=-90, clip=]{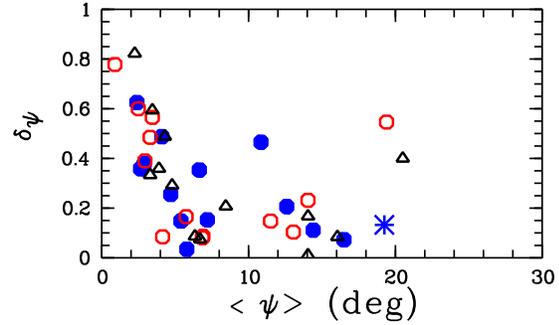}
\caption{Relative warp asymmetry $\delta_{\psi}$ versus mean warp angle
$\langle \psi \rangle$. Solid (blue) circles -- $g$ filter, 
open (red) cirles -- $r$ filter, open triangles -- $i$ filter. Blue asterisk
shows characteristics of the famous ``integral sign'' galaxy UGC~3697 (\citealp{ann2007}).}
\label{asym}
\end{figure}

Fig.~\ref{asym} shows dependence of the relative asymmetry $\delta_{\psi}$ 
on the average warp angle. As one can see, for larger mean warp angles 
$\langle \psi \rangle$ we have less 
relative asymmetry $\delta_{\psi}$ (Fig.~\ref{asym}). This anti-correlation was noted
earlier by \citet{clsb2002} (see fig.~1i in their work). Most likely,
this dependence does not have physical sense and just reflects the definition
of $\delta_{\psi}$: assuming that $A_\psi = |\psi_N - \psi_S|$ is restricted
within the narrow region, we can expect that 
$\delta_{\psi} \propto \langle \psi \rangle^{-1}$. (In fact, for most galaxies 
$A_{\psi} \leq 5^{\rm o}$ -- see Fig.~\ref{hist}b.) 

\subsection{Dark haloes and optical warps}
\label{halo_char}
According to most theoretical models, dark haloes play a significant or even decisive
role in the fate of warps. In order to verify the effect of dark halo, we
calculate the ratio of the dynamical mass of our galaxies to their
stellar mass. The dynamical mass is defined as the total mass enclosed within
the sphere of radius $r = 4\,h_r$: M$_{tot} = 4\,h_r\,V_{2.2}^2/G$. 
The stellar mass was estimated according to \citet{bell2003} using the $g - r$
colours and $r$ band luminosities of galaxies. The total (dynamical)
mass includes the stellar mass of the galaxy (M$_*$) and the dark halo (M$_h$).
Therefore, M$_{tot}$/M$_*$ = (M$_h$ + M$_*$)/M$_*$ = M$_h$/M$_*$ + 1.

\begin{figure}
\includegraphics[width=12cm, angle=-90, clip=]{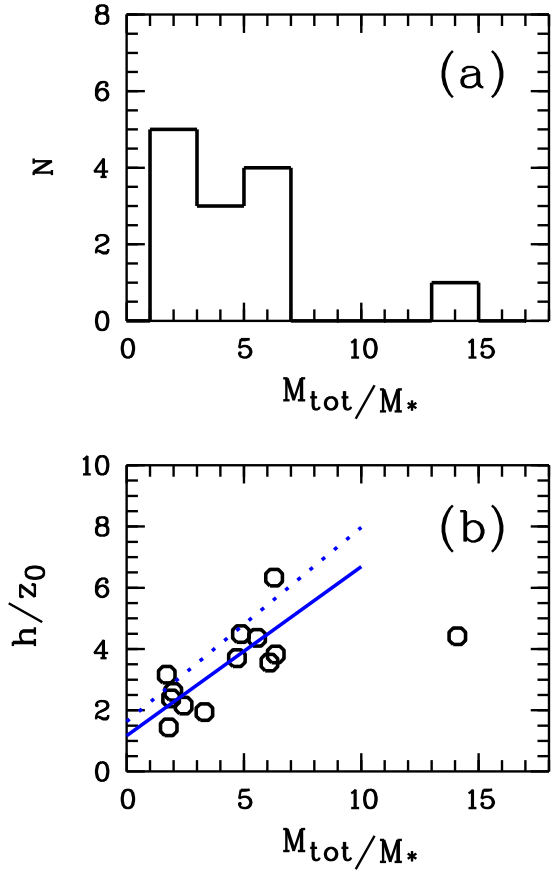}
\caption{(a) Distribution of the sample galaxies over the ratio of dynamical
mass to stellar mass.
(b) The ratio of $h/z_0$ in the $r$ filter as a function of the ratio
of dynamical mass to stellar mass. Solid line shows linear regression for
the sample galaxies (excluding MCG~+06-22-041), dotted line represents the same
relation in the $J$ filter according to \citet{msr2010}.}
\label{hz}
\end{figure}

Most of the galaxies in the sample (12 of 13) show the M$_{tot}$/M$_*$ ratio 
in the range from 1.7 to 6.4 (Fig.~\ref{hz}a) with the mean value 
$\langle$M$_{tot}$/M$_* \rangle = 3.9 \pm 1.9$. These values are in 
agreement with what was obtained by \citet{msr2010} for the sample of 
edge-on spirals with $JHK$ photometry (see fig.~13a in their work). 
(MCG~+06-22-041 is the most dark-matter dominated galaxy in the sample with
M$_{tot}$/M$_* \approx 14$. Rotation curve of this galaxy steeply rises up
to the last observational point (Fig.~\ref{rc}). Also, MCG~+06-22-041 has unusually blue
optical colours -- see Table~\ref{Table1}.) Fig.~\ref{hz}b illustrates that the sample galaxies 
follow a trend for the disc flattening with the relative contribution of 
the dark halo -- more dark-dominated galaxies are, on average, 
thinner (e.g. \citealp{zasov2002}).

\begin{figure}
\includegraphics[width=12cm, angle=-90, clip=]{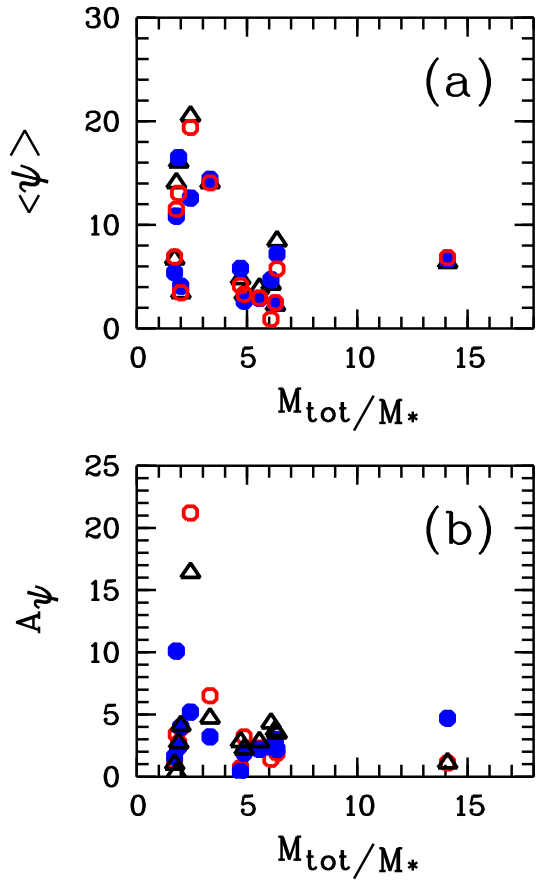}
\caption{Mean warp angle $\langle \psi \rangle$ (a) and asymmetry of warp 
$A_\psi = |\psi_N - \psi_S|$ (b) versus the ratio of dynamical mass to stellar mass. 
Symbols as in Fig.~\ref{asym}. $\langle \psi \rangle$ and $A_\psi$ values are in degrees.}
\label{am}
\end{figure}

We have investigated possible correlations of the warp parameters with
M$_{tot}$/M$_*$ ratio. The most promising trends we found are shown in Fig.~\ref{am}a,b.
As one can see that the mean warp angle $\langle \psi \rangle$ and the 
warp asymmetry $A_\psi$ both decrease with the rise of the dark halo contribution. 
Large and strongly asymmetric warps are more common among 
galaxies with relatively less massive haloes. (Several related correlations
were presented in \citealp{clsb2002}.)

\section{Conclusions}
\label{Conclusions}
We have performed detailed photometric and kinematic study of the 13
edge-on spiral galaxies with warped stellar discs. The galaxies were selected
solely on the basis of their optical morphology and, therefore, they are {\it a priori} 
with notable integral-shape warp. As it turned out, most of them reside 
in dense spatial environments -- in pairs, groups, clusters -- and, hence,
tidal interaction  (current or past) with companions  may be possible
mechanism for the origin of stellar warps in our sample.

Our main conclusions are as follows:

(i) The sample galaxies demonstrate wide distribution of the warp angles $\psi$,
with maximum values of $\psi \sim 20^{\rm o}$ (Fig.~\ref{hist}a).

(ii) On average, stellar warps start at a projected distances 
of (2--3)\,$h$, i.e. near or just beyond the maximum of the rotation 
curve of a self-gravitating exponential disc (Table~\ref{Table6}).

(iii) Stronger warps have on average a smaller projected starting
point (Sect.~\ref{warp_char}).

(iv) Warps show notable asymmetry in all the sample galaxies. Typically, 
asymmetry reaches several degrees in $\psi$, about 50 per cent of the exponential 
scale length in $R_w$, and about 0.5 mag in $\mu_w$ (Tables~\ref{Table3} and \ref{Table7}). 

(v) Apparently, as the dark halo becomes more and more massive compared 
to the stellar disc, it prevents the formation of very strong and 
asymmetric warps (Fig.~\ref{am}).

In order to investigate the formation of stellar warps, \citet{kim2014} recently
presented a set of $N$-body simulations of fly-by encounters
between galaxies. They have found that fly-bys can excite integral-shape
warps in galaxies, and such induced warps can survive for a few billion years.
In \citet{kim2014} simulations, the maximum warp angle reaches about
25$^{\rm o}$, and warps are often non-symmetric. These results are quite 
consistent with our observational data.
Significant asymmetry of the projected starting points of warps (Sect.~\ref{warp_char})
could probably arise due to the underlying asymmetry in the dark halo
potential (\citealp{saha2009}), which could be induced by the galaxy-galaxy
interactions.

Our results are based on a small sample of galaxies, and they can only 
be regarded as preliminary. Also, these results can be biased due
to the misidentification of strongly inclined spirals as warped stellar
discs.  \cite{rc1998} simulated this projection effect and found that no
more than $\approx$15\% of integral-shaped warps could actually be spiral arms.
Careful study of the sample galaxies showed that, with a certain probability, in 
2 of 13 galaxies the global warp can be explained as inclined spiral arms
(these galaxies are UGC~6882 and SDSS~J153538.63+464229.5). Thus, the fraction
of possible false warps in the sample is in general agreement with the 
\cite{rc1998} results. We verified the position of the studied characteristics of 
these two galaxies in our figures and found that they do not bias our
conclusions.

A further extended studies of warped galaxies 
in different spatial environments 
will help us to better understand this common but still puzzling phenomenon.

\section*{Acknowledgments}

The observations at the 6-meter BTA telescope were carried out with the financial support of the Ministry of Education and Science of the Russian Federation (agreement No. 14.619.21.0004, project ID RFMEFI61914X0004).
This work was partly supported by the Russian Foundation for Basic Researches (grant number 14-02-810).
Aleksandr  Mosenkov is a beneficiary of a mobility grant from the Belgian Federal Science Policy Office. Alexei Moiseev 
is grateful for the financial support a  grant  from the President of the Russian Federation (MD-3623.2015.2). 
We thank  Sergey Dodonov,  Dmitri Oparin, Roman Uklein and Oleg Egorov, who performed the significant part of  SCORPIO/SCORPIO-2 observations and especially Victor Afanisiev  for his great contribution to spectroscopy at the 6-m telescope. Also we thank Tatiana Briukhareva for her help in spectral data reduction.
We are thankful to the referee for useful comments.

Funding for the SDSS has been provided by the Alfred
P. Sloan Foundation, the Participating Institutions, the
National Science Foundation, the U.S. Department of Energy,
the National Aeronautics and Space Administration, the Japanese Monbukagakusho, the Max Planck Society, and
the Higher Education Funding Council for England. The
SDSS Web Site is http://www.sdss.org/.

We acknowledge the usage of the HyperLeda database.
This research has made use of the NASA/IPAC Extragalactic Database (NED) which is operated by the Jet Propulsion Laboratory, California Institute of Technology, under
contract with the National Aeronautics and Space Administration.

\bibliographystyle{mn2e}
\bibliography{art}

\label{lastpage}

\end{document}